\documentstyle[12pt]{article}
\def\boxriv#1#2#3#4#5{}
\newskip\humongous \humongous=0pt plus 1000pt minus 1000pt
  \newif\ifdtup

\def\ltap{\raisebox{-.4ex}{\rlap{$\sim$}} \raisebox{.4ex}{$<$}}

\def\frac#1#2{ {{#1} \over {#2} }}

\def\abar{{\bar \alpha_S}}
\def\abn{{{\bar \alpha_S} \over N}}
\def\as{\alpha_S}
\def\msbar{\overline{\mbox {\rm MS}}}
\def\kper{k_{\perp}}
\def\cf{{\cal F}}

\def\ga{\gamma}
\def\si{\sigma}
\def\al{\alpha}
\def\ep{\epsilon}
\def\bk{{\mbox{\bf k}}}
\def\bq{{\mbox{\bf q}}}

\def\rt){\right)}
\def\lt({\left(}
\def\rq]{\right]}
\def\lq[{\left[}

\def\figcap{\section*{Figure Captions\markboth
        {FIGURECAPTIONS}{FIGURECAPTIONS}}\list
        {Figure \arabic{enumi}:\hfill}{\settowidth\labelwidth{Figure
999:}
        \leftmargin\labelwidth
        \advance\leftmargin\labelsep\usecounter{enumi}}}
 \relax
\def\np#1#2#3{Nucl.\ Phys.\ B#1 (19#3) #2}
\def\pl#1#2#3{Phys.\ Lett.\ #1B (19#3) #2}

\def\prep#1#2#3{Phys.\ Rep.\ #1 (19#3) #2}
\def\prl#1#2#3{Phys.\ Rev.\ Lett.\ #1 (19#3) #2}

\catcode`\@=11
\def\ps@headings{\def\@oddfoot{}\def\@evenfoot{}
\def\@oddhead{\hbox{}\hfill --\thepage{}-- \hfill}
\def\@evenhead{\@oddhead}
\def\subsectionmark##1{\markboth{##1}{}}
}

\ps@headings

\catcode`\@=12

\relax

\relax
\newcounter{sect}
\renewcommand{\theequation}{\thesect.\arabic{equation}}
\begin{document}

\renewcommand{\thefootnote}{\fnsymbol{footnote}}

\begin{titlepage}
\begin{flushright}
     Cavendish-HEP-94/01 \\   April 1994
     \end{flushright}
\vspace*{5mm}
%\par \vskip 10mm
\begin{center}
{\Large \bf \boldmath High-Energy Factorization and \\
Small-$x$ Deep Inelastic Scattering \\
Beyond Leading Order\footnote{Research supported in part
by the EC Programme ``Human Capital and Mobility", Network ``Physics at High
Energy Colliders", contract CHRX-CT93-0537 (DG 12 COMA).}} \\
%\end{center}
\vspace*{1cm}

        \par \vskip 5mm \noindent
        {\bf S. Catani}\\
        \par \vskip 3mm \noindent
        INFN, Sezione di Firenze\\
        Largo E. Fermi 2, I-50125 Florence, Italy\\
        \par \vskip 5mm \noindent
        and \\
        \par \vskip 5mm \noindent
        {\bf F. Hautmann}\\
        \par \vskip 3mm \noindent
        Cavendish Laboratory\\
        Department of Physics, University of Cambridge\\
        Madingley Road, Cambridge CB3 0HE, UK\\

\par \vskip 1cm

\end{center}
%\par \vskip 2mm
\vspace*{1cm}

\begin{center} {\large \bf Abstract} \end{center}
\begin{quote}
High-energy factorization in QCD is investigated beyond leading order
and its relationship to the factorization theorem of mass singularities
is established to any collinear accuracy. Flavour non-singlet observables are
shown to be regular at small $x$ order by order in perturbation theory. In the
singlet sector, we derive the relevant master equations for the space-like
evolution of gluons and quarks. Their solution enables us to sum
next-to-leading corrections to the small-$x \, $ behaviour of
quark anomalous dimensions and deep inelastic scattering coefficient functions.
We present results in both $ \msbar $ and DIS factorization schemes.
\end{quote}
\vspace*{2cm}
\end{titlepage}

\setcounter{equation}{0}
\setcounter{sect}{1}
\setcounter{footnote}{0}

\noindent {\bf 1. Introduction  }
\vskip 0.3 true cm

Hadronic processes at large transferred momentum $p_{t}$ are
accurately investigated by using perturbative QCD. The comparison
between theoretical predictions and jet physics data from
high-energy colliders has enhanced our confidence in perturbative
QCD up to the 10\% accuracy level [\ref{EPS}]. The reason for this
success is that not only the non-perturbative (higher-twist)
contributions vanish as powers of $\Lambda /p_{t}$ in the
hard-scattering regime $p_{t} \gg \Lambda$ ($\Lambda$ being the QCD
scale), but mainly the fact that logarithmic corrections to the
na\"{\i}ve parton model (i.e. lowest-order perturbative QCD) are
systematically computable and, in most  cases, known as a power
series expansion in the `small' (due to asymptotic freedom) running
coupling $\as (p_{t}^{2}) \sim (\beta_{0} \,\ln\,
p^{2}_{t}/\Lambda^{2})^{-1}$.

Precise quantitative tests of QCD and searches for new physics at
present and future hadron colliders are nonetheless carried out at
an increasingly large centre-of-mass energy $\sqrt{S}$, thus
opening up a new kinematic region characterized by small values
of the ratio $x = p^{2}_{t}/S \,$ ($S \gg p^{2}_{t}$). In this
small-$x$ regime, our capability to make perturbative QCD
predictions decreases. The reason is twofold. First, the parton
densities $f(x,p^{2}_{t})$ of the incoming hadrons (which are
necessary as inputs in any perturbative calculation) are poorly
known at very small values of $x$. Second, the perturbative
expansion is slowly (or badly) convergent because of the presence
of logarithmic corrections of the type $\as^{n} (p^{2}_{t})
\ln^{m}x$: higher powers of $\as (p^{2}_{t})$ associated with multiple
hard-jet production can  indeed be compensated by large enhancing
factors ${\ln} (S/p^{2}_{t})$. Note that this second point also
affects the first one. In the case that a reliable set of small-$x$
parton densities can be extracted from a single experiment, we
should be able to predict accurately their perturbative evolution
with $p^{2}_{t}$.

The discussion above makes clear that our understanding of QCD in
the small-$x$ regime has still to be improved. In particular, on
the theoretical side, it is highly desirable to keep under
control and to evaluate reliably the QCD component which can be
computed perturbatively. This can be done by combining the
customary perturbative approach to hard-scattering processes with
an improved perturbative expansion which systematically sums
classes (leading, next-to-leading, and so on) of small-$x$ logarithmic
contributions to all orders in $\as$.

At present, the QCD multiparton matrix elements have been computed
to double logarithmic accuracy in the small-$x$ region
[\ref{FAD}-\ref{Muel}]. They can be used to study the structure of
hadronic final states in small-$x$ processes, thus predicting some
new distinctive features [\ref{Ciaf}-\ref{CFM}] such as the increase of
particle (jet) multiplicity and the suppression of large rapidity
gaps. Only some phenomenological investigations have been carried
out [\ref{MW}], and more detailed analyses are needed. In
this paper we do not consider the issue of the structure of the
final states but rather  concentrate on the evaluation of
higher-order corrections for total cross sections.

The leading high-energy contributions to total cross sections are
single-logarithmic terms $(\as {\ln} \, x)^{n}$ (higher powers
of $\ln x$ cancel in this case) due to multiple gluon exchanges in
the $t$-channel. In the case of hard processes which are directly
coupled to gluons in the na\"{\i}ve parton  model, these leading
logarithmic contributions can be resummed to all orders in
perturbation theory by using the high-energy or
$k_{\perp}$-factorization theorem [\ref{CCH}-\ref{LRSS}]. The basic
idea [\ref{CCH}] is to replace the collinear (or parton pole) factorization
by gluon Regge pole factorization.

Considering, for
instance, the simplest small-$x$ process initiated by hard-gluon
interactions at the Born level, namely the heavy-flavour
photoproduction process
\begin{equation}
\label{phot}
\gamma(p_1) + h(p) \to Q(p_3) + {\bar Q}(p_4) + X \;\;,
\end{equation}
one can write the total cross section in the following factorized
form [\ref{CCH}]
\begin{equation}
\label{ktfac}
4 M^2 \;\si(x, M^2) = \int d^2\bk \int_x^1 {{dz} \over z}
\; {\hat \si}(x/z, \bk^2/M^2,  \as(M^2)) \;{\cal F}(z, \bk) \;\;.
\end{equation}
Here the heavy flavour mass $M \,$ ($M \gg \Lambda$)  defines the hard
scale of the process and $x\equiv 4M^{2}/S \,$ ($S \simeq 2 p_{1} \cdot p
\gg M^{2}$).

In Eq.~(\ref{ktfac})  $\hat\si$ is the basic
high-energy hard cross section for the subprocess $\gamma + g(k)
\rightarrow Q {\bar Q}$, computed to the lowest order in $\as$ as a
function of the transverse momentum $\bk$ of the incoming off-shell
(essentially transverse $k \simeq z p + k_{\perp}, \, k^{2}\simeq -
\bk^{2}$) gluon $g(k)$. On the other hand, ${\cal F}(z, \bk)$ is
the unintegrated gluon density of the incoming hadron $h(p)$ and
is related to the customary gluon density $f_{g}(x,M^{2})$ via
$k_{\perp}$-integration
\begin{equation}
\label{xfg}
x f_{g}(x,M^{2}) \sim \int_{0}^{M^{2}} d^{2} \bk \;\;{\cal F}(x, \bk) \;\;.
\end{equation}
Therefore the $\kper$-dependent factorization in
Eq.~(\ref{ktfac}) reduces to the leading-order collinear
factorization [\ref{CSS}] for $S \gg M^{2} \gg k_{\perp}^{2}$.
However it holds also for $S \gg k_{\perp}^{2} \sim M^{2}$, thus
controlling all the logarithmically-enhanced terms $(\as\ln
x)^{n}$ associated with hard-gluon radiation in the final state.

The resummation of the leading $\ln x$-contributions follows from
noticing that the hard cross section $\hat \si$ is well-behaved
at high energy (i.e. ${\hat \si}(x, \bk^{2}/M^{2}, \as) \sim x$
modulo $\ln x$-terms, for $x\rightarrow 0$). Therefore the large
perturbative corrections $(\as\ln x)^{n}$ in the cross section
(\ref{ktfac})
are generated precisely by the $k_{\perp}$-integration from the
ones in the gluon density  ${\cal F}(x, \bk)$, as given by the
Balitskii-Fadin-Kuraev-Lipatov (BFKL) equation [\ref{BFKL}]
\begin{equation}
\label{calF}
{\cal F}(x, \bk) \sim \frac{1}{2\pi} \, e^{-\lambda \,\ln x} \,
(\bk^{2})^{-\frac{1}{2}} \;\; ,
\end{equation}
\begin{equation}
\label{x=4}
\lambda=4 \, C_{A} \, \frac{\as}{\pi} \, \ln 2 \;\; .
\end{equation}
Inserting Eq.~(\ref{calF}) into Eq.~(\ref{ktfac}), one obtains
the following perturbative result for the total cross section
at very high energy [\ref{CCH}]
\begin{equation}
\label{4M2}
4M^{2} \si(x,M^{2}) \sim x^{-\lambda} \;(M^{2})^{\frac{1}{2}}
\;\;h(1/2) \;\;,
\end{equation}
\begin{equation}
\label{h(1/2)}
h(1/2) \equiv \frac{1}{2} \int_{0}^{\infty}
\frac{d \bk^{2}}{\bk^{2}} \left( \frac{\bk^{2}}{M^{2}} \right)
^{\frac{1}{2}} \int_{0}^{1} \frac{dx}{x} \;\;{\hat\si}(x, \bk^2/M^2,
%\frac{\bk^{2}}{M^{2}}
\as) \;\;.
\end{equation}

The main features of Eqs.~(\ref{4M2}), (\ref{h(1/2)}), derived from
the $\kper$-factorization formula (\ref{ktfac}), are the
following. The total cross section increases at high energy
with a universal (process-independent) power behaviour
$S^{\lambda}$, $ \, \lambda $ being the intercept of the
perturbative QCD pomeron in Eq.~(\ref{x=4}).
This result is the consequence of the very steep behaviour
(\ref{calF}) of the gluon density\footnote{Note, however, that
if the gluon density has a non-perturbative component steeper
than the perturbative one, the former dominates over the latter at high
energy.} at small $x$ and large $\kper$. The normalization
of the total cross section instead depends on the process and
the process-dependent factor $h(1/2)$ derives from the
detailed and calculable transverse-momentum dynamics of the
hard subprocess.

The resummation of the leading-order contributions at high
energy is a crude (although mandatory) approximation. This is
true from both the theoretical (the perturbative QCD pomeron
violates unitarity to leading order) and phenomenological
(terms of relative order $\as$ are systematically neglected)
sides. The evaluation of subleading contributions is therefore
relevant $i)$ to include corrections necessary to restore unitarity at
asymptotic energies and $ii)$ to estimate the accuracy and  set
the limits of applicability of the leading-order formalism.

Unitarization effects have been extensively studied in
Refs.~[\ref{uni1}-\ref{uni4}]. Although a systematic calculational approach
based on first principles is still missing,
the likely
conclusion emerging from these studies is that the full
restoration of unitarity can be achieved only after the
inclusion of higher-twist corrections.

As regards the leading-twist contributions to hard-scattering
processes, a calculational program of high-energy logarithms
based on Regge behaviour is being pursued by Fadin and Lipatov
[\ref{FL}], and the evaluation of the two-loop correction to
the BFKL kernel now seems feasible.

In this paper we follow a different approach towards the
computation of subleading corrections at high energy. We show
how the $\kper$-factorization theorem can be extended beyond
leading order in a consistent way with {\em all-order}
(leading-twist) collinear factorization. This allows us to set up
a systematic logarithmic expansion both for hard coefficient
functions and parton anomalous dimensions. Moreover, once these
quantities have been computed to a certain logarithmic
accuracy, they can  unambiguously be supplemented with
non-logarithmic (finite-$x$) contributions exactly calculable
to any fixed order in perturbation theory. Note, also, that
a further advantage of this approach is that the effects of
the running  coupling  can be included exactly (at
least in principle), thus avoiding the infrared instabilities
encountered in phenomenological attempts [\ref{uni1},\ref{CK},\ref{Dur}]
to extend
the BFKL equation beyond leading order. Obviously, with the
present attitude, we abandon any demand to predict the
{\em absolute} behaviour of the cross sections at very high energies
(very small $x$) because higher-twist corrections are
systematically neglected. However, since logarithmic scaling violations
are systematically under control, we think that such an
approach can be useful in making quantitative phenomenological
predictions at high (but finite) energies and large transferred
momenta [\ref{EKL}].

The outline of the paper is as follows. In Sect.~2 we first recall the general
framework of leading-twist collinear factorization. Then, on the basis of
power counting arguments, we show how the high-energy factorization
of Ref.~[\ref{CCH}] can be extended beyond leading order and consistently
matched with all-order collinear factorization, in terms of resummed
anomalous dimensions and coefficient functions. In Sect.~3 we start our
calculational program in dimensional regularization by deriving the master
equation for the high-energy behaviour of the gluon forward scattering
amplitudes and computing the ensuing anomalous dimensions and normalization
factor in the (modified) minimal subtraction scheme. The analogous calculation
for the quark channel is performed in Sect.~4. Here, we obtain an algebraic
equation for the quark anomalous dimensions to next-to-leading logarithmic
accuracy $\as (\as \ln x)^n$, and present its explicit solution up to the
six-loop order. In Sect.~5 we turn our attention to the calculation of the
coefficient functions for deep inelastic lepton-hadron scattering. We compute
both the longitudinal and transverse coefficient functions by resumming
the logarithmic contributions $\as (\as \ln x)^n$. We also consider the
all-order generalization of the DIS factorization scheme and obtain explicit
resummed expressions for the corresponding quark anomalous dimensions with
next-to-leading logarithmic accuracy. Section 6 is devoted to summarizing our
main results,
and may also serve as a guide  for the reader mostly interested
in extracting perturbative QCD results for phenomenological applications.
The definition of singlet and non-singlet parton densities and
a few mathematical details are left to appendices A to C.

Some
of the results derived in this paper have already been presented
in Refs.~[\ref{HEF},\ref{QAD}].

\vskip 1.5 true cm

\setcounter{equation}{0}
\setcounter{sect}{2}
\setcounter{footnote}{0}

\noindent {\bf 2. QCD factorizations }
\vskip 0.3 true cm
%\ref{QCDf}

This section is devoted to setting up the formal basis of the
high-energy factorization. We start in Subsect.~2.1 by introducing the
factorization theorem of collinear singularities and defining the parton
densities and coefficient functions. Then in Subsect.~2.2 we recall the proof
of this factorization theorem to leading-twist order in the context of
dimensional regularization. This formal apparatus is used in Subsect.~2.3 to
develop a simple power counting at high energy and to show how high-energy
factorization can be carried out consistently with all-order collinear
factorization. The high-energy factorization formulae are $\kper$-dependent
and can be considered as the generalization of collinear factorization in
terms of {\em unintegrated} parton densities and {\em off-shell} coefficient
functions. Alternatively, high-energy factorization can be compared with
Regge factorization [\ref{Regge}]. The gluon Green functions and the two-gluon
irreducible kernels entering as building blocks in the high-energy
factorization discussed in Subsect.~2.3 are the analogue of the reggeon
trajectory and its residue. However, this analogy has to be taken with caution
since it does not properly
account for the
 issue  of
 factorization of
collinear singularities. Finally, in Subsect.~2.4, we show how the high-energy
factorization leads to resummed anomalous dimensions and coefficient functions.

\vskip 0.5 true cm
\noindent {\it 2.1 Hard processes and parton densities}
\vskip 0.2 true cm

The factorization theorem of collinear (mass) singularities
%\cite{old,CFP,CSS}
[\ref{old},\ref{CFP},\ref{CSS}]
states that, in a general hard collision
(i.e. a scattering process involving a large transferred
momentum $<p_{t}^{2}> = Q^{2} \gg \Lambda^{2}$) of incoming
hadrons, all long-distance (non-perturbative) effects can be
factorized into universal (process-independent) parton
densities thus leading to a perturbatively calculable
dependence on the hard scattering scale $Q^{2}$.

Considering, for the sake of simplicity, the case of a single
incoming hadron (like the heavy-flavour photoproduction
process (\ref{phot}) or deep inelastic lepton-hadron scattering),
one can write the dimensionless cross section $F(x,Q^{2}) \sim
Q^{2} \si (x,Q^{2})$ as follows $(x \equiv Q^{2}/S)$
\begin{equation} F(x,Q^{2}) = \sum_{a} \int_{x}^{1} dz \;
C_{a}(x/z; \as (\mu^{2}_{F}),  Q^{2}/\mu^{2}_{F}) \;
f_{a}(z, \mu^{2}_{F}) \;\;.
\label{F(x}
\end{equation}
Here $C_{a}$ are the process-dependent coefficient functions,
$f_{a}$ are the parton densities ($a=q_{i}, {\bar q}_{i}, g, \; i=1,
\dots , N_{f}, \; N_{f}$ being the number of flavours) of the
incoming hadron and $\mu^{2}_{F}$ is an arbitrary factorization
scale such that $\mu^{2}_{F} \gg \Lambda^{2}$. The observable
$F$ is independent of $\mu^{2}_{F}$.
Correspondingly the
$\mu^{2}_{F}$-dependence of $C_{a}$ on the r.h.s. of
Eq.~(\ref{F(x}) is  exactly cancelled by that of $f_{a}$.
Moreover, both the $\as$-dependence of the coefficient functions
and the scale dependence of the parton densities are computable
as a power series expansion in $\as$. In particular, the parton
densities fulfil the renormalization group evolution equations
\begin{equation}
\frac{d \, f_{a}(x, \mu^{2})}{d \ln \mu^{2}} =  \sum_{b} \int_{x}^{1}
\frac{dz}{z} P_{ab}(\as (\mu^{2}), z) \;f_{b}(x/z,\mu^{2}) \;,
\label{dfa}
\end{equation}
where $P_{ab} (\as ,z)$ are generalized Altarelli-Parisi
splitting functions which are computable in QCD perturbation
theory:
\begin{equation}
P_{ab} (\as ,x) = \sum^{\infty}_{n=1} \left( \frac{\as}{2\pi}
\right)^{n} P_{ab}^{(n-1)}(x)\;.
\label{Pab}
\end{equation}
The corresponding expansion for the coefficient function is
\begin{equation}
C_{a} (x; \as , Q^{2}/\mu^{2}) = (\as)^{p} \left[ C_{a}^{(0)}(x) +
\sum^{\infty}_{n=1} \left( \frac{\as}{2\pi}
\right)^{n} C_{a}^{(n)} (x; Q^{2}/\mu^{2}) \right] \;\;,
\label{Ca}
\end{equation}
where the integer power $p$ depends on the process.

The $x$-dependence of the factorization formula (\ref{F(x}) and
of the evolution equations (\ref{dfa}) can be diagonalized by
introducing the $N$-moments of the cross section
\begin{equation}
F_{N} (Q^{2}) = \int^{1}_{0} dx \;x^{N-1} F(x,Q^{2})
\label{FN}
\end{equation}
and the analogous moments of any other function of $x$. It is
also convenient to define the rescaled parton densities $\tilde
f_{a}$
\begin{equation}
{\tilde f}_{a} (x,\mu^{2}) \equiv x f_{a} (x,\mu^{2}) \;\;,
\label{tildef}
\end{equation}
and the anomalous dimension matrix $\gamma_{ab,N}$ \footnote{Our
definition differs from the standard one in which $\gamma_{N} =
P_{N}$.}
\begin{equation}
\gamma_{ab,N}(\as ) \equiv \int^{1}_{0} dx \;x^{N} \;P_{ab}(\as ,x) =
P_{ab,N+1}(\as )\;\;.
\label{gamma}
\end{equation}
In the $N$-moment space, Eqs.~(\ref{F(x}),(\ref{dfa}) respectively
read as follows
\begin{equation}
F_{N}(Q^{2}) = \sum_{a} \, C_{a,N}( \as (\mu^{2}_{F}),
Q^{2}/\mu^{2}_{F} ) \;{\tilde f}_{a,N}(\mu^{2}_{F}) \;\;,
\label{FN2}
\end{equation}
\begin{equation}
\frac{d \, {\tilde f}_{a,N} (\mu^{2})}{d \ln \mu^{2}} = \sum_{b}
\gamma_{ab,N}(\as (\mu^{2})) \, {\tilde f}_{b,N}(\mu^{2}) \;\;.
\label{fracd}
\end{equation}

Eq.~({\ref{F(x}}) is correct apart from terms vanishing as inverse
powers of $Q^{2}$ at high $Q^{2}$ (higher-twist corrections). In
this regime, a consistent perturbative use of the collinear
factorization formula (\ref{F(x}) requires the knowledge of
$\{P_{ab}^{(0)}, C_{a}^{(0)}\}$ (leading order),
$\{P_{ab}^{(1)}, C_{a}^{(1)}\}$ (next-to-leading order) and so on.
Therefore, QCD predictions for the cross section are usually
obtained by computing in fixed-order perturbation theory both the
splitting functions and the coefficient functions. The splitting
functions up to two-loop accuracy have been known for a long time
[\ref{CFP}-\ref{Floratos}].
Next-to-leading order coefficient
functions have been computed for most
processes
[\ref{Stirl}],
%\cite{stirl},
and, in the cases of deep inelastic lepton-hadron
scattering (DIS)
%\cite{DIS}
[\ref{NDIS}] and Drell-Yan process
%\cite{DY}
[\ref{DY}], the next-to-next-to-leading terms $C_{a}^{(2)}$ are also known.

As discussed in Sect.~1,
% \ref{intro},
higher-order contributions to
the cross section and, hence, to  splitting and coefficient functions
are logarithmically enhanced at small $x$. More precisely, in the
small-$x$ limit we have \footnote{According to our definition, the
lowest-order coefficient function $C^{(0)}$ is normalized in such
a way that $C^{(0)}(x) \sim x$, modulo $\ln x $ corrections, at
small $x$.}
\begin{equation}
P_{ab}^{(n-1)}(x) \sim \frac{1}{x} \left[ \ln^{n-1} x + {\cal O}(\ln
^{n-2}x) \right] \;\;,
\label{Pabn}
\end{equation}
\begin{equation}
C_{a}^{(n)}(x) \sim \ln^{n-1}x+ {\cal O}(\ln^{n-2} x) \;\;.
\label{Can}
\end{equation}
This small-$x$ behaviour corresponds, in $N$-moment space, to
singularities for $N\rightarrow 0$ in the form
\begin{equation}
\gamma_{ab,N} {(\as)} = \sum^{\infty}_{k=1} \left[
\left(\frac{\as}{N}\right)^{k} A^{(k)}_{ab} + \as
\left(\frac{\as}{N}\right)^{k} B^{(k)}_{ab} + \dots \right] \;,
\label{gam}
\end{equation}
\begin{eqnarray}
C_{a,N}(\as , Q^{2}/\mu^{2}) &=& \as^{p} \,C_{a,N=0}^{(0)}
\left[ 1+\sum^{\infty}_{k=1}
\left(
\frac{\as}{N}\right)^{k} D_{a}^{(k)}(Q^{2}/\mu^{2}) \right. \nonumber \\
&+& \left. \as \left(
\frac{\as}{N}\right)^{k} E_{a}^{(k)}(Q^{2}/\mu^{2})
+ \dots \right] .
\label{Ca,N}
\end{eqnarray}
These singularities may spoil the convergence of the perturbative
expansions (\ref{Pab}) and (\ref{Ca}) at small $x$ (high energy).
Nonetheless one can consider an improved perturbative expansion
obtained by resumming the {\em leading}
$(A_{ab}^{(k)},D_{a}^{(k)})$, {\em next-to-leading}
$(B_{ab}^{(k)},E_{a}^{(k)})$,
%$\dots$
etc.,
coefficients in the
high-energy regime. Once these coefficients are known, they can
be combined
%\cite{HERA}
[\ref{HERA}]
with Eqs.~(\ref{Pab}) and (\ref{Ca})
(after subtracting the resummed logarithmic terms in order to
avoid double counting), to obtain a prediction throughout the
region of $x$
where
$\as \ln (1/x)
%\lesssim
\ltap
1 \,$
(or $\as /N
%\lesssim
\ltap
1$),
which is much larger than the domain $\as \ln (1/x) \ll
1$ where the $\as$-perturbative expansions (\ref{Pab}),
(\ref{Ca}) are applicable.

As outlined in Sect.~1
%\ref{intro}
and discussed in detail in
%Ref.~\cite{CCH}
Ref.~[\ref{CCH}]
(see also Sect.~5),
%\ref{cinque}),
the
$\kper$-factorization formula (\ref{ktfac}) allows one to resum
the leading high-energy contributions $(\as \ln x)^{n}$
($(\as /N)^{n}$) both for the anomalous dimensions and for the
coefficient function. By comparing Eqs.~(\ref{ktfac}) and
(\ref{F(x}) and considering the relation (\ref{xfg}) between
unintegrated and full parton densities, we  see that the
$\kper$-factorization is, in a sense, more general than the
collinear factorization in Eq.~(\ref{F(x}). The latter is recovered
after the transverse momentum integration in the former. This
relationship can be made explicit in a simple way to leading order
%all \cite
[\ref{CCH}]. However, beyond leading order, a careful analysis
of the low-$\kper$ integration region in (\ref{ktfac}) is needed in
order to disentangle non-perturbative and higher-twist effects from
the perturbative ones.

This issue can be investigated from a formal viewpoint by
considering cross sections at parton level instead of  hadronic
cross sections. In the partonic cross section, the
non-perturbative contributions show up in terms of collinear
divergences. Once these divergences are properly regularized, they
can be factorized and subtracted by a
procedure
of renormalization
of the bare parton densities. This procedure can be
carried out to any collinear accuracy, although the ensuing
anomalous dimensions and coefficient functions are no longer
separately regularization/factorization scheme independent beyond
one-loop level.

The point we want to address in the following is that the
$\kper$-factorization can be consistently implemented beyond
leading order without spoiling the all-order factorization of
collinear singularities. In particular, the resummation of
next-to-leading contributions at high energy can be performed by
properly taking care of the factorization scheme dependence of the
splitting and coefficient functions. To this end, let us first
recall the formal basis of collinear factorization.

\vskip 1 true cm
\noindent {\it 2.2 Collinear factorization}
\vskip 0.2 true cm

In order to present a formal derivation of the factorization
theorem of collinear singularities, we follow the technique
developed by Curci, Furmanski and Petronzio
%\cite{CFP}.
[\ref{CFP}].
The dimensionless cross section $F(x,Q^{2})$ in eq.~(\ref{F(x}) is
first expressed in terms of partonic cross sections and parton
distributions in the form \footnote{Here and in the following we
use a symbolic notation in which the product $C=AB$ of two kernels
$A$ and $B$ understands the integration over the intermediate
momenta (or the corresponding product in $N$-moment space) and the
sum over the intermediate parton species and their spin and
colour indices.} (Fig.~1)
\begin{equation}
F= F^{(0)}(\dots , p) \;{\tilde f}^{(0)}(p, \dots) + [ F^{(0)}_{4}(\dots
; p_{1},p_{2}) \;{\tilde f}^{(0)}_{4}(p_{1},p_{2}; \dots)+ \dots ] \;\;,
\label{FF0}
\end{equation}
where, in the limiting case of on-shell partons $(p^{2}_{i}=0)$,
the first term on the r.h.s. picks out the leading-twist
contribution we are interested in. Nonetheless, the partonic cross
section $F^{(0)}$ is collinear divergent in the on-shell limit, so
that both $F^{(0)}$ and ${\tilde f}^{(0)}$ have to be regarded as
properly regularized `bare' quantities.

We use the standard procedure of dimensional regularization, in
which the bare cross section $F^{(0)}$ is evaluated in
$n=4+2 \, \varepsilon$ space-time dimensions, considering $(n-2)$
helicity states for gluons and 2 helicity states for quarks. The
corresponding dimensional-regularization scale is denoted by
$\mu$. Once $F^{(0)}$ has been renormalized (using, for instance,
the $\msbar$ renormalization scheme), collinear singularities are
automatically regularized and show up as single poles in $1/
\varepsilon$. The factorization theorem allows one to subtract
these poles from $F^{(0)}$ and factorize them (to all orders in
$\as$) into process-independent transition functions $\Gamma$,
according to
\begin{equation}
F^{(0)}= C \, \Gamma \;\;,
\label{F(0)}
\end{equation}
where the coefficient function $C$ is finite for $\varepsilon
\rightarrow 0$. Using the transition functions $\Gamma$ to define
the `physical' parton densities $\tilde f$
\begin{equation}
{\tilde f} = \Gamma {\tilde f}^{(0)} \;,
\label{tilf}
\end{equation}
one then recovers the factorization formula (\ref{FN2}) by
performing the limit $\varepsilon \rightarrow 0$.

The factorization procedure leading to eq.~(\ref{F(0)}) is
simplified if we evaluate the gauge-invariant partonic cross
section $F^{(0)}$ in a physical gauge. Denoting by
%$p^{\mu} = P(1,\underline{0}, 1)$
$p^{\mu} = P(1,{\mbox {\bf 0}}, 1)$
the incoming parton momentum (Fig.~2), we
introduce the following Sudakov parametrization for any other
momentum $k$
\begin{equation}
k^{\mu} = z p^{\mu} + \kper^{\mu} + \frac{k^{2}+
{\bk}^{2}}{z} \frac{{\bar p}^{\mu}}{2p\cdot {\bar p}} \;\;\;, \;\;
\kper^{\mu} = (0,{\bk},0) \;\;\;, \;\;  {\bar p}^{\mu} = P(1,
{\mbox {\bf 0}}, -1) \;\;,
\label{kmu}
\end{equation}
and we choose the axial gauge ${\bar p}\cdot A=0$, where the sum over the
gluon helicities is given by the polarization tensor
\begin{equation}
d^{\mu \nu}(k) = - g^{\mu \nu} +
\frac{k^\mu {\bar p}^\nu + {\bar p}^\mu k^\nu}{{\bar p}\cdot k} \;\;.
\label{polten}
\end{equation}
Then we consider
the expansion of $F^{(0)}$ in terms of kernels $C^{(0)}$ and
$K^{(0)}$ which are two-particle irreducible (2PI) in the
$t$-channel:
\begin{equation}
F^{(0)} = C^{(0)} (1+K^{(0)}+K^{(0)}K^{(0)}+\dots) \equiv C^{(0)}
\;{\cal G}^{(0)}\;,
\label{F0}
\end{equation}
\begin{equation}
{\cal G}^{(0)} = 1+K^{(0)}+K^{(0)}K^{(0)}+\dots = \frac{1}{1-K^{(0)}}\;.
\label{G(0)}
\end{equation}
In the axial gauge in which we are working, the 2PI amplitudes are
free from mass singularities [\ref{2pi}]. Therefore all the collinear
divergences originate from the integrations over the momenta coming
out from the kernels $K^{(0)}$ and connecting them to each other in the
(process-independent) bare Green function ${\cal G}^{(0)}$. The
factorization formula (\ref{F(0)}) can now be realized by
introducing a suitable projection operator ${\cal P}_{C}$ which
decouples $C^{(0)}$ and ${\cal G}^{(0)}$ in the spin indices and extracts
the singular part of the $d^{n}k$ integrals (i.e. poles in
$\varepsilon$) thus decoupling $C^{(0)}$ and ${\cal G}^{(0)}$ also in
momentum space.

For each kernel, one can write the decomposition $K^{(0)} =
(1-{\cal P}_{C}) K^{(0)} + {\cal P}_{C}K^{(0)},$ where all the singularities
are due to the second term on the r.h.s. Applying this procedure
in an iterative way, one obtains
\begin{equation}
{\cal G}^{(0)} = {\cal G} \, \Gamma
\label{G0G}
\end{equation}
where
%$G$
all the $\varepsilon$-poles have been subtracted from
the `renormalized' Green function $\cal G$:
\begin{equation}
{\cal G}=\frac{1}{1-(1-{\cal P}_{C})K^{(0)}} =
1+(1-{\cal P}_{C})K^{(0)}+(1-{\cal P}_{C})[K^{(0)}(1-{\cal P}_{C})K^{(0)}]
+\dots \;,
\label{Gfrac}
\end{equation}
and associated with the transition function $\Gamma$
\begin{equation}
\Gamma = \frac{1}{1-({\cal P}_{C}K)} = 1+({\cal P}_{C}K)+({\cal P}_{C}K)
({\cal P}_{C}K)+\dots
\;, \label{Gamma}
\end{equation}
\begin{equation}
K \equiv K^{(0)} \;{\cal G} \;.
\label{KKG}
\end{equation}
The coefficient function $C$ in eq.~(\ref{F(0)}) is thus
identified with
\begin{equation}
C=C^{(0)} \, {\cal G} \;\;.
\label{CCG}
\end{equation}

In order to show the consistency of collinear and
$\kper$-factorization (see Sect.~2.3), we have to recall the
form of the projection operator ${\cal P}_{C}$ [\ref{CFP}]. Let us denote
its action on helicity and momentum space respectively by $P^{(s)}_{C}$
and $P^{(\varepsilon)}_{C}$, so that ${\cal P}_{C} =
P^{(\varepsilon)}_{C} \otimes P^{(s)}_{C}$. If $A$ and $B$ are
two kernels connected by a parton of momentum $k$ (Fig.~3), the action on
the helicity space is
\begin{equation}
A \;P_{C}^{(s)} \;B = A(\dots , k)_{\dots \alpha ' \beta '}
\left(\frac{1}{2} k\!\!\!/ \right) _{\alpha ' \beta '} \;\;
\left(\frac{\bar p\!\!\!/}{2{\bar p}\cdot k} \right) _{\alpha\beta}
B _{\alpha\beta\dots}(k,\dots) \;,\label{APcB}
\end{equation}
when the connecting parton is a quark and
\begin{equation}
A \;P_{C}^{(s)} \;B = A(\dots , k)_{\dots \mu^{\prime}\nu^{\prime}}
\frac{d^{\mu^{\prime}\nu^{\prime}}(k)}{n-2} \;\; (-g^{\mu\nu})
B_{\mu\nu\dots}(k,\dots) \;,
\label{APB}
\end{equation}
when the connecting parton is a gluon. The operator
$P_{C}^{(\varepsilon)}$ sets $k^{\mu}=zp^{\mu}$ on the left-hand
side $(A)$, performs the $dk^{2} d^{n-2} {\bk}$ integration
up to the factorization scale $\mu_{F}^{2}$ on the
right-hand side $(B)$, and extracts the ensuing poles in $\varepsilon$.

Note from Eqs.~(\ref{APcB}),(\ref{APB}) that $P_{C}^{(s)}$
acts on the left by performing the average over the parton
helicities in {\em $n$-dimensions}.
Note also that $P_{C}^{(\varepsilon)}$ is not (at least in
principle) unambiguously defined. The factorization of
Eqs.~(\ref{F(0)}),(\ref{G0G}) in terms of collinear-finite and
divergent (for $\varepsilon \rightarrow 0$) contributions can
still be achieved if $P_{C}^{(\varepsilon)}$ extracts not only the
$\varepsilon$-poles but also any finite contribution for
$\varepsilon \rightarrow 0$. This leads to the
factorization-scheme dependence of both anomalous dimensions and
coefficient functions. The factorization scheme is completely
specified once $P_{C}^{(\varepsilon)}$ has been uniquely
defined. Equivalently, one can specify the explicit (and finite)
$\varepsilon$-dependence of the transition functions $\Gamma$. One
of the most commonly used scheme is the {\em modified minimal
subtraction} $(\msbar)$ scheme, in which the transition functions
have the form
%\cite{CFP}
[\ref{CFP}]
\begin{equation}
\Gamma_{ab,N} (\as(\mu^{2}_F/\mu^{2})^{\varepsilon}, \varepsilon) =
\left[ P_{\alpha} \exp \left( \frac{1}{\varepsilon}
\int_{0}^{\as(\mu_F^{2}/\mu^{2})^{\varepsilon}S_{\varepsilon}}
\frac{d\alpha}{\alpha} \;\gamma_{N}(\alpha) \right) \right]_{ab}  \;\;.
\label{lceiabN}
\end{equation}
In this equation we have reintroduced the explicit dependence on
the parton labels and $N$-moment indices. The symbol $P_{\alpha}$
denotes, as usual, the path-ordered integration of the anomalous
dimension matrix $\gamma_{ab,N}$. Note the presence of the
$\varepsilon$-finite factor $S_{\varepsilon} = \exp \{ -
\varepsilon [ \psi (1) + \ln 4 \pi ]\} \;(\psi(z)$ is the Euler
$\psi$-function), which characterizes the $\msbar$-scheme. The
only formal simplification we have used in Eq.~(\ref{lceiabN}) is
that we have considered the case of a fixed coupling constant
$\as$. As shown in the following, this simplification is sufficient
for the purposes of the present paper.

\vskip 1 true cm
\noindent {\it 2.3 Power counting and factorization at high energy}
\vskip 0.2 true cm

The expansion in 2PI kernels introduced in the previous
subsection is particularly useful to discuss the high-energy
behaviour. High-energy (or small-$x$) logarithmic contributions
are indeed generated by multiple gluon exchanges in the
$t$-channel. Therefore we are led to consider kernels which are
two-gluon irreducible (2GI).

For the parton cross section $F^{(0)}_{a} \;(a=q_{i}, {\bar
q}_{i},g)$ , we single out the part which is 2GI by selecting
the first (starting from above in Fig.~4a) two-gluon intermediate state.
 Considering the small-$N$ limit in $N$-moment space, the
2GI kernel behaves as $\, \as^{p} \, (1+\as + \dots)$, where the first
term corresponds to the tree-approximation and the remaining terms
stand for corrections which are subleading at high energy. The
large perturbative contributions $(\as /N)^{k}$ are thus
generated precisely by  $k$-integration from the ones in the
gluon Green functions ${\cal G}^{(0)}_{ga} \;(a=q_{i}, {\bar
q}_{i},g)$. In particular, since flavour non-singlet parton cross
sections (App.~A) get no contribution from pure-gluon intermediate states,
we can immediately conclude that {\em non-singlet} anomalous
dimensions and coefficient functions are {\em regular} for
$N\rightarrow 0$ order by order in $\as$.

A decomposition similar to that for $F^{(0)}_{a}$ can be performed
also for the (flavour singlet) quark Green function (Fig.~4b).
Since the 2GI kernel behaves in this case as $\, \as \, (1+\as +
\dots)$, we see that the quark anomalous dimensions contribute to
{\em next-to-leading} terms $\as (\as /N)^{k}$ in the
high-energy limit.

Note that the expansion in 2GI kernels is more general than that
in 2PI kernels. The 2GI kernels in Fig.~4a and Fig.~4b can still
be expanded respectively as $(C^{(0)}_{a} + \sum_{b \neq g}
C^{(0)}_{b} K^{(0)}_{ba} + \dots)$ and $(K^{(0)}_{ba} +
\sum_{c \neq g} K^{(0)}_{bc}K^{(0)}_{ca} + \dots )$. However,
only the tree-level approximations for $C^{(0)}_{a}$ and
$K^{(0)}_{ba} \;(b\neq g)$ contribute to leading order in $\as /N$.
A similar simplification does not occur for the gluon Green
function ${\cal G}^{(0)}_{ga}$ because the gluon kernels $K^{(0)}_{ga}$
contain terms of the type $\, (\as / N)^{k} \, $ to any order $k$
(modulo dynamical cancellations) in $\as$.

The expansion in 2GI kernels described so far allows one a simple
power counting at high energy. The next step towards high-energy
factorization consists of decoupling the 2GI kernels and the
gluon Green functions with $\as /N$ fixed. This factorization is
conceptually different from the collinear one, where, roughly
speaking, one expands in $\as$ (or $\varepsilon$) with $\as
/\varepsilon$ fixed. In particular, we cannot perform the collinear
limit $k^{2} \rightarrow 0$ to any fixed order in $\as$, because
small-$x$ contributions $\, (\as /N)^{k} \, $ are associated with any
value of $k$ [\ref{CCH}]. On the other side, we do not want to spoil
the collinear factorization, so that the high-energy factorization
has to be valid for any value of $\varepsilon$ (i.e., in any
number of space-time dimensions). We are going to show that the
$\kper$-factorization procedure introduced in Ref.~[\ref{CCH}] can
be carried out consistently with all-order collinear factorization.

The high-energy limit of the product $A_{g} {\cal G}^{(0)}_{ga}$,
$ \,  A_{g} \, $
being the 2GI kernel involved in the decomposition of Fig.~4, was
discussed in detail in [\ref{CCH}]. Considering the decomposition
of $A_{g}$ in Lorentz-invariant amplitudes, it was shown that the
leading high-energy behavior can be extracted via $\kper$-factorization,
i.e. by inserting into the two-gluon intermediate state a
$\kper$-dependent projection operator ${\cal P}_{H}$ as follows
\begin{equation}
A_{g} \;{\cal G}^{(0)}_{ga} = A_{g} \;{\cal P}_{H} \;{\cal G}^{(0)}_{ga}
+ \dots \;\;.
\label{AgG}
\end{equation}
As in the case of the collinear projector ${\cal P}_{C}$ in
Sect.~2.2,
%\ref{ColFac},
we introduce the notation ${\cal P}_{H} =
P^{(\varepsilon)}_{H} \otimes P^{(s)}_{H}$, where $P^{(s)}_{H}$
and $P^{(\varepsilon)}_{H}$ denote respectively the action of the
high-energy projector ${\cal P}_{H}$ on helicity and momentum space.
$P^{(s)}_{H}$ acts as follows
\begin{equation}
A_{g} \;P^{(s)}_{H} \;{\cal G}^{(0)}_{ga} = A(\dots , k)^{\dots \mu ' \nu
'}_{g}
\frac{{\kper}_{ \mu '} {\kper}_{ \nu '}}{{\bk}^{2}} \;(-g_{\mu\nu})
{\cal G}^{(0)\mu\nu}_{ga} (k,p) \;\;,
\label{AgP}
\end{equation}
whilst $P^{(\varepsilon)}_{H}$ sets $k^{\mu} = zp^{\mu} +
\kper^{\mu}$ on the left-hand side $(A_{g})$ and integrates the
right-hand side $({\cal G}^{(0)}_{ga})$ over the invariant mass $k^{2}$ at
fixed $\kper$. Note that the
$\kper$-dependence is left unaffected by ${\cal P}_{H}$ and, in
particular, the 2GI kernel $A_{g}$ has to be evaluated with an
incoming off-shell (essentially transverse $k^{2} =
-{\bk}^{2}$) gluon.

Equation (\ref{AgG}) generalizes the $\kper$-factorization formula
(\ref{ktfac}) to $n=4+2 \, \varepsilon$ space-time dimensions. The key
point, however, is not just the formal resemblance between
Eqs.~(\ref{AgG}),(\ref{AgP}) and (\ref{ktfac}), but rather the fact that
${\cal P}_{H}$
selects the correct high-energy behaviour in any number of
dimensions. We mean that, for instance, in performing the
approximation (\ref{AgG}), we are not neglecting any contribution
of order $(\as /N)^{k} \cdot \varepsilon$ with respect to the
four-dimensional case. This statement is a consequence of the fact
that ${\cal P}_{H}$ is a `true' projection operator:
\begin{equation}
{\cal P}_{H}^{2} = {\cal P}_{H} \;,
\label{PHPH}
\end{equation}
and fulfils the property:
\begin{equation}
{\cal P}_{H} \supseteq {\cal P}_{C}\;, \;\;\;\; ({\cal P}_{H} =
{\cal P}_{C} \;\; { \mbox {\rm iff}} \;\; \kper = 0) \;.
\label{PHPC}
\end{equation}
Eqs.~(\ref{PHPH}) and (\ref{PHPC}) are self-evident for the
momentum space components
$P_{H}^{(\varepsilon)},P_{C}^{(\varepsilon)}$ and follows from the
simple relations ($< >_{\phi}$ denotes the average over the $n-3$
azimuthal angles in the transverse momentum space)
\begin{equation}
(-g_{\mu\nu})
\frac{\kper^{\mu}\kper^{\nu}}{{\bk}^{2}} = 1 \;, \;\;\;\;
< \frac{\kper^{\mu}\kper^{\nu}}{{\bk}^{2}} >_{\phi}
\;\;\; \stackrel{\kper \rightarrow 0}{=}
\;\frac{d^{\mu\nu}(k=zp)}{n-2} \label{-g} \;,
\end{equation}
for the spin components $P_{H}^{(s)},P_{C}^{(s)}$.

Eq.~(\ref{PHPC}) guarantees the consistency between high-energy
factorization and collinear factorization. Due to Eq.~(\ref{PHPC})
we can first perform the high-energy approximation in
Eq.~(\ref{AgG}) and then proceed to the all-order factorization of
collinear singularities by applying iteratively the collinear
projector ${\cal P}_{C}$, as described in Sect.~2.2.
%\ref{ColFac}.

\vskip 1 true cm
\noindent {\it 2.4 Resummation at high energy}
\vskip 0.2 true cm

In order to describe how in practice this procedure works  and leads
to the high-energy resummation, let us
consider the
case involving a 2GI kernel which is collinear safe. In particular, we refer to
the
heavy-flavour photoproduction process (\ref{phot})
already introduced in Sect.~1.
%\ref{intro}.
The corresponding 2GI
kernel $A_{g}$ to lowest order in $\as$ is given in Fig.~5: the
dashed line denotes the incoming on-shell photon and the full
lines correspond to the heavy-flavour pair produced in the final
state. Considering the case of an incoming-gluon partonic state
and using the high-energy approximation in Eq.~(\ref{AgG}), we
immediately obtain the factorized formula (\ref{ktfac}) (in
$n=4+2 \, \varepsilon$ dimensions), which we can rewrite in the
$N$-moment space as follows
\begin{equation}
\label{ktfac2}
4 M^2 \;\si_{N}(M^2) = \int d^{2+2\varepsilon} {\bk}\;\;
{\hat \si}_{N}({\bk}^{2}/M^{2}, \as (M^2/\mu^2)^{\varepsilon} ;
\varepsilon ) \;{\cal F}^{(0)}_{N}({\bk}; \as, \mu,
\varepsilon) \; {\tilde f}^{(0)}_{g,N} (\mu, \varepsilon)\;.
\end{equation}
Here ${\hat \si} \sim A_{\mu\nu}(k) \;\kper^{\mu} \kper^{\nu} /
{\bk}^{2}$, $\, {\tilde f}^{(0)}_{g}$ is the bare gluon
distribution and, according to the action of the high-energy
projector ${\cal P}_{H}$, we have introduced the $\kper$-dependent gluon
Green function
\begin{equation}
{\cal F}^{(0)}(z,{\bk}; \as, \mu, \varepsilon) =
\int \frac{dk^{2}}{2(2\pi)^{4+2\varepsilon}} \; (-g_{\mu\nu}
{\cal G}^{(0)\mu\nu}_{gg} (k,p)) \;.
\label{calF0}
\end{equation}

Equation (\ref{ktfac2}) has to be regarded as the bare (and collinear
regularized) version of the factorization formula (\ref{ktfac}),
in the sense that it still contains collinear poles in
$\varepsilon$ which, according to Eq.~(\ref{G0G}), can be
factorized in the form
\begin{equation}
{\cal F}^{(0)}_{N}({\bk}; \as, \mu, \varepsilon) =
\frac{\gamma_{gg,N}(\as)}{\pi {\bk}^{2}}
\;{\tilde R}_{N}({\bk},\mu_{F},\as; \mu, \varepsilon)
\;\; \Gamma_{gg,N}(\as(\mu^{2}_{F} /\mu^{2})^{\varepsilon},
\varepsilon)\;,
\label{calF0N}
\end{equation}
where the $\msbar$-scheme gluon transition function is given by
\footnote{We limit ourselves to the case of a fixed coupling
constant $\as$.
%This is sufficient to the leading accuracy $(\as/N)^{k}$, because
In fact,
the running coupling effects lead to subleading
contributions $ \as (\as / N )^k $ and can thus be neglected in the present
leading-logarithmic analysis.}
\begin{equation}
\Gamma_{gg,N}(\as ,\varepsilon) = \exp \left\{
\frac{1}{\varepsilon} \int^{\as S_{\varepsilon}}_{0}
\frac{d\alpha}{\alpha} \;\gamma_{gg,N}(\alpha) \right\} \;.
\label{Gam}
\end{equation}
The function ${\tilde R}_{N}$ in Eq.~(\ref{calF0N}) has no
$\varepsilon$-poles order by order in $\as$, and cannot depend on the
dimensional regularization scale $\mu$ in the limit $\varepsilon
\rightarrow 0$. Moreover, since the l.h.s. of Eq.~(\ref{calF0N})
is independent of the factorization scale $\mu_{F}$, the only
${\bk}$-dependence of ${\tilde R}_{N}$ allowed by
dimensional arguments, for $\varepsilon = 0$, is the following
\begin{equation}
{\tilde R}_{N}({\bk}, \mu_{F}, \as; \mu, \varepsilon =0)
= R_{N}(\as) \;({\bk}^{2}/\mu^{2}_{F})^{\gamma_{gg,N}(\as)}\;.
\label{RN}
\end{equation}

The reduced cross section $\hat \si$ in Eq.~(\ref{ktfac2}) is
collinear safe because it corresponds to a 2PI kernel
of the type $\, C^{(0)} \,$ in Eq.~(\ref{F0}).
Therefore, after  using the transition function
$\Gamma_{gg,N}$ in (\ref{calF0N}) to `renormalize' the bare gluon
density ${\tilde f}^{(0)}$ as in Eq.~(\ref{tilf}), we can safely
perform the $\varepsilon \rightarrow 0$ limit in
Eq.~(\ref{ktfac2}) and obtain:
\begin{equation}
4M^{2} \;\si_{N}(M^{2}) = C_{N} (\as , M^{2}/ \mu^{2}_{F})
\;\; {\tilde f}_{g,N} (\mu^{2}_{F})\;,
\label{4M2si}
\end{equation}
\begin{equation}
C_{N} (\as , M^{2}/ \mu^{2}_{F}) = h_{N}(\gamma_{gg,N}(\as))
\;R_{N}(\as) \;(M^{2}/\mu^{2}_{F})^{\gamma_{gg,N}(\as)} \;\;,
\label{CN}
\end{equation}
where the process-dependent part $h_{N}$ of the coefficient
function $C_{N}$ is given by the following $\kper$-transform of the
hard cross section $\hat \si$:
\begin{equation}
h_{N}(\gamma) \equiv \gamma \int^{\infty}_{0} \frac{d
{\bk}^{2}}{{\bk}^{2}}\left(\frac{\bk^{2}}{M^{2}}\right)^\gamma
\;\; {\hat \si}_{N}({\bk^{2}}/M^{2},\as; \varepsilon = 0) \;.
\label{hN}
\end{equation}

The result in Eq.~(\ref{CN}) gives the resummed expression
(including the dependence on the factorization scale
$\mu^{2}_{F}$) for the coefficient function $C_{N}$ to the leading
order $(\as /N)^{k}$, provided the gluon anomalous dimensions
$\gamma_{gg,N}$ and the process-independent function $R_{N}$ are
known to the same accuracy. A similar result was first derived
in Ref.~[\ref{CCH}]. The only difference with respect to [\ref{CCH}]
is that Eq.~(\ref{CN}) takes into account the explicit dependence
on the process-independent but {\em factorization-scheme
dependent} function $R_{N}$. This point is essential for precise
phenomenological predictions at finite energies, when one has to
combine the resummed coefficient function (\ref{CN}) with
fixed-order non-logarithmic contributions computed in a well
defined factorization scheme of collinear singularities. Moreover
this issue is relevant to extending the high-energy resummation to
subleading orders, where a corresponding scheme dependence of the
anomalous dimensions comes into play (see, for instance, Sect.~5).

Note, however, that the presence of the process independent factor $R_N$ in
Eq.~(\ref{CN}) is no longer relevant (at least, to leading order) if one limits
oneself to considering only ratios of cross sections. In this case $R_N$ and
the factorization scale dependence cancel in the ratio, thus leading to an
absolute prediction in terms of hard scales and $h_N(\gamma)$ functions of the
type in Eq.~(\ref{hN}) [\ref{CCH}].

The high-energy contributions $(\as /N)^{k}$ are embodied in
Eqs.~(\ref{AgG}),(\ref{ktfac2}) through the $\kper$-integration of
the gluon Green function ${\cal G}^{(0)}_{ga}$. As discussed in
[\ref{CCH}] and recalled in Sect.~1, in the case of four
space-time dimensions the resummation of the leading terms $(\as
/N)^{k}$ in the gluon density (anomalous dimensions) is
accomplished by the BFKL equation [\ref{BFKL}]. The analogous
master equation in $n=4+2 \, \varepsilon$ dimensions, which is
necessary to compute both the anomalous dimensions and the
function $R_{N}$, is derived and discussed in the following
section.

\vskip 1 true cm

\setcounter{equation}{0}
\setcounter{sect}{3}
\setcounter{footnote}{0}

%\vskip 3 true cm
\noindent {\bf 3. The gluon Green functions }
\vskip 0.3 true cm

The leading high-energy behaviour of the gluon Green functions
%$ \, {\cal G}_{g a}^{(0)} \,$ ($ a = q_i, {\bar q}_i , g $)
$ \, {\cal G}^{(0)}_{g g } (k, p) \,$ and
\linebreak
$ \, {\cal G}^{(0)}_{g q } (k, p) \,$
can  be easily derived by generalizing the soft-gluon insertion technique
in Refs.~[\ref{Ciaf},\ref{CFM},\ref{CFMO}] to the case of $ \, n \, $
dimensions. In the present paper we do not repeat all the detailed
calculations described in [\ref{Ciaf},\ref{CFMO}], but we simply sketch the
main steps and properties which are necessary for the $ \, n$-dimensional
generalization.

The starting observation is that the high-energy contributions
$ \, \left( \as  / N \right)^k \,$
to the gluon Green function are produced by radiation of soft gluons,
that is real and virtual gluons carrying a very small fraction $ \, x \, $
of the longitudinal momentum $ \, p \, $ of the incoming parton.
Therefore, in the soft-gluon approach, the matrix element
$ \, M^{(k+1)} (k , p) $, contributing to
$ \, {\cal G}^{(0)} \,$ to the  $ \, (k+1)-$loop order,
is obtained from
$ \, M^{(k)} \,$ by the insertion of an additional
(real or virtual) soft gluon with momentum
$ \, q $.
Using the  soft approximation for vertices and propagators, this
 insertion can  in turn be factorized, leading to the recurrence relation:
\begin{eqnarray}
\label{softinsert}
{| M^{(k+1)} (k , p) |}^2 &=& g_s^2 \,
\left\{
[ M^{(k)}(k+q, p) ]^{\dagger} \;
[J^{(R)}_{soft}(q)]^2  \;
{ M^{(k)}  } (k+q , p) \right.
\nonumber\\
&-& \left.
[ M^{(k)}(k, p) ]^{\dagger} \;
[J^{(V)}_{soft}(q)]^2 \;
{ M^{(k)}  } (k , p)
\right\}   \;\;.
\end{eqnarray}

The explicit expressions for the real and virtual soft-gluon currents
$ \, {J^{(R)}_{soft}}  \, $ and
$\, {J^{(V)}_{soft}} $
can be found in Refs.~[\ref{Ciaf},\ref{CFMO}]. The point we want to
emphasize here is that they do not explicitly depend on the number
$ \, n \, $ of space-time dimensions in which the soft-gluon factorization
is carried out. As a matter of fact, the $ \, n$-dependence of the matrix
element $ \, M^{(k+1)} \,$ can only be due to the spin structure of the
vertices (the scalar propagators
$ \, i / ( q^2 + i \, \ep ) \, $
are the same in any number of dimensions).  It turns out that a
{\em single} (essentially {\em eikonal}) helicity
flow dominates to leading-logarithmic
accuracy in the high-energy limit. Thus, the relevant QCD matrix elements
are the same as for $ \, n = 4 $ and the only difference comes from the
$ \, n$-dimensional phase space over which Eq.~(\ref{softinsert}) has to be
integrated.

Using the recurrence relation (\ref{softinsert}),  performing the sum
over the number $ \, k \,$ of  loops and following exactly the same steps as
in Refs.~[\ref{Ciaf},\ref{CFMO}], one obtains an integral equation for the
gluon Green function
$ \, {\cal G}^{(0)} $.
More precisely, considering the gluon density
%in $ \, k_{\perp} \,$
in Eq.~(\ref{calF0}),
one finds ($\abar \equiv C_A \, \as / \pi $, $ \, C_A = N_c \,$ being the
number of colours)
%[\ref{HEF}]
\begin{eqnarray}
\label{maeq}
{\cf}^{(0)}_N(\bk; \as, \mu,\varepsilon) &=& \delta^{(2+2\varepsilon)}(\bk)
+ \frac{\abar}{N} \int
\frac{d^{2+2\varepsilon} \bq}{(2\pi \mu)^{2\varepsilon}} \;
\frac{1}{\pi {\bq}^2} \;
\nonumber \\
&\cdot& \left\{ \cf^{(0)}_N(\bk- \bq; \as , \mu,\varepsilon) -
\frac{\bk \cdot (\bk - \bq)}{(\bk - \bq)^2} \;\cf^{(0)}_N(\bk; \as,  \mu ,
\varepsilon)
\right\} \;\;.
\end{eqnarray}

Some comments are in order.
In the case of
$ \, n = 4 \,$ dimensions ($ \varepsilon = 0 $), after azimuthal average over
$\, \bq $,
 Eq.~(\ref{maeq}) reproduces the
BFKL equation [\ref{BFKL},\ref{BCM}]. Moreover, the integrand (not the
 phase space) of the homogeneous term in Eq.~(\ref{maeq}) is exactly the same
as for the BFKL equation. As discussed above, this is a straightforward
consequence of the dynamical dominance of a single gluon polarization to
the present accuracy. Note, however, an essential difference with respect to
the BFKL equation: the full kernel of Eq.~(\ref{maeq}) is not scale invariant.
Indeed, scale invariance is broken by the dimensional regularization
procedure and the eigenfunctions of the kernel are no longer simple powers
$\, (\bk^2)^{\ga}$.
{}From a physical viewpoint this means that whilst small and large
transverse momentum regions contribute equally to the BFKL equation,
they are now weighted asymmetrically. The breaking of scale invariance in
Eq.~(\ref{maeq}) allows one to regularize the collinear singularities
in terms of $ \, \varepsilon$-poles. Once these poles have been
factorized and subtracted, as in Eqs.~(\ref{calF0N}),(\ref{RN}),
one can perform the $\, \varepsilon  \to 0 \,$ limit
and recover scale invariance at the expenses of an additional dependence
on the factorization scale as dictated by the leading-twist behaviour of the
parton densities (cross sections).

A further consequence of the lack of scale invariance in the kernel of the
gluon Green function is that no simple technique is available to
diagonalize the integral equation (\ref{maeq}). However, it is possible
(see App.~B) to solve it as a formal power series in $ \, \as \, $
with $ \varepsilon$-dependent coefficients. The result reads as follows
\begin{equation}
\label{sersol}
{\cf}^{(0)}_N(\bk; \mu, \as, \varepsilon) = \delta^{(2+2\varepsilon)}(\bk) +
{ {\Gamma (1+ \varepsilon)} \over
{ ( \pi \, \bk^2 )^{1 + \varepsilon}} } \,
\sum_{k=1}^\infty
\, \left[ \abn  \, S_\varepsilon \,
{{e^{ \varepsilon \, \psi (1 )}} \over {\Gamma ({1 +  \varepsilon})}}  \,
\left( {\bk^2 \over \mu^2} \right)^\varepsilon
\, \right]^k \, c_k (\varepsilon) \;\;\;,
\end{equation}
where we have explicitly introduced the $\, \msbar$-scheme factor
$ \, S_\varepsilon = \exp \left[ - \varepsilon \, \left( \psi (1)
+ \ln 4 \pi \right) \right]$,
and the  coefficients
$\, c_k  \,$ are defined by the recurrence relation
\begin{equation}
\label{recurs}
c_{1}(\varepsilon) = 1  \;\;, \;\;\;\; \;\;
c_{k + 1}(\varepsilon) = c_{k} (\varepsilon) \, { { I}}_k
(\varepsilon)  \;\;\;\; ( k \geq 1) \;\;,
\end{equation}
with ($ \Gamma (z) \,$ is the Euler $\,  \Gamma$-function)
\begin{equation}
\label{calcI}
{ { I}}_k ( \varepsilon) =   { 1 \over \varepsilon} \,
\frac{\Gamma^2 (1 + \varepsilon)}{ \Gamma ({1} + 2 \, \varepsilon)}
\left[
\frac{\Gamma (1 + 2 \, \varepsilon) \, \Gamma ( k \, \varepsilon) \,
\Gamma (1 - k \, \varepsilon) }
{ \Gamma \left( (1 + k) \,
\varepsilon \right) \,\Gamma \left( 1 + (1-k) \, \varepsilon \right)}
-  \Gamma ( 1 + \varepsilon)  \, \Gamma (1 - \varepsilon) \right]  \;\;.
\end{equation}

The gluon Green function  (\ref{sersol})
is formally a distribution acting on the transverse-momentum space
in $ \, n - 2 = 2 + 2 \, \varepsilon \,$ dimensions.
Following the  procedure of factorization
of  collinear singularities described
in Sect.~2.2, one can extract from
$ \, {\cal F}_N^{(0)} \,$
the transition function
$ \, \Gamma_{g g , \, N} \,$ as in Eq.~(\ref{calF0N}), thus leading to a
`renormalized' gluon density which, for finite $ \, \varepsilon $,
is a very
cumbersome $ \, k_{\perp}$-distribution. Therefore, it is more convenient to
introduce the gluon Green function
$ \, G^{(0)} \,$
integrated up to the  factorization scale  $ \, \mu_F^2 = Q^2 \,$:
\begin{equation}
\label{integrG}
{G}_{g g , \, N}^{(0)}(\as \,(Q^2/\mu^2)^\varepsilon , \,
\varepsilon)
\equiv
\int
d^{2+2\varepsilon} \bk
\; {\cf}^{(0)}_N(\bk; \as, \mu, \varepsilon) \; \Theta ( Q^2 - \bk^2) \;\;.
\end{equation}
Note that $ \, G^{(0)} \,$
does not depend on  $ \, \as \,$ e $ \, Q^2 /  \mu^2 \,$ independently,
but only on the  product
$ \, \as ( Q^2/\mu^2)^\varepsilon $. Performing the
 $ \, k_{\perp}$-integration of Eq.~(\ref{sersol}) we find
\begin{equation}
\label{sersolG}
{G}_{g g , \, N}^{(0)}(\as \,(Q^2/\mu^2)^\varepsilon , \varepsilon)
= 1 + \sum_{k=1}^\infty
\, \left[ \abn
\, S_\varepsilon \,
{{e^{ \varepsilon \, \psi (1 )}} \over {\Gamma ({1 +  \varepsilon})}}  \,
\left( {Q^2 \over \mu^2} \right)^\varepsilon
\, \right]^k \, {1 \over {k \, \varepsilon}} \, c_k (\varepsilon) \;\;\;.
\end{equation}

Introducing explicitly the parton indices, the
%factorized expression (\ref{G0G})
collinear factorization in Eq.~(\ref{G0G})
reads
$ \,
{G}_{g g , \, N}^{(0)} = \sum_{a}
{G}_{g a , \, N}
{\Gamma}_{a g , \, N} $.
However, from the power counting in Sect.~2.3, we know that the quark
anomalous dimensions $ \, \ga_{ q a , \, N} \,$
(and hence the corresponding transition functions
$ \, \Gamma_{ q a , \, N} $)
are subleading at high energy. Therefore, to leading order in
$ \, \left( \as / N \right)^k $, we can write
\begin{equation}
\label{renggcap2}
{G}_{g g , \, N}^{(0)}(\as (Q^2/\mu^2)^\varepsilon, \varepsilon) =
{G}_{g g , \, N}(\as (Q^2/\mu^2)^\varepsilon, \varepsilon) \;
{\Gamma}_{g g , \, N}(\as (Q^2/\mu^2)^\varepsilon, \varepsilon) \;,
\end{equation}
where $ \, {\Gamma}_{g g , \, N}  \,$ is given by Eq.~(\ref{Gam}),
in terms of the gluon  anomalous dimension
$ \, {\gamma}_{g g , \, N}  $.
By comparing  Eqs.~(\ref{sersolG}) and (\ref{renggcap2}),
one can compute $ \, {\gamma}_{g g , \, N}  \,$
and the  `renormalized' Green function $ \, {G}_{g g , \, N}  $
for any value of $ \, \varepsilon$. Moreover, from
Eqs.~(\ref{integrG}) e (\ref{calF0N}) it follows that the function
$ \, R_N \,$ in Eq.~(\ref{RN}) is related to the
$\, \varepsilon \to 0 \,$ limit of $ \, {G}_{g g , \, N}  \,$
\begin{equation}
\label{Rcoeff}
R_N (\as) = {G}_{g g , \, N}(\as (Q^2/\mu^2)^\varepsilon ,
\varepsilon)|_{\varepsilon = 0}
\;\;\;.
\end{equation}

The recurrence factor $ I_n (\varepsilon) \,$  behaves like
$ \, 1 / \varepsilon $ for $ \, \varepsilon \to 0 $.
Therefore, in agreement with Eqs.~(\ref{Gam}) and (\ref{renggcap2}), the power
 series expansion (\ref{sersolG}) has at most a single
$\, \varepsilon$-pole for any power of  $ \, \as $. More  precisely,
for $ \, \varepsilon \to 0 $ we have
\begin{equation}
\label{Ismalleps}
  {  { I}}_k ( \varepsilon)  \simeq   { 1 \over  {k \, \varepsilon}}
\left( 1 + {\cal O}(\varepsilon^3) \right)
\;\; \;\;\;\;\; ( \varepsilon \to 0 )
\;\;,
\end{equation}
and correspondingly
\begin{eqnarray}
\label{Gperturb}
&\,&{G}_{g g , \, N}^{(0)}(\as (Q^2/\mu^2)^\varepsilon, \varepsilon)
= 1 + \sum_{k=1}^\infty
\, \left[ {1 \over \varepsilon} \,\abn
\, S_\varepsilon \,
{{e^{ \varepsilon \, \psi (1 )}} \over {\Gamma ({1 +  \varepsilon})}}
\,
\left( {Q^2 \over \mu^2} \right)^\varepsilon
\, \right]^k \, {1 \over {k {\rm !} }} \,
\left( 1 + {\cal O}(\varepsilon^3) \right)
\nonumber\\
&\,& \simeq \exp
 \left[ {1 \over \varepsilon} \,\abn
\, S_\varepsilon \,
{{e^{ \varepsilon \, \psi (1 )}} \over {\Gamma ({1 +  \varepsilon})}}  \,
\left( {Q^2 \over \mu^2} \right)^\varepsilon
\, \right]
\, \left( 1 +
{\cal O}\left( \left( \abn \right)^3 \right)+
%\right)
{\cal O}\left( {1 \over \varepsilon} \left( \abn \right)^4 \right) \,
\right)  \;\;.
\end{eqnarray}
Comparing  Eqs.~(\ref{renggcap2}) e (\ref{Gperturb}), we  see that the slow
departure of $I_k(\varepsilon)$
%(\ref{Ismalleps})
from its leading-pole  approximation
$ \, I_k(\varepsilon)  \simeq 1/k\varepsilon \,$
gives rise
to subleading collinear corrections,
for both $ \, {\gamma}_{g g , \, N}  \,$ and $ \, R_N $,
which are of relative order
$\, \as^3 \, $
in the high-energy regime:
\begin{equation}
\label{cubic}
 \, {\gamma}_{g g , \, N}  = \abn \,
\, \left[ 1 + {\cal O}\left( \left( \abn \right)^3\right)
\right]
 \;\;\;,\;\;\; R_N  =
1 + {\cal O}\left(\left( \abn \right)^3\right) \;\;.
\end{equation}
Terms of  relative order $\, \as \, $ and $\, \as^2 \, $ are
present in the various Feynman diagrams contributing to the gluon  Green
function but they cancel
in the sum.
This cancellation, which is automatically embodied in the master
equation (\ref{maeq}), does no longer occur in higher perturbative orders.
The resummed expressions for
$ \, {\gamma}_{g g , \, N}  \,$
and
$ \, R_N $ to the leading accuracy $ \, (\as / N )^n \,$
are derived in App.~B. The results are the following.

The dominant contribution
$ \, \ga_{N} (\as) \,$
to the gluon anomalous dimensions
\begin{equation}
\label{glulead}
  \ga_{gg, N} =  \ga_{N} (\as) + {\cal O}( \as(\as/N)^k )
\end{equation}
is obtained by solving the implicit equation (Fig.~6)
\begin{equation}
\label{implbfkl}
1 = \abn \; \chi\left( \ga_N (\as) \right)  \;\;,
\end{equation}
where the characteristic function $ \chi(\ga) $ is expressed in
terms of the Euler $ \psi$-function
\begin{equation}
\label{cfmo57}
\chi(\gamma) = 2 \psi(1) - \psi(\gamma) - \psi(1-\gamma) =
{ 1\over \ga} \, \left[ 1+ \sum_{k  = 1}^{\infty} 2 \;
\zeta (2 \, k + 1) \, \ga^{2 \, k + 1} \right]
\end{equation}
and $\, \zeta(n) \,$  is the  Riemann $ \zeta$-function.
The solution of Eq.~(\ref{implbfkl}) in power series of the coupling constant
gives\footnote{The explicit values of the coefficients $g_n$,
up to $n = 14$, can be found in Ref.~[\ref{CFM}].}
$(\zeta(3) \simeq 1.202, \; \zeta(5) \simeq 1.037)$:
%\begin{eqnarray}
%\label{pertbfkl}
%\ga_N (\as) &=& \sum_{n=1}^{\infty} g_n \left( \abn \right)^n \nonumber \\
%&=& \abn + 2 \,
%\zeta (3 ) \left( \abn \right)^4 + 2 \,
%\zeta (5 ) \left( \abn \right)^6 + O \left( \abn \right)^7 \;\;.
%\end{eqnarray}
\begin{equation}
\label{pertbfkl}
\ga_N (\as)= \sum_{n=1}^{\infty} g_n \left( \abn \right)^n
= \abn + 2 \,
\zeta (3 ) \left( \abn \right)^4 + 2 \,
\zeta (5 ) \left( \abn \right)^6 + {\cal O}\left(\left( \abn \right)^7
\right) \;.
\end{equation}

The function $ \, R_N(\as) \, $  is given by
\begin{eqnarray}
\label{rn}
R_N(\as) &=& \left\{ \frac{\Gamma(1-\ga_N) \; \chi(\ga_N)}{\Gamma(1+\ga_N)
\;[-\ga_N \;\chi^{\prime}(\ga_N)]} \right\}^{\frac{1}{2}} \nonumber \\
&\cdot& \exp \left\{ \ga_N \;\psi(1) + \int_0^{\ga_N} d\ga
\;\frac{\psi^{\prime}(1) - \psi^{\prime}(1-\ga)}{\chi(\ga)} \right\} \;\;,
\end{eqnarray}
where $ \chi $ and $ \chi^{\prime} $ are the characteristic
function in Eq.~(\ref{cfmo57}) and its first derivative, respectively. The
$\, \as $-dependence of the r.h.s. in
Eq.~(\ref{rn}) is implicit in that of the gluon anomalous dimension
$ \, \ga_N = \ga_N (\as) $.
The first perturbative terms are $(\zeta(4)= 2 \zeta^2(2)/5 \simeq 1.082)$:
\begin{equation}
\label{rpert}
R_N(\as) = 1 + \frac{8}{3} \zeta(3) \left(\abn \right)^3 - \frac{3}{4} \zeta(4)
\left( \abn \right)^4 + \frac{22}{5} \zeta(5) \left( \abn \right)^5 +
{\cal O}\left( \left( \abn \right)^6 \right) \;\;.
\end{equation}
Some comments are in order.

The result in Eqs.~(\ref{glulead}),(\ref{implbfkl}) for the gluon anomalous
dimension is
exactly the celebrated  BFKL anomalous dimension [\ref{BFKL}]. In the present
paper this result has been derived by consistently carrying out  the
procedure of factorization of the collinear singularities in
dimensional regularization. However it is
worth noting
that the same
expression for the anomalous dimension can be obtained by using
alternative and less sophisticated regularization prescriptions of the
collinear singularities (for instance, considering the Eq.~(\ref{maeq})
directly in $ \, n = 4 \,$ dimensions and regularizing it by keeping the
incoming gluon slightly off-shell, via the replacement $ \delta^{(2)}
( \bk ) \to \delta ( \bk^2 - \mu^2) / \pi \, $ in the inhomogeneous term
[\ref{CFMO},\ref{BCM}]). The reason for this has to be traced back  to
the property of the kernel in the master equation (\ref{maeq}) of being
collinear regular for
$ \, \varepsilon \to 0$: the collinear divergences in the solution of
Eq.~(\ref{maeq}) originate only from the fact that the inhomogeneous term
is not sufficiently smooth for $ \, \bk \to 0$. As a result, in the
small-$N $ limit, the gluon anomalous dimensions to leading accuracy
$ \, ( \as / N )^k $ are {\em regularization/factorization scheme independent}
(and related via Eq.~(\ref{implbfkl}) to the eigenvalues $ \, \chi (\ga) \, $
of the BFKL kernel) within a wide class of schemes.
This class includes all
the schemes which do not introduce pathologically singular terms of the type
$ \, \as^k / N^{k + p} \; (p \geq 1) \, $ in the perturbative calculation
at high energy
\footnote{This feature
of the gluon anomalous dimensions
 was first pointed out by T. Jaroszewicz [\ref{jaro}].},
 or, more precisely, which do not violate the high-energy
power counting discussed in Sect.~2.3.

The BFKL anomalous dimension in Eqs.~(\ref{implbfkl}),(\ref{pertbfkl})
departs rather slowly from its one-loop contribution. However for very
small values of $ \, x  $, corresponding to sizeable values of $ \, \abar / N
\sim {\cal O}(1)  $, $ \, \ga_N \,$ increases quite fast and for $ \, \abar / N
=
( 4 \, \ln 2 )^{-1} \,$ reaches the saturation value $ \ga = 1 / 2 \, $
at which $ \, \chi ( \ga ) \,$ has a minimum (Fig.~6). For still larger
values of $ \, \abar / N \,$ there are two complex conjugate branches of
 $ \, \ga_N $, coming from the pinching with the symmetrical solution
of Eq.~(\ref{implbfkl}) at $ \ga = 1 - \ga_N$. Therefore the resummation
of the singular terms $ \,( \as / N )^k \, $ builds up a stronger singularity
at $ \, N = \lambda \equiv 4 \, \abar \, \ln 2 \simeq 2.65 \, \as$
[\ref{BFKL}].
As discussed in Sect.~1, this branch point singularity (known as the
{\em perturbative QCD pomeron}) is responsible for the steep behaviour
$ \, x^{ - \lambda} \, $ of the gluon density
$ \, {\tilde f}_{g} ( x , \mu^2) $,
generated by the perturbative QCD evolution at high scale $ \, \mu^2 $.

The function $ \, R_N ( \as) \, $
in Eq.~(\ref{rn}), on the other hand, depends on the factorization scheme
more than the gluon anomalous dimensions. For
instance, going from the $ \, \msbar$-scheme result (\ref{rn})  to the
$ \, {\mbox {\rm MS}} \, $ scheme (in this scheme the transition function
$ \, \Gamma_{g g } \, $ is obtained by setting $ \, S_\varepsilon = 1 \, $
on the r.h.s. of Eq.~(\ref{Gam})), $ \, R_N ( \as) \, $  has to be multiplied
by the factor $ \, \exp \left[ - \ga_N ( \as) \left( \psi(1) + \ln 4 \pi
\right) \right] $. This scheme dependence of $ \, R_N ( \as) \,$  has to be
compensated  in physical observables by subleading contributions of order
$ \, \as ( \as / N )^k \, $ in the anomalous dimensions.

Since $ \, R_N ( \as) \,$ is related to the $ \, \varepsilon  \to 0 \, $
limit of the renormalized gluon Green function $ \, G_{g g , N } \, $
(see Eq.(\ref{Rcoeff})), it can be regarded as the normalization factor of
the perturbative QCD pomeron. Note, in particular, that
 $R_N (\as) \, $ is singular at the saturation value $\ga=1/2$ of the
BFKL anomalous dimension $ \, \ga_N ( \as)$:
\begin{equation}
\label{rsing}
R_N(\as) \simeq {\mbox {\rm const.}} \left( \frac{1}{1-2\ga_N(\as)}
\right)^{\frac{1}{2}} \;\;, \;\;\; \;\; \;\;\;\; (\ga_N \rightarrow 1/2) \;\;.
\end{equation}
This behaviour is related to the
branch point singularity of the BFKL anomalous dimension at $N=
4 \, \abar  \ln 2$
and signals the ultimate failure of the
leading-twist approach to the mass
singularity factorization and the onset of the multi-Regge factorization at
extreme energies (see Ref.~[\ref{CCH}] for a more detailed discussion of this
issue).

We have so far considered the leading high-energy behaviour of the gluon
Green function
$ \, {\cal G}_{g g }^{(0)} $. In order to complete our
discussion on the gluon channel, we conclude this section by examining the
Green function
$ \, {\cal G}_{g q }^{(0)} $ with an incoming quark
or antiquark.

Let us denote by
$ \, {\cal F}_{ q }^{(0)} \,$  the $\, k_{\perp}$-dependent Green function
defined by replacing
$ \, {\cal G}_{g g }^{(0)} $ by
$ \, {\cal G}_{g q }^{(0)} $  on the r.h.s. of Eq.~(\ref{calF0}). The
corresponding master equation to leading accuracy $ (\as / N )^k \, $ is the
following
\begin{eqnarray}
\label{maeqiniq}
{\cf}^{(0)}_{q , \, N}(\bk; \as, \mu,\varepsilon) &=&  \frac{C_F}{C_A}  \;
 \frac{\abar}{N} \;
\frac{1}{(2\pi \mu)^{2\varepsilon}} \;\frac{1}{\pi {\bk}^2} \;
+ \frac{\abar}{N} \int
\frac{d^{2+2\varepsilon} \bq}{(2\pi \mu)^{2\varepsilon}} \;
\frac{1}{\pi {\bq}^2} \;
\nonumber \\
&\cdot& \left\{ \cf^{(0)}_{q , \,N}(\bk- \bq; \as , \mu,\varepsilon) -
\frac{\bk \cdot (\bk - \bq)}{(\bk - \bq)^2} \;\cf^{(0)}_{q , \,N}
(\bk; \as,  \mu ,\varepsilon)
\right\} .
\end{eqnarray}

Comparing Eq~.(\ref{maeqiniq}) and (\ref{maeq}), we see that
$ \, {\cal F}_{ q , \, N }^{(0)} \,$
and
$ \, {\cal F}_{ N }^{(0)} \,$
fulfil the same integral equation, apart from a different inhomogeneous term.
In particular, it is straightforward to check that the solution of
Eq.~(\ref{maeqiniq}) can be expressed in terms of the pure-gluon
Green function
$ \, {\cal F}_{ N }^{(0)} \,$  as follows
\begin{equation}
\label{quagluF}
{\cf}^{(0)}_{q , N}(\bk; \as, \mu,\varepsilon) =
 \frac{C_F}{C_A}  \; \left[
{\cf}^{(0)}_N(\bk; \as, \mu,\varepsilon) -  \delta^{(2+2\varepsilon)}(\bk)
\right] \;\;.
\end{equation}
The $ \, k_{\perp}$-integrated Green function
$ \, { G}_{ g q }^{(0)} \,$ is thus given by
\begin{eqnarray}
\label{integrGiniq}
{G}_{g q , \, N}^{(0)}(\as (Q^2/\mu^2)^\varepsilon , \varepsilon)
&\equiv&
\int d^{2+2\varepsilon} \bk
\; {\cf}^{(0)}_{q , \, N}(\bk; \as, \mu, \varepsilon) \; \Theta ( Q^2 - \bk^2)
\nonumber\\
&=&
 \frac{C_F}{C_A}  \; \left[
{G}_{g g , \, N}^{(0)}(\as (Q^2/\mu^2)^\varepsilon , \varepsilon)
- 1 \right] \;\;.
\end{eqnarray}

The result (\ref{integrGiniq}) allows us to compute the anomalous
dimensions $ \, \ga_{g q , \, N} \, $ by performing simple algebraic
manipulations.   We have first to factorize from
$ \, {G}_{g q , \, N}^{(0)} \, $ the collinear singularities according to
Eq.~(\ref{G0G}). Introducing explicitly the parton indices,
$ \,
{G}_{g q , \, N}^{(0)} = \sum_{a}
{G}_{g a , \, N}
{\Gamma}_{a q , \, N} $, and using the fact that the quark transition
functions $ \, \Gamma_{q b} \,$ are subleading at high energy, we obtain:
\begin{eqnarray}
\label{rengqcap2}
{G}_{g q , \, N}^{(0)}(\as (Q^2/\mu^2)^\varepsilon , \varepsilon) &=&
{G}_{g g , \, N}(\as (Q^2/\mu^2)^\varepsilon , \varepsilon) \;
{\Gamma}_{g q , \, N}(\as (Q^2/\mu^2)^\varepsilon , \varepsilon) \nonumber \\
&+& {G}_{g q , \, N}(\as (Q^2/\mu^2)^\varepsilon , \varepsilon) \;.
\end{eqnarray}
Inserting Eqs.~(\ref{rengqcap2}) and (\ref{renggcap2}) into
Eq.~(\ref{integrGiniq}) we get the identity
\begin{equation}
\label{idiniq}
C_A \left[ {G}_{g g , \, N}( \as  , \varepsilon )
\,{\Gamma}_{g q , \, N}( \as  , \varepsilon )
+ {G}_{g q , \, N}( \as  , \varepsilon ) \right] =
C_F \; \left[ {G}_{g g , \, N}( \as  , \varepsilon )
\,{\Gamma}_{g g , \, N}( \as  , \varepsilon ) - 1 \right] \;,
\end{equation}
which we can rewrite in the following form:
\begin{eqnarray}
\label{idiniqbis}
&\,& \left[ C_A \, {G}_{g q , \, N}( \as  , \varepsilon )
- C_F \; \left( {G}_{g g , \, N}( \as  , \varepsilon )
 - 1 \right) \right]
\; { 1 \over {
 {G}_{g g , \, N}
 ( \as  , \varepsilon ) }} \nonumber \\
&\,& = C_F \; \left[
{\Gamma}_{g g , \, N}( \as  , \varepsilon )
 - 1 \right] - C_A \,{\Gamma}_{g q , \, N}( \as  , \varepsilon ) \;.
\end{eqnarray}
We now notice that, order by order in perturbation theory,
the renormalized Green functions $ \, G_{a b } \, $ are regular for
$ \, \varepsilon \to 0 $ whilst the transition functions $\Gamma_{ab}$ are
series
of poles. Therefore the only solution to Eq.~(\ref{idiniqbis}) is
\begin{equation}
\label{solidiniq1}
 {G}_{g q , \, N}
 ( \as  , \varepsilon )
= \frac{C_F}{C_A}  \; \left[
{G}_{g g , \, N}
 ( \as  , \varepsilon )
 - 1 \right] \;\;\;,
\end{equation}
\begin{equation}
\label{solidiniq2}
   {\Gamma}_{g q , \, N}( \as  , \varepsilon )  =
 \frac{C_F}{C_A}  \; \left[{\Gamma}_{g g , \, N}( \as  , \varepsilon )
 - 1 \right] \;\;\;.
\end{equation}
In particular, since to leading accuracy $ \, ( \as / N )^k \,$ we have
\begin{equation}
\label{trangq}
\Gamma_{g q, N} (\as  , \varepsilon) \;
=   \frac{1}{\varepsilon} \int_0^{\as \, S_\varepsilon}
\frac{d\al}
{\al} \, \left[
\ga_{g q,N}(\al) \, +
\ga_{g g,N}(\al) \, \Gamma_{g q,N}(\al, \varepsilon) \right] \;\;,
\end{equation}
\begin{equation}
\label{trangg}
\Gamma_{g g,N}(\as  , \varepsilon) \;
=  1 +  \frac{1}{\varepsilon} \int_0^{\as }
\frac{d\al}
{\al} \, \ga_{g g,N}(\al) \, \Gamma_{g g,N}(\al, \varepsilon) \;\;,
\end{equation}
from Eq.~(\ref{solidiniq2}) we obtain:
\begin{eqnarray}
\label{bfkliniq}
   {\gamma}_{g q , \, N}
 ( \as  )  &=&
 \frac{C_F}{C_A}  \;
{\gamma}_{g g , \, N}
 ( \as  ) + {\cal O}(\as (\as/N)^k )
\nonumber\\
&=&
 \frac{C_F}{C_A}  \;
{\gamma}_{ N}   ( \as  ) + {\cal O}(\as(\as/N)^k ) \;,
\end{eqnarray}
$ \, {\gamma}_{ N}   ( \as  ) \,$ being the BFKL anomalous dimension.
We  see that the coefficients of the leading terms
$ \, ( \as / N )^k \,$ in
$\, {\gamma}_{g a,N } \; \;\,( a = q, g ) \, $ are equal, apart from an
overall factor given by the ratio of the colour charges of the
initial-state parton $ \, a  $.

\vskip 1 true cm

\newpage

\setcounter{equation}{0}
\setcounter{sect}{4}
\setcounter{footnote}{0}

%\vskip 3 true cm
\noindent {\bf 4. The quark Green functions }
\vskip 0.2 true cm

{}From the high energy power counting discussed in Sect.~2.3, we know that
the gluon channel dominates to leading logarithmic accuracy in the
small-$x \,$ regime. Beyond leading order, however, the quark sector has to be
considered on an equal footing with the gluon sector. Actually, from a
phenomenological viewpoint, the knowledge of the next-to-leading quark
anomalous dimensions may be more relevant than that of the corresponding
corrections to the gluon anomalous dimensions. The reason for this is that
the most accurate information on small-$x \, $ parton densities is coming out
from HERA data [\ref{DATA}] on deep inelastic structure functions, which
couple directly to quarks (and not gluons).

In order to compute the (flavour-singlet) quark anomalous dimensions we
consider the Green function
$\, {\cal G}_{q a}^{(0)} \,$:
\begin{equation}
\label{2GIquagreenshort}
{\cal G}_{q a}^{(0)} =  \sum_b \,
K^{(0) }_{q b} \,{\cal G}^{(0)}_{b a}
\;\;,
\end{equation}
or, more precisely, its expansion in 2GI kernels (Fig.~4b),  and we apply
the high-energy factorization formula (\ref{AgG}). We thus arrive at the
analogue of the
$\, k_{\perp}$-factorization formula (\ref{ktfac2}):
\begin{equation}
\label{quagrektfaceps1}
{\cal G}_{q g}^{(0) \; \al \beta} ( q, p)
= \int d^{2+ 2 \, \varepsilon} \bk
\int_0^1 {{dy} \over y}
\;
\left(
{\hat K}^{(0) \; \al \beta}_{\mu \nu} (q, k) \,
{ \frac{ k_{\perp}^{\mu} \, k_{\perp}^{\nu} }{\bk^2} }
\right)_{| \, k = y p
+ k_{\perp}}
{\cal F}^{(0)} ({y} , \bk; \as, \mu, \varepsilon) \;\;\,,
\end{equation}
where
$\, {\hat K}^{(0)} \,$ (Fig.~7) is the 2GI kernel
$\, { K}^{(0)}_{q g } \,$
to the lowest order in $ \, \as \, $
($\mu , \, \nu \, $ and
$\, \al , \, \beta \, $ are respectively the spin indices of the incoming
gluon and  outgoing quark). Moreover, since we are interested in the
anomalous dimensions rather than the coefficient function, it is also
convenient to apply the collinear projector $ \, {\cal P}_C \,$ and
consider (as in the gluon sector) the Green function
$ \, { G}^{(0)}_{q g} \,$ integrated over $ \, q_{\perp} \, $ and the
invariant mass $ \, q^2 \, $ up to the factorization scale $ \, \mu_F^2 =
Q^2 \,$ (we use the Sudakov parametrization $ \, q^{\mu} = x \, p^{\mu} +
q_{\perp}^{\mu} + ( q^2 + {\bq}^2 ) \,
{\bar p}^{\mu} /
2 \, x \, p \cdot {\bar p}\;$)
\begin{eqnarray}
{\cal P }_C \;{\cal G}^{(0)}_{q g} &\equiv&
{ G}^{(0)}_{q g}(x, \as (Q^2/\mu^2)^\varepsilon, \varepsilon)
\nonumber\\
&=&
\int \frac{dq^{2} \, d^{2 + 2 \varepsilon} \bq}{2(2\pi)^{4+2\varepsilon}}
\left(\frac{\bar p\!\!\!/}{2{\bar p}\cdot q} \right) _{\alpha\beta}
{\cal G}^{(0) \; \al \beta}_{qg}(q,p) \,\; \Theta\!\left( Q^2 - |q^2|
\right) \;.
\label{Gdritto}
\end{eqnarray}
Inserting Eq.~(\ref{quagrektfaceps1}) into Eq.~(\ref{Gdritto}) and taking the
$ \, N$-moments we get
\begin{equation}
\label{quagrektfaceps2}
G_{q g , \, N }^{(0)}(\as (Q^2/\mu^2)^\varepsilon , \varepsilon)
= \int d^{2+ 2 \, \varepsilon} \bk
\,\; {\hat K}_{ q g , \, N }(\bk^2/Q^2, \as (Q^2/\mu^2)^\varepsilon;
\varepsilon) \;
{\cal F}^{(0)}_{N}(\bk; \as, \mu, \varepsilon) \,\;\;.
\end{equation}

The factorization formula
(\ref{quagrektfaceps2})
relates the quark Green function to the $ \, k_{\perp}$-distribution
$ \, {\cal F}^{(0)} \,$  in the same way as  Eq.~(\ref{ktfac2}) relates
the heavy-flavour cross section to the gluon density. The
{\em off-shell}  kernel
$\, {\hat K}_{  q g , \, N} $, obtained from
$\, {\hat K}^{(0) } (q, k) \,  $ after integration over $ \, q \,$ and
explicitly calculated in App.~C, is indeed regular for $ \, N \to 0 $
so that the resummation of the singular terms $ \, (\as / N )^k \,$
in $\, { G}_{  q g , \, N}^{(0)} \,$ is achieved
by $ \, k_{\perp}$-integration of the corresponding terms in
$ \, {\cal F}^{(0)} $. Note, however, an essential physical difference
with respect to Eq.~(\ref{ktfac2}). Unlike the hard cross section
$ \, {\hat \sigma}_N $, the off-shell kernel
$\, {\hat K}_{  q g , \, N} \,$ is {\em not} collinear safe. The
collinear singularity arises from the integration of the splitting process
$ \, g \to q \, \bar q \, $ in Fig.~7b and shows up
as an $ \, \varepsilon$-pole in the $ \, n$-dimensional on-shell case
$\, \bk^2 = 0 \,$:
\begin{eqnarray}
\label{apkernel1}
{\hat K}_{ q g}(z, \bk^2 = 0, \as (Q^2/\mu^2)^\varepsilon; \varepsilon) &=&
z \,  {\as \over {2 \, \pi}}
\, S_\varepsilon \,
{{e^{\varepsilon \, \psi(1)}} \over {\Gamma(1 + \varepsilon)}} \,
 \left( {{(1-z)Q^2} \over \mu^2} \right)^{ \varepsilon}
\nonumber\\
&\cdot& {1 \over \varepsilon} \,
T_R \, { {(1 - z)^2 + z^2 + \varepsilon } \over {1+ \varepsilon}} \;\;
%P^{(0)}_{q  g} ( z, \varepsilon)
\;, \;
\end{eqnarray}
or, alternatively, as a logarithmic divergence in the on-shell limit
$\, \bk^2 \to 0 \,$ for the $ \, $ four dimensional case:
\begin{eqnarray}
\label{apkernel2}
{\hat K}_{ q g}\!\left( z, {\bk^2}/{Q^2}, \as , \varepsilon = 0 \right)
&=&\!z
\;  {\as \over {2 \, \pi}} \; \Theta\!\left( Q^2 - z \, \bk^2 \right)
\\
&\cdot&\!T_R \left[
\left( (1-z)^2 + z^2 \right) \ln {{Q^2} \over {z \, \bk^2}}
- \left( 1 - 6 z (1-z) \right) \left( 1 -  { {z \, \bk^2}\over Q^2} \right)
\right] \;. \nonumber
\end{eqnarray}
Note that the coefficients of both the $ \, \varepsilon$-pole and
$ \, \ln (
 {{Q^2} / {  \bk^2}} ) $-term are correctly proportional to the
$ \, n$-dimensional Altarelli-Parisi splitting function
$\, P^{(0)}_{ q g} \,$ in one-loop order:
\begin{equation}
\label{apsplieps}
P^{(0)}_{ q g} (z; \varepsilon) =
T_R \, { {(1 - z)^2 + z^2 + \varepsilon } \over {1+ \varepsilon}} \;\;.
\end{equation}
Actually, the full off-shell kernel
$\, {\hat K}_{  q g , \, N} \,$ in Eq.~(\ref{quagrektfaceps2})
can consistently be interpreted as the integral of a generalized
{\em off-shell} (and positive definite!) splitting function
$\, {\hat P}^{(0)}_{ q g} $. The explicit calculation in App.~C gives
\begin{eqnarray}
\label{K0res}
{\hat K}_{ q g}(z, {\bk^2}/Q^2, \as (Q^2/\mu^2)^\varepsilon; \varepsilon) &=&
z \; \Theta\!\left( Q^2 - z \, \bk^2 \right)  \,
\int_{0}^{(1-z) (Q^2 - z {\bf k}^2)} \,
\frac{d {\tilde \bq}^2}{ {\tilde \bq}^2}
 \left( {{{\tilde \bq}^2} \over \mu^2} \right)^{ \varepsilon}
\nonumber\\
&\cdot& {\as \over {2 \, \pi}}
\, S_\varepsilon \,
{{e^{\varepsilon \, \psi(1)}} \over {\Gamma(1 + \varepsilon)}} \,
 {\hat P}^{(0)}_{ q g}(z , \bk^2/{ \tilde \bq}^2; \varepsilon) \;,
\;
\end{eqnarray}
where the splitting function is
\begin{equation}
\label{K0res1}
{\hat P}^{(0)}_{ q g}(z , \bk^2/{ \tilde \bq}^2; \varepsilon) =
T_R \,
\left(
{{{\tilde \bq}^2} \over { {\tilde \bq}^2 + z \, (1-z) \, \bk^2 }}
\right)^2
\, \left[
{ {(1 - z)^2 + z^2 + \varepsilon } \over {1+ \varepsilon}} \, +
4 \, z^2 \, (1-z)^2 \,
 { \bk^2 \over {{ \tilde \bq}^2}}
\right] \;,
\end{equation}
and $ \, {\tilde \bq} = \bq - z \, \bk \, $ denotes the boost-invariant
(along the $ \, k$-direction) transverse momentum transferred in the
splitting process $ \, g \to q \bar q \,$ of Fig.~7b.

The fact that
the kernel
$\,{\hat K}_{ q g} \,$ is not collinear safe
prevents us from taking
the
$ \varepsilon \to 0 \, $ limit in
Eq.~(\ref{quagrektfaceps2})
even after
having factorized
the transition function $ \, \Gamma_{g g , \, N}
\,$ as in Eqs.~(\ref{ktfac2}),(\ref{calF0N}). The non-polynomial
$\, \varepsilon$-dependence
of $\,{\hat K}_{ q g} \, $
(due to
both $ \,
 {\hat P}^{(0)}_{ q g} \,$ and the phase space
integration
in Eq.~(\ref{K0res})),
as much as
the off-shell dependence of
$\, {\hat P}^{(0)}_{ q g} $, are thus responsible for non-trivial
(and factorization scheme dependent) quark anomalous dimensions
$ \, \gamma_{q g , \, N} \,$ in higher perturbative orders.

Using the power series solution (\ref{sersol}) for the gluon
distribution $ \, {\cal F}_{N}^{(0)} \,$ and performing the
$ \, k_{\perp}$-integration in
Eq.~(\ref{quagrektfaceps2}),
we can
determine
the
corresponding power series expansion for the quark Green function
$\, { G}^{(0)}_{ q g , \, N} $. Considering the $ \, N \to 0 \, $ limit
for
$\,{\hat K}_{ q g ,\, N} \,$
(i.e., neglecting
subdominant corrections at high energy)
we find (see App.~C)
\begin{eqnarray}
\label{quagresersol}
&~& {G}^{(0)}_{ q g , \, N}(\as (Q^2/\mu^2)^\varepsilon, \varepsilon) =
{\as \over {2 \, \pi}} \, T_R \,
\, S_\varepsilon \,
{{e^{\varepsilon \, \psi(1)}} \over {\Gamma(1 + \varepsilon)}} \,
\left( {Q^2 \over \mu^2} \right)^{ \varepsilon} \,
{1 \over \varepsilon} \,
{{4 + \varepsilon} \over {( 3 + \varepsilon) \, ( 2 + \varepsilon)}}
\;\;\;\;\;\;\;
\nonumber\\
&~& \cdot \left\{
1 + \sum_{k=1}^\infty
\, \left[ \abn
\, S_\varepsilon \,
{{e^{\varepsilon \, \psi(1)}} \over {\Gamma(1 + \varepsilon)}} \,
\left( {Q^2 \over \mu^2} \right)^\varepsilon
\, \right]^k \, {1 \over {k \, \varepsilon}} \, d_k (\varepsilon)
\,  \right\} +
 {\cal O}\left( \as^2 (\as/N)^k \right) \;\;\,,
\end{eqnarray}
where the coefficients
$ \, d_k (\varepsilon) \,$ are expressed in terms of
$ \, c_k (\varepsilon)  \,$ in Eq.~(\ref{recurs}) as follows
\begin{equation}
\label{dneps}
d_k (\varepsilon) =
{1 \over {k + 1}} \, { { 4 + (1-3k) \varepsilon } \over
{ 4 + \varepsilon}} \,
{ { \Gamma (1 + k \, \varepsilon) \, \Gamma(1 - k \, \varepsilon) \,
\Gamma ( 1 + \varepsilon) \,
\Gamma( 4 + \, \varepsilon) } \over { \Gamma ( 1 + (1+k) \varepsilon ) \,
\Gamma(4 + (1-k) \varepsilon ) } } \,
c_k (\varepsilon)
 \;\;\;.
\end{equation}

On the other side, the
 procedure of factorization of the collinear singularities
in Eq.~(\ref{G0G}) gives
$ \,
{G}_{q g , \, N}^{(0)} = \sum_{a}
{G}_{q a , \, N}
{\Gamma}_{a g , \, N} $, which, neglecting terms of order
  $ \as^2 ( \as / N )^k  $, can be written as
\begin{eqnarray}
\label{renqgcap4}
{G}_{q g , \, N}^{(0)}(\as (Q^2/\mu^2)^\varepsilon , \varepsilon)
&=& {G}_{q g , \, N}(\as (Q^2/\mu^2)^\varepsilon , \varepsilon)
\; {\Gamma}_{g g , \, N}(\as (Q^2/\mu^2)^\varepsilon , \varepsilon)
\nonumber \\
&+& {\Gamma}_{q g , \, N}(\as (Q^2/\mu^2)^\varepsilon , \varepsilon) \;\;\;,
\end{eqnarray}
\begin{equation}
\label{tranqg}
\Gamma_{q g,N}(\as, \varepsilon) \;
=   \frac{1}{\varepsilon} \int_0^{\as \, S_\varepsilon}
\frac{d\al}
{\al} \, \ga_{q g,N}(\al) \; \Gamma_{g g,N}(\al, \varepsilon) \;\;.
\end{equation}
Remember that $ G_{q g , \, N}$ is finite for $ \varepsilon \to 0 $
order by order in $ \as $.
The structure of the $ \varepsilon$-poles on the r.h.s. of
Eq.~(\ref{renqgcap4}) thus defines  $ G_{q g , \, N}$ and the
transition function $  \Gamma_{q g , \,N} \, $ uniquely. It follows that,
 comparing the expansion
(\ref{quagresersol})
with Eqs.~(\ref{renqgcap4}),(\ref{tranqg}) and using the known result for
$\, \Gamma_{g g,N} \,$ (i.e. the leading-order gluon anomalous
dimension (\ref{glulead})), we can  compute the anomalous dimension
 $ \, \ga_{q g , \, N } \, $ (and the renormalized Green function
$ \, {G}_{q g , \, N}  $) order by order in perturbation
theory.

We have explicitly performed this calculation up to the sixth order and
the result for the quark anomalous dimensions in the $ \, \msbar \,$
scheme reads
\begin{eqnarray}
\label{qms}
\ga_{qg,N}(\as) &=& \frac{\as}{2\pi} T_R \;\frac{2}{3}
\left\{ 1 + \frac{5}{3} \abn +
\frac{14}{9} \left(\abn \right)^2 +
\left[\frac{82}{81}+ 2 \, \zeta (3)    \right] \,
 \left(\abn \right)^3
+ \left[ {122 \over 243}
\right.
\right.
\nonumber\\
&+& \left.
\left.
 {25 \over 6} \, \zeta (3) \right] \,
 \left(\abn \right)^4
+ \left[ {146 \over 729}+{14 \over 3}\,\zeta (3)+2\,\zeta (5) \right] \,
 \left(\abn \right)^5
+ {\cal O}\!\left( \left(\abn \right)^6 \right) \right\} \nonumber \\
&\simeq& \frac{\as}{2\pi} T_R \;\frac{2}{3} \left\{ 1 + 1.67 \abn +
1.56 \left(\abn \right)^2 +
3.42 \left(\abn \right)^3 \right.
\nonumber\\
&+& \left.
5.51 \left(\abn \right)^4 +
7.88 \left(\abn \right)^5 +
{\cal O}\!\left( \left(\abn \right)^6 \right) \right\} \;\;.
\end{eqnarray}
The calculation is straightforward but already very cumbersome
to this
order in perturbation theory\footnote{Note that, because of the cancellations
involved in the gluon sector (see Eqs.~(\ref{Ismalleps}),(\ref{cubic})), the
first three coefficients in the perturbative expansion (\ref{qms}) are
entirely related to the $\varepsilon$-dependence of the off-shell kernel
${\hat K}_{ q g , \, N }$.}, and we are not able to provide an explicit
resummed formula for $ \, \ga_{q g} \, $ to all orders (some all-order
features of $\ga_{qg}$ are presented in App.~C). However we emphasize that this
is just an (open) algebraic problem  related  to the use of the $ \, \msbar \,$
factorization scheme, that is the highly non-trivial
$ \, \varepsilon$-dependence
of the coefficients $ \, d_k ( \varepsilon) \, $ in Eq.~(\ref{dneps}) and the
non-local structure of the $\, \varepsilon$-poles in Eq.~(\ref{tranqg}).
The series
(\ref{quagresersol}) for the quark Green function
indeed contains all the necessary information on the anomalous dimensions
$ \, \ga_{q g , \, N} \, $ to {\em any} perturbative order
  $ \,   \as ( \as / N )^k   $. In the next section we show that, choosing a
different factorization scheme of mass singularities, we can
explicitly resum
all the next-to-leading terms
  $ \,   \as ( \as / N )^k  \, $ in
$ \, \ga_{q g , \, N}  $.

The coefficients of the first two terms in the curly bracket of
Eq.~(\ref{qms}) agree with the known one- and two-loop anomalous
dimensions in the $ \, \msbar \,$ scheme [\ref{FP},\ref{Floratos}].  The
higher-order contributions can be used to estimate the effect of these
small-$x \, $ corrections at intermediate values of $ \, x $. In particular,
the $ \, {\cal O}(\as^3) \,$ term in Eq.~(\ref{qms}) can be combined with the
existing $ \,{\cal O}(\as^2) \,$ calculations of the coefficient functions
for the DIS [\ref{NDIS}] and Drell-Yan [\ref{DY}] processes in order to
check the stability of the fixed-order perturbative expansion in the
$ \, x$-range  accessible at present. Note also that, unlike the case of the
gluon anomalous dimensions (\ref{glulead}),(\ref{pertbfkl}), all the
perturbative coefficients in Eq.~(\ref{qms}) are non-vanishing
(and positive). Therefore in the quark sector one may expect [\ref{QAD}]
(and finds [\ref{EKL}]) an earlier departure from the fixed-order perturbative
behaviour.

The flavour-singlet anomalous dimension
$ \, \ga_{q q , \, N }^S \, $ (App.~A)
starts in order $ \, \as^2 $ and its resummed expression at high
energy is related in a simple way to that for
$ \, \ga_{q g , \, N}  $, namely
\begin{equation}
\label{qqq}
\ga_{qq,N}^S(\as) = \frac{C_F}{C_A} \left[ \ga_{qg,N}(\as) - \frac{\as}{2\pi}
\,T_R \,\frac{2}{3} \right] +
{\cal O}(\as^2 \left( \as/N)^k \right) \;\;.
%\nonumber
\end{equation}

The derivation of Eq.~(\ref{qqq}) is similar to that of the
analogous Eq.~(\ref{bfkliniq}) for the gluon sector.
One has to apply the high-energy factorization procedure of Sect.~2.3 to
the Green function $ \, G_{q q }^{S \; (0) } \,$ (here, the superscript
$ \, S \, $ denotes the fact that the incoming and outgoing quarks
carry a different flavour), thus obtaining the following
$ \, k_{\perp}$-factorization formula
\begin{equation}
\label{ktfaconcemore}
G_{q q , \, N}^{S \; (0)}(\as (Q^2/\mu^2)^\varepsilon , \varepsilon)
= \int d^{2+ 2 \, \varepsilon} \bk
\,\; {\hat K}_{ q g , \, N}(\bk^2/Q^2, \as (Q^2/\mu^2)^\varepsilon;
\varepsilon)
\; {\cal F}^{(0)}_{q , \, N }(\bk; \as, \mu, \varepsilon)  \;\;.
\end{equation}
This equation differs from Eq.~(\ref{quagrektfaceps2})
only by the replacement of the pure-gluon $ \, k_{\perp}$-distribution
$\, {\cal F}^{(0)}_{ N } \,$
by $\, {\cal F}^{(0)}_{q , \,  N } \,$ (see Eq.(\ref{maeqiniq})). Therefore,
using Eq.~(\ref{quagluF}), one can relate the two Green functions
$ \, G_{q q }^{S \; (0) } \,$  and
$ \, G_{q g }^{ (0) }\; $:
\begin{equation}
\label{Sqq1}
G_{q q , \, N}^{S \; (0)}
\left( \as \left( { Q^2 \over  \mu^2}\right)^\varepsilon , \varepsilon \right)
% \, G_{q q }^{S \; (0) }
% ( \as  , \varepsilon )
=
 \frac{C_F}{C_A}  \; \left[
{G}_{q g , \, N}^{(0)}
\left( \as \left( { Q^2 \over  \mu^2} \right)^\varepsilon ,
\varepsilon \right)  - {\hat K}_{q g , \, N}
\left( \bk= 0 , \as \left( { Q^2 \over  \mu^2}\right)^\varepsilon ,
\varepsilon \right) \right] \;\;. \;\;
\end{equation}
The factorization of collinear singularities in
$ \, G_{q q , \, N}^{S \; (0)} \,$ is achieved by the following
high-energy relations
\begin{equation}
\label{Sqq2}
G_{q q , \, N}^{S \; (0)}
 ( \as  , \varepsilon ) =
{\Gamma}_{q q , \, N}^{S }  ( \as  , \varepsilon )
+
G_{q q , \, N}^{S \; }
 ( \as  , \varepsilon )
+
G_{q g , \, N}
 ( \as  , \varepsilon ) \,
{\Gamma}_{g q , \, N}
 ( \as  , \varepsilon ) \;,
\end{equation}
\begin{equation}
\label{Sqq3}
{\Gamma}_{q q , \, N}^{S }  ( \as  , \varepsilon )
=   \frac{1}{\varepsilon} \int_0^{\as \, S_\varepsilon}
\frac{d\al}
{\al} \, \left[
\ga_{q q,N}^{S}(\al) \, +
\ga_{q g,N}(\al) \, \Gamma_{g q,N}(\al, \varepsilon) \right] \;\;.
\end{equation}
Using Eqs.~(\ref{Sqq2}),(\ref{Sqq3}),(\ref{solidiniq2}) one can express
$ \, G_{q q }^{S \; (0)} \,$ as a function of
$ \, G_{q q }^{S } $,
$ \, \ga_{q q }^{S } $,
$ \, G_{q g } $,
$ \, \ga_{q g } $,
$ \, \Gamma_{g g } $. Analogously, by means of
Eqs.~(\ref{renqgcap4}),(\ref{tranqg})
one can express
$ \, G_{q g }^{ (0)} \,$ as a function of
$ \, G_{q g } $,
$ \, \ga_{q g } $,
$ \, \Gamma_{g g } $. Inserting the expressions obtained in this way
into Eq.~(\ref{Sqq1}), one ends up with the equation
\begin{eqnarray}
\label{Sqq4}
&~& G_{q q , \, N}^{S } ( \as  , \varepsilon ) +
   \frac{1}{\varepsilon} \int_0^{\as \, S_\varepsilon}
\frac{d\al}{\al} \, \ga_{q q,N}^{S}(\al)
\nonumber\\
&~& = \frac{C_F}{C_A} \,
\left[
G_{q g , \, N} ( \as  , \varepsilon )
+   \frac{1}{\varepsilon} \int_0^{\as \, S_\varepsilon}
\frac{d\al}{\al} \,
\ga_{q g,N}(\al) -
{\hat K}_{q g , \, N}
( \bk= 0 , \as; \varepsilon )
  \right] \;\;, \;\;
\end{eqnarray}
from which the result (\ref{qqq}) follows.

It is worth noting
that the all-order relation
 (\ref{qqq}) between
$ \, \ga_{q q , \, N }^S \, $ and
$ \, \ga_{q g , \, N}  $, as well as  its leading-order
analogue  (\ref{bfkliniq}),
have been derived using only algebraic identities at high energy, with no
reference to the details of the dimensional-regularization prescription.
In particular they remain  true using dimensional reduction [\ref{DR}],
a dimensional regularization scheme which explicitly respects
supersymmetric Ward identities. In this sense, Eqs.~(\ref{bfkliniq}) and
(\ref{qqq}) can be considered as high-energy limits of the supersymmetry
identity
$ \, \ga_{q q }  +
 \ga_{g q }  =
  \ga_{q g }  + \ga_{g g }  \,$ valid in
$\, N=1 \,$ supersymmetric Yang-Mills theory, i.e. for $\, C_F = T_R = C_A $.
It follows that in the supersymmetric case the gluon anomalous dimensions
$\,  \ga_{g g }  \,$ and
$\,  \ga_{g q }  \,$ coincide also at the next-to-leading level
$\, \as (\as / N)^k $. This property may be useful as a technical tool
to check and simplify the calculation of the (still unknown) next-to-leading
corrections in the gluon sector.

\vskip 1.3 true cm

\setcounter{equation}{0}
\setcounter{sect}{5}
\setcounter{footnote}{0}

\noindent {\bf \boldmath 5. Deep inelastic scattering at small $ \, x $ }
\vskip 0.3 true cm

\vskip 0.5 true cm
\noindent {\it 5.1 Structure functions and parton densities}
\vskip 0.2 true cm

The cross section for deep inelastic lepton-hadron scattering  is given
in terms of the customary structure functions $ \, F_i ( x , Q^2 ) \, $
($ i = 1,2,3 $). Here $\, Q^2 \,$ denotes the square of the momentum
transferred by the scattered lepton and $ \, x \, $ is the Bjorken variable.
In the following we only consider   the scattering process occurring
through the exchange of a single photon. As far as  the hadronic
component is concerned, this approximation simply amounts to neglecting
$ \, F_3 $,
which is a non-singlet structure function and, hence, non-singular at
small $\, x \, $ (see Sect.~2.3). We also present our results in terms of the
structure functions $ \, F_2 \, $ and $ \, F_L  $, $ \, F_L (x, Q^2 ) =
 F_2 (x, Q^2 )  - 2 x F_1 (x, Q^2 ) \, $ being the longitudinal
structure function.

In the na\"{\i}ve parton model the DIS structure functions are related
to the
parton densities of the incoming hadron as follows ($e_i \, $ are the quark
charges)
\begin{equation}
\label{parmod2}
F_2 ( x , Q^2 ) = \sum_{i = 1}^{N_f} e_i^2 \left[
{\tilde f}_{q_i} (x , Q^2) +
{\tilde f}_{{\bar q}_i} (x , Q^2)
\right] \;\;\,,
\end{equation}
\begin{equation}
\label{parmodL}
F_L ( x , Q^2 ) = 0  \;\;\,.
\end{equation}
Taking into account perturbative QCD corrections in leading-twist order
and using the factorization theorem of mass singularities,
Eqs.~(\ref{parmod2}),(\ref{parmodL}) become
\begin{equation}
\label{disstrfun}
F_i ( x , Q^2 ) = {1 \over N_f} \, \left( \sum_{j = 1}^{N_f} e_j^2 \right)
F_i^S ( x , Q^2 )
+ F_i^{NS} ( x , Q^2 ) \;\;\;, \;\;\; (i = 2 , L) \;,
\end{equation}
where the singlet and non-singlet components
$ F^S \, $ and
$\, F^{NS} \,$
are given by
\begin{eqnarray}
\label{Fsing}
F_i^S (x,Q^{2}) &=&
\int_{x}^{1} {{d\, z} \over z}  \; \left[
C_{i}^S(z; \as(\mu^{2}_{F}),  {Q^{2}}/{\mu^{2}_{F}})
\,{\tilde f}_{S}(x/z, \mu^{2}_{F})  \right.
\nonumber\\
&+& \left.
C_{i}^g(z; \as(\mu^{2}_{F}), {Q^{2}}/{\mu^{2}_{F}})
\,{\tilde f}_{g} (x/z, \mu^{2}_{F}) \right] \;\;,
\end{eqnarray}
\begin{equation}
\label{Fnonsing}
F_i^{NS} (x,Q^{2}) =
\int_{x}^{1} {{d\, z} \over z}  \;
C_{i}^{NS}(z; \as(\mu^{2}_{F}), {Q^{2}}/{\mu^{2}_{F}})
\,
\sum_{j = 1}^{N_f} e_j^2
\,{\tilde f}_{q_j}^{(+)} (x/z, \mu^{2}_{F})
\;\;.
\end{equation}
Here
$ \, {\tilde f}_{S} \,$ and
$ \, {\tilde f}_{q_j}^{(+)} \,$
are respectively  the singlet and non-singlet quark densities
defined in App.~A, and $ \, C_i^{A} \, $ ($A= S, \, g ,  \, NS $) are the
coefficient functions\footnote{The coefficient functions in
Eqs.~(\ref{Fsing}),(\ref{Fnonsing}) are normalized according to our notation
in Sect.~2.1 (see, in particular, Eq.~(\ref{Can})). This normalization differs
from that often used in the literature [\ref{NDIS}].}
computable as power series in $ \, \as ( \mu^2_F) $.
Note that, because of the na\"{\i}ve parton model relation
(\ref{parmod2}), we have
$ \, C_2^{A} (z) = \delta ( 1- z) + {\cal O}( \as) \, $ for
$\, A= S,  \, NS $, and
$ \, C_2^{g} (z) = {\cal O}( \as)  $. Therefore also the gluon density $ \,
{\tilde f }_g \,$
contributes to $ \, F_2 \, $ beyond the lowest order. Moreover,
since
$ \, C_L^{A}  = {\cal O}( \as) $ is not vanishing, the Callan-Gross
relation (\ref{parmodL}) is violated.

At present, the coefficient functions $ \, C_i^A \, $ are completely known
up to ${\cal O}( \as^2) \, $ [\ref{NDIS},\ref{CDIS}]. The non-singlet
coefficients
$ \, C_i^{NS} \,$ are not enhanced by $ \, \ln x \, $ terms in higher orders.
The all-order resummation of the logarithmic contributions
$ \, \as^{n+2} {\ln}^n x \, $ (or $\, \as (\as / N )^{n+1} \,$ in the
$ \, N$-moment space) to
$ \, C_i^{g} \,$ and
$ \, C_i^{S} \,$ is performed in the next subsection.

\vskip 1 true cm
\noindent {\it 5.2 Coefficient functions}
\vskip 0.2 true cm

The coefficients functions $ \, C_i^A \,$ are evaluated starting from the
expression $ \, F_i = $
\linebreak
$\sum_{a} F_{i \,a}^{(0)} {\tilde f}_{a}^{(0)} \,$
of the hadronic structure functions $ \, F_i \,$ in terms of the partonic
structure functions
$ \, F_{i \,a}^{(0)} $, and then performing the collinear factorization
as in Eq.~(\ref{F(0)}). Considering first the gluon structure functions
$ \, F_{i \,g}^{(0)} $, we have
\begin{equation}
\label{Fgluhadr}
F_{i \, g}^{(0)} =
{ 1 \over N_f} \, \left( \sum_{j = 1}^{N_f} e_j^2 \right) \, \left[
C_{i}^{g} \, \Gamma_{g g } +
C_{i}^{S} \, 2 \, N_f \, \Gamma_{q g }
\right]
\;\;.
\end{equation}
On the other side, from the high-energy factorization in Sect.~2.3
(see Fig.~4a), we obtain
\begin{eqnarray}
\label{ktfacdis}
 F_{i \, g}^{(0)}(x; \as (Q^2/\mu^2)^\varepsilon,\varepsilon)  &=&
{ 1 \over N_f} \, \left( \sum_{j = 1}^{N_f} e_j^2 \right)
\int d^{2+2\varepsilon} {\bk} \\
&\cdot&
\int_x^1 \frac{d\, z}{z}\;
{\hat \si}_{i}^{g}(z, {{\bk}^{2}}/{Q^{2}}, \as (Q^2/\mu^2)^\varepsilon;
\varepsilon) \;{\cal F}^{(0)}_{N}({x/z}, {\bk}; \as , \mu ,
\varepsilon) \;\;, \nonumber
%{\tilde f}^{(0)}_{g,N} (\mu , \varepsilon)\;.
\end{eqnarray}
where
$ \;{\cal F}^{(0)} \,$
is the $ \, k_{\perp}$-dependent gluon Green function in Eq.~(\ref{calF0}),
and
$ \, {\hat \si}_{i}^{g} \,$ are obtained by applying the high-energy
projector $ \, {\cal P}_{H} \,$  to  the lowest-order $ \, q \bar q \, $
contribution to the $\,  \ga^* g \to \ga^* g \, $ absorptive part
$ \, A_{\mu \nu} \,$ (Fig.~5), as in Eq.~(\ref{AgP}). The off-shell cross
sections $ \, {\hat \si}_{i}^{g} \,$ have been explicitly evaluated in
Refs.~[\ref{CCH}] ($i = 2$) and [\ref{HERA}] ($i = L$)
for the case of massive quarks and $ \, n = 4 \,$ dimensions. The
generalization  to the massless case and $ \, n = 4 + 2 \, \varepsilon \, $
dimensions
is straightforward.

Eq.~(\ref{ktfacdis}) is the dimensionally-regularized version of the
$ \,  k_{\perp}$-factorization formula  (\ref{ktfac}). We can thus properly
address the issue of collinear-singularity factorization. The main point
to be noticed is that the off-shell cross sections
$ \, {\hat \si}_{L}^{g} \,$ and
$ \, {\hat \si}_{2}^{g} \,$ have a different collinear behaviour. As a
consequence of the Callan-Gross relation (\ref{parmodL}) (i.e.,
the fact that on-shell partons do not couple directly to a
longitudinally polarized photon),
$ \, {\hat \si}_{L}^{g} \,$ is collinear safe. Its on-shell and $ \,
\varepsilon = 0 \, $ limit is indeed finite:
\begin{equation}
\label{limitL}
{\hat \si}_{L}^{g}(z, {\bk = 0 }, \as (Q^2/\mu^2)^\varepsilon; \varepsilon)
= {\as \over { 2 \, \pi}} \, S_\varepsilon  \,
\left({Q^2 \over \mu^2}\right)^\varepsilon \,
N_f \, T_R \, \left[ 8 \, z^2 \, ( 1 - z)
+ {\cal O}( \varepsilon) \right]
\;\;.
\end{equation}
On the contrary,
$ \, {\hat \si}_{2}^{g} \,$ is not collinear safe and  its on-shell limit
has an  $ \, \varepsilon$-pole proportional to the lowest-order
Altarelli-Parisi splitting function $ \, P^{(0)}_{q g } \, $ in
Eq.~(\ref{apsplieps}):
\begin{equation}
\label{limit2}
{\hat \si}_{2}^{g}(z, {\bk = 0 }, \as (Q^2/\mu^2)^\varepsilon; \varepsilon)
= {\as \over { 2 \, \pi}} \, S_\varepsilon  \,
\left({Q^2 \over \mu^2}\right)^\varepsilon \,
N_f \, {1 \over \varepsilon} \,
 \left[ z
P^{(0)}_{q g } ( z; \varepsilon)
+ {\cal O}( \varepsilon) \right]
\;\;.
\end{equation}
Therefore, the factorization structures
(\ref{ktfacdis}) for
$\,  F_{L \, g}^{(0)} \, $ and
$\,  F_{2 \, g}^{(0)} \, $ are respectively analogous to
Eq.(\ref{ktfac2})
for the heavy-flavour case and
Eq.(\ref{quagrektfaceps2})
for the quark Green function. The factorization of collinear singularities
has to be carried out accordingly.

We start considering the longitudinal structure function. In the high-energy
limit, both the coefficient function $ \, C_L^{S} \, $ and the transition
function $ \, \Gamma_{q g} \, $ are of order $ \, \as ( \as / N )^k \,$
($k \geq 0$). Therefore from Eq.~(\ref{Fgluhadr})
we have
\begin{equation}
\label{rglong}
F_{L \, g}^{(0)} =
{ 1 \over N_f} \, \left( \sum_{j = 1}^{N_f} e_j^2 \right) \, \left[
C_{L}^{g} \, \Gamma_{g g } +
{\cal O}\!\left( \as^2 \, (\as/N)^k \right)
\right]
\;\;,
\end{equation}
and we see that, in order to compute $ \, C_L^g \, $, we have to factorize
$\, \Gamma_{g g } \, $ on the r.h.s. of Eq.~(\ref{ktfacdis}). Using the
factorization formula (\ref{calF0N}) and proceeding as in Sect.~2.4, we
arrive at the following expression for the $ \, N$-moments of the
longitudinal coefficient function
\begin{equation}
C_{L , \, N}^g(\as , Q^{2}/\mu^{2}_{F}) =
h_{L , \, N} \left( \gamma_{N} (\as) \right) \;
R_{N}(\as) \;(Q^{2}/\mu^{2}_{F})^{\gamma_{N}(\as)} +
{\cal O}( \as^2 (\as/N)^k) \;\;,
\label{CLNg}
\end{equation}
where the
function $h_{L , \, N} (\ga ) \, $
 is given in terms of the
off-shell cross section  $\, {\hat \si}_L^g \, $:
\begin{equation}
h_{L , \, N}(\gamma) = \gamma \int^{\infty}_{0} \frac{d
{\bk}^{2}}{{\bk}^{2}}\left(\frac{\bk^{2}}{Q^{2}}\right)^\gamma
\; {\hat \si}_{L , \, N}^{g}(\bk^{2}/Q^{2}, \as; \varepsilon = 0)\;.
\label{hLN}
\end{equation}

The explicit evaluation of the function
$ \, h_{L , \, N } ( \ga) \, $
requires the knowledge of the off-shell cross section $ \, {\hat \si}_L^g \,$
in $\, n = 4 \,$ dimensions. The analogous calculation for massive quarks
has been performed in great detail in Refs.~[\ref{CCH},\ref{HERA}]. Therefore
in this paper we limit ourselves to presenting only the final result.
%for $ \, h_{L , \, N } ( \ga) $.
In particular, since the off-shell cross section
 $ \, {\hat \si}_L^g \,$
in Eq.~(\ref{ktfacdis}) vanishes uniformly in $ \, k_{\perp} \,$ in the
high-energy limit $ \,  z \to 0 \,$ (as discussed in Sect.~2.3, this property
follows from the fact that only 2GI kernels contribute to
 $ \, {\hat \si}_L^g $), the function
$ \, h_{L , \, N } ( \ga) \, $ is weakly $ \, N$-dependent:
\begin{equation}
\label{weakly}
 \, h_{L , \, N } ( \ga) =
 \, h_{L  } ( \ga) \left( 1 + {\cal O}(N) \right)
\;\;\;, \;\;\; (\,  N \to 0 \,)
\;\;.
\end{equation}
In order to evaluate the
dominant
terms
$ \,  \as \, ( {\as / N } )^k \,$
in Eq.~(\ref{CLNg}), we can thus set $ \, N = 0 \, $ in
Eqs.~(\ref{hLN}),(\ref{weakly}). Computing explicitly the corresponding
function
$ \, h_{L  } ( \ga) \, $ ,
we find
\begin{equation}
\label{longres}
 \, h_{L  } ( \ga) = { \as \over { 2 \, \pi}} \, N_f \, T_R \,
{ { 4 \, ( 1 - \ga) } \over { 3 - 2  \ga }} \, { {
\Gamma^3 ( 1 - \ga) \, \Gamma^3 ( 1 + \ga) } \over {
\Gamma ( 2 - 2 \ga) \, \Gamma ( 2 + 2 \ga) } }
\;\;.
\end{equation}

%Eq.~(\ref{CLNg}) is the resummed expression for  $C_L^g$ we were looking for.
The results in Eqs.(\ref{CLNg}),(\ref{longres}) give the resummed
expression for the coefficient function $ \, C_{L ,\, N}^g \, $ in the
$ \, \msbar \,$ scheme to the logarithmic accuracy
$ \,  \as \, ( {\as / N } )^k $,
including  the corresponding
dependence on the factorization scale $ \, \mu_F \,$.
The resummation effect is incorporated in
(\ref{CLNg}) through the $ \, (\as/N)$-dependence of the BFKL
anomalous dimension (\ref{implbfkl}) and the $ \, \ga$-dependence
of $ \, R_N \, $ and
$ \, h_{L , \, N } \, $ as given by Eqs.~(\ref{rn}) and
(\ref{hLN}).

Let us now consider the structure function $ \, F_2 \,$. As noted
above, the pattern of collinear singularities in Eq.~(\ref{ktfacdis})
is similar to that in the corresponding Eq.~(\ref{quagrektfaceps2}) for the
quark Green function. Eq.~(\ref{ktfacdis}) thus contains all the
relevant information  on both the DIS coefficient function $ \, C_2^g \, $
and the quark anomalous dimensions $ \, \ga_{q g } $. As a matter of fact,
since in the small-$N \, $ limit $ \, C_{2 , \, N}^S = 1 +
{\cal O}(\as (\as/N)^k) $, from Eq.~(\ref{Fgluhadr})
we obtain the analogue of Eq.~(\ref{renqgcap4}):
\begin{eqnarray}
\label{rg2}
F_{2 \,g,N}^{(0)}(\as (Q^2/\mu^2)^\varepsilon, \varepsilon)
&=&
{ 1 \over N_f} \left( \sum_{j = 1}^{N_f} e_j^2 \right) \left[
C_{2 , \, N}^{g}(\as (\mu_F^2/\mu^2)^\varepsilon, {Q^2/\mu^2_F}; \varepsilon)
\, \Gamma_{g g ,\, N}(\as (\mu_F^2/\mu^2)^\varepsilon, \varepsilon)
\right.
\nonumber\\
&+& 2 \, N_f \,  \left.
\, \Gamma_{q g ,\, N }(\as (\mu_F^2/\mu^2)^\varepsilon, \varepsilon) +
{\cal O}(\as^2 (\as/N)^k ) \right] \;\;.
\end{eqnarray}
Starting from Eq.~(\ref{ktfacdis}) and factorizing the collinear
singularities according to Eq.~(\ref{rg2}), one can compute
$\, C_{2 , \, N}^{g} \,$ and
$ \, \gamma_{q g ,\, N } \,$ order by order in perturbation theory.

Obviously, this algebraic problem is by no means simpler than that
encountered in Sect.~4 for the evaluation of
$ \, \gamma_{q g ,\, N } $ and we are not able to provide an explicit
resummed formula for $ \, C_{2 , \, N}^g $. Nonetheless, Eq.~(\ref{ktfacdis})
can be used in a very simple way for deriving an all-order relation between
$\, C_{2 , \, N}^{g} \,$ and
$ \, \gamma_{q g ,\, N } $.

The main observation is that, setting $ \, \mu^2_F = Q^2 \,$ on the r.h.s.
of Eq.~(\ref{rg2}) and performing the derivative with respect to $ \, Q^2 $,
we obtain a factorized structure similar to Eq.~(\ref{rglong}):
\begin{eqnarray}
\label{rg2deriv}
&~&{\partial \over { \partial \ln Q^2} }
F_{2 \,g,N}^{(0)}(\as (Q^2/\mu^2)^\varepsilon , \varepsilon)
= { 1 \over N_f} \, \left( \sum_{j = 1}^{N_f} e_j^2 \right) \\
%\nonumber\\
&~& \cdot
\left[
\, \gamma_{g g ,\, N}(\as (Q^2/\mu^2)^\varepsilon )
\, C_{2 , \, N}^{g}(\as (Q^2/\mu^2)^\varepsilon, 1; \varepsilon)
+ 2 \, N_f \,
\, \gamma_{q g ,\, N }(\as (Q^2/\mu^2)^\varepsilon, \varepsilon)
 \right.
\nonumber\\
&~& + \left.  \varepsilon \, \as
{\partial \over { \partial \as } }
\, C_{2 , \, N}^{g}(\as (Q^2/\mu^2)^\varepsilon, 1; \varepsilon )
\right] \,
\, \Gamma_{g g ,\, N}(\as (Q^2/\mu^2)^\varepsilon , \varepsilon) +
{\cal O}(\as^2 (\as/N)^k )  \;\;. \nonumber
\end{eqnarray}
Here, all the collinear singularities are factorized into the gluon
transition function
$\, \Gamma_{g g \, } \,$
whereas  the term in the square bracket,
much like
$ \, C_L^g \, $ in Eq.~(\ref{rglong}), is finite as $ \, \varepsilon \to 0 $.
Stated differently, the off-shell kernel $ \, \partial {\hat \si}_2^g /
\partial \ln Q^2 \,$ in Eq.~(\ref{ktfacdis}) (although not
$ \, {\hat \si}_2^g \,$ itself)
is collinear safe, and its $ \, \varepsilon = 0 \,$ limit is related to the
linear combination $ \, \ga_{gg} C_2^g + 2 N_f \ga_{qg}$. Therefore,
proceeding as in the case of the longitudinal structure function, we obtain
\begin{eqnarray}
\label{lincomb}
 \ga_{N} (\as) \,  C_{2 , \, N}^g(\as, {Q^2/\mu^2_F} = 1 )
+ 2 N_f \;\ga_{qg , \,N }( \as)  &=&
h_{2 ,\,N}\left(  \ga_{N} (\as) \right) \, R_N (\as)
\nonumber\\
&+& {\cal O}(\as^2 ({\as/N})^k )  \;\;,  \;\;\;
\end{eqnarray}
where  $ \, R_N \,$ is given in Eq.~({\ref{rn}),
$\, \ga_{N} (\as) \, $ is the BFKL anomalous dimension and the function
$\, h_{2 ,\,N} (\ga) \,$
is
\begin{equation}
h_{2 , \, N}(\gamma) = \gamma \int^{\infty}_{0} \frac{d
{\bk}^{2}}{{\bk}^{2}}\left(\frac{\bk^{2}}{Q^{2}}\right)^\gamma
\; { {\partial } \over {\partial \ln Q^2}}
{\hat \si}_{2 , \, N}^{g}({\bk^{2}}/{Q^{2}},\as;
\varepsilon = 0)\;.
\label{h2N}
\end{equation}

The off-shell kernel
$\;  \partial
{\hat \si}_{2 , \, N}^{g}
/ {\partial \ln Q^2} \,$
in $ \, n = 4 \, $ dimensions has been computed in Ref.~[\ref{CCH}] for the
case of massive quarks. Performing the massless limit, we find
\begin{equation}
\label{weakly2}
 \, h_{2 , \, N } ( \ga) =
 \, h_{2  } ( \ga) \left( 1 + {\cal O}(N) \right)
\;\;\;, \;\;\; (\,  N \to 0 \,)
\;\;,
\end{equation}
\begin{equation}
\label{2res}
 \, h_{2  } ( \ga) = { \as \over { 2 \, \pi}} \, N_f \, T_R \,
{ { 2 \, ( 2 + 3 \ga  - 3 \ga^2) } \over { 3 - 2  \ga }} \, { {
\Gamma^3 ( 1 - \ga) \, \Gamma^3 ( 1 + \ga) } \over {
\Gamma ( 2 - 2 \ga) \, \Gamma ( 2 + 2 \ga) } }
\;\;.
\end{equation}

The results in Eqs.~(\ref{lincomb}),(\ref{weakly2}),(\ref{2res}) provide
the explicit resummation of the contributions
$ \,  \as \, ( \as / N  )^k \,$
to the coefficient function
$\,   C_{2 , \, N}^g \,$
in terms of
$ \, h_{2  } ( \ga_N) \,$
and the quark anomalous dimensions
$\, \ga_{qg , \,N }$. In particular, using the six-loop expression
(\ref{qms}) for
$\, \ga_{qg , \,N } $, one can compute
$\,   C_{2 , \, N}^g \,$
up to the five-loop order. The dependence of
$\,   C_{2 , \, N}^g \,$
on the factorization scale $ \, \mu^2_F \,$ is given by
\begin{eqnarray}
\label{C2N}
{C_{2 , \, N}^{g}}(\as , {Q^2/\mu_F^2})
&=&
{C_{2 , \, N}^{g}}(\as , {Q^2/\mu_F^2} =1)
\;
\left(  {Q^2 \over \mu_F^2} \right)^{\ga_N(\as)}
\nonumber\\
&+& 2 \, N_f \,
\, \frac{\gamma_{q g ,\, N } (\as )}{ \gamma_{ N }(\as )} \, \left[   \,
\left(  {Q^2 \over \mu_F^2} \right)^{\ga_N(\as)}
-1 \, \right]
+{\cal O}(\as^2 (\as / N )^k ) \;\;. \;\;
\end{eqnarray}

The singlet coefficient functions
$ \, C_L^S $,$\, C_2^S \, $ in Eq.~(\ref{Fsing}) can be evaluated starting
from parton structure functions $ \, F_{ i \, q }^{(0)} \,$ with an
incoming quark. These structure functions fulfil
a $ \, k_{\perp}$-factorization formula similar to Eq.~(\ref{ktfacdis}) with
the replacement $ \, {\cal F}^{(0)} \to
{\cal F}_q^{(0)} $,
$ \, {\cal F}_q^{(0)} \,$ being the $ \, k_{\perp}$-distribution with an
incoming quark in Eq.~(\ref{maeqiniq}). One can apply the algebraic
manipulations analogous to those used in Sects.~3 and 4 for evaluating
$\, \ga_{gq  }\,$ and
$\, \ga_{qq  }^S$, thus obtaining the following colour charge relations
\begin{equation}
\label{cpsL}
C_{L , \, N}^{PS}
\left( \as , {Q^2 \over \mu_F^2}  \right) =
\frac{C_F}{C_A} \left[
{C_{L , \, N}^{g}}
\left( \as , {Q^2 \over \mu_F^2}  \right)
- \frac{\as}{2\pi}  \,
N_f \,T_R \,\frac{4}{3} \right] +
{\cal O} \left( \as^2 ( \as/N)^k \right) \;\;,
\end{equation}
\begin{equation}
\label{cps2}
C_{2 , \, N}^{PS}
\left( \as , {Q^2 \over \mu_F^2}  \right) =
\frac{C_F}{C_A} \left[
{C_{2 , \, N}^{g}}
\left( \as , {Q^2 \over \mu_F^2}  \right)
- \frac{\as}{2\pi}  \,
N_f \,T_R \,\frac{2}{3} \,(1 + 2 \, \ln {Q^2 \over \mu_F^2} )
\right] +
{\cal O}\left(\as^2 ( \as/N)^k \right) \;\;.
\end{equation}
Here, for the sake of convenience,  we have introduced the pure-singlet
coefficient functions
$\, C_{i}^{PS} =
 C_{i}^{S} -
 C_{i}^{NS} $. They have the same singular small-$x \,$ behaviour as the
singlet functions $ \,
 C_{i}^{S} \,$ and start in $ {\cal O}( \as^2 ) \,$ in perturbation theory.

The perturbative expansions of the resummed results derived in this section
   read:
\begin{eqnarray}
\label{CMSL}
&~&{C_{L , \, N}^{g}}(\as , {Q^2/\mu_F^2} = 1 )
= \frac{\as}{2\pi} T_R N_f { 4 \over 3}
\left\{ 1 - \frac{1}{3} \abn +
\left[\frac{34}{9}-  \, \zeta (2)    \right]
\left(\abn \right)^2 + \left[ - \frac{40}{27} \right. \right.
\nonumber
\\
&~&+
\left.
{1 \over 3} \, \zeta (2) +
{8 \over 3} \, \zeta (3)    \right] \,
 \left(\abn \right)^3 +
\left[  \frac{1216}{81}-
{34 \over 9} \, \zeta (2) -
{14 \over 9} \, \zeta (3)  - 6 \, \zeta (4)  \right] \,
 \left(\abn \right)^4
\nonumber\\
&~&\left. + \, {\cal O}\left(\left(\abn \right)^5\right) \right\}
%\nonumber
\\
&~&\simeq \frac{\as}{2\pi} T_R N_f \frac{4}{3} \left\{ 1 - 0.33 \abn +
2.13 \left(\abn \right)^2 +
% \right. \nonumber\\ &+& \left.
2.27 \left(\abn \right)^3 +
0.43 \left(\abn \right)^4 \right. \nonumber \\
&~&+ \left.
{\cal O}\left(\left(\abn \right)^5\right) \right\} \;\;, \nonumber
\end{eqnarray}
\begin{eqnarray}
\label{CMS2}
&~& {C_{2 , \, N}^{g}}(\as , {Q^2/\mu_F^2} = 1 )
= \frac{\as}{2\pi} T_R N_f \;{ 2 \over 3}
\left\{
1 + \left[\frac{43}{9}- 2 \, \zeta (2) \right] \abn
+ \left[  \frac{1234}{81}- {13 \over 3} \, \zeta (2)
\right. \right.
\nonumber \\
&~& + \left.
%\left.
{4 \over 3} \, \zeta (3)    \right] \,
 \left(\abn \right)^2 +
\left[  \frac{7412}{243}- {71 \over 9} \, \zeta (2) +
{89 \over 9} \, \zeta (3) - 12
 \, \zeta (4)    \right] \,
 \left(\abn \right)^3
\nonumber\\
&~& +
%\left.
\left[  \frac{50012}{729}- {466 \over 27} \, \zeta (2) +
{910 \over 27} \, \zeta (3) - {28 \over 3} \, \zeta (2) \, \zeta(3) - 26
 \, \zeta (4)  + {24 \over 5} \, \zeta (5)   \right] \,
 \left(\abn \right)^4
\nonumber\\
&~& \left. + \, {\cal O}\!\left(\left(\abn \right)^5\right) \right\}  \\
%&~& + \left.
%{\cal O}\!\left(\left(\abn \right)^4\right) \right\} \nonumber \\
&~& \simeq \frac{\as}{2\pi} T_R N_f \frac{2}{3} \left\{ 1 + 1.49 \abn +
9.71 \left(\abn \right)^2 +
%\right. \nonumber\\ &+& \left.
16.43 \left(\abn \right)^3 +
39.11 \left(\abn \right)^4 \right. \nonumber \\
&~&+ \left.
{\cal O}\!\left(\left(\abn \right)^5\right) \right\} \;\;.
\nonumber
\end{eqnarray}
The first two coefficients in Eqs.~(\ref{CMSL}),(\ref{CMS2}) agree
with those recently computed in Refs.~[\ref{NDIS},\ref{Sanchez}]. We regard
this agreement as a non-trivial check of our results. Note also that
the three- and four-loop coefficients (and the five-loop coefficient in
Eq.~(\ref{CMS2})!) are substantially larger than the
two-loop ones. We thus argue that the higher-order contributions
computed in this paper may have a phenomenological relevance already at the
values of $x$ accessible at the HERA $ \, e p$-collider [\ref{DATA}].

Concluding this subsection, we point out that $ \, k_{\perp}$-factorization
formulae similar to Eq.~(\ref{ktfacdis}) have  recently been used [\ref{Dur}]
with the phenomenological aim of relating the original  BFKL equation
[\ref{BFKL}] to the DIS structure functions.

\vskip 1 true cm
\noindent {\it 5.3 The DIS factorization scheme}
\vskip 0.2 true cm

In the previous sections we have repeatedly noted that the parton
densities are not physical observables. Indeed they depend on the
regularization/factorization scheme used for removing the parton level
collinear singularities. This freedom in defining the parton densities
means that, starting from the $ \, \msbar \, $  densities
$ \, {\tilde f}_a $, one can introduce a new set
$ \, {\tilde f}_a^{\prime} \,$ of parton densities via an invertible
transformation
\begin{equation}
\label{invtbl}
 \, {\tilde f}_{a, \, N}^{\prime}  (\mu^2)
= \sum_b U_{a b ,\,  N}(\as(\mu^2)) \;{\tilde f}_{b, \, N}(\mu^2) \;\;.
\end{equation}
 Obviously a similar transformation applies to the coefficient functions,
in order to leave the physical cross section unchanged. The
evolution
of the new parton densities
with $ \, \mu^2 \,$
is controlled by
the new anomalous dimension matrix
\begin{equation}
\label{newandim}
\ga_{a b , \, N}^{\prime} =
\left[
\beta (\as) \, \left( \as \, {\partial \over {\partial \as}} U \right) \,
U^{-1} + U \, \ga \, U^{-1} \right]_{a b , \, N} \;\;.
\end{equation}

The transformation matrix $ \, U_{ a b } \,$ has a power series
expansion in $ \, \as \, $ such that
$\, U_{a b ,\,  N}(\as)$
$ = \delta_{ a b} + {\cal O}( \as) $, and has
to fulfil the following physical constraints :

i) flavour and charge conjugation invariance
\begin{eqnarray}
\label{chavour}
U_{q_ig}=U_{{\bar q}_ig} \equiv U_{qg} \;&,& \;\;
U_{gq_i}=U_{g{\bar q}_i} \equiv U_{gq} \;
\nonumber\\
U_{q_iq_j}=U_{{\bar q}_i{\bar q}_j} &\equiv&
\left( \delta_{ij} - { 1 \over N_f} \right) \, U_{qq}^{NS} +
{1 \over { 2 \, N_f}} \,
\left( U^{(V)} + U_{SS} \right) \;, \\
U_{q_i{\bar q}_j}=U_{{\bar q}_iq_j} &\equiv&
\left( \delta_{ij} - { 1 \over N_f} \right) \,U_{q{\bar q}}^{NS}  -
{1 \over { 2 \, N_f}} \,
\left( U^{(V)} - U_{SS} \right)  \;\;; \nonumber
\end{eqnarray}

ii) fermion number conservation
\begin{equation}
\label{fermnum}
U_{q_iq_j , \, N=0} - U_{{ q}_i{\bar q}_j , \, N=0}
= \delta_{i j}
\end{equation}
or, equivalently, $ \,
U_{qq , \, N=0}^{NS} =   U_{{ q}{\bar q} , \, N=0}^{NS} =
U_{N = 0}^{(V)} / 2 $;

iii) longitudinal momentum conservation
\begin{equation}
\label{longcons}
\sum_{a} \, U_{a b , \, N=1} = 1 \;\;\;.
\end{equation}
Eq.~(\ref{chavour}) is the analogue of Eq.~(\ref{adpr}) in App.~A for the
anomalous dimensions $ \, \ga_{ a b } $. In particular,
%the constraint i)
it guarantees that the flavour singlet and non-singlet sectors are
decoupled in any regularization/factorization scheme. The matrix
components $\,
U_{qq }^{NS} , \,   U_{{ q}{\bar q} }^{NS} , \,
U^{(V)}  \,$ introduced in Eq.~(\ref{chavour}) act on the flavour non-singlet
parton densities, whilst
 $\,
U_{SS } , \,   U_{{ q}{g} } , \,
U_{gq } , \,   U_{{ g}{g} }  \,  $
control the transformation on the singlet sector.

Higher-order QCD calculations for hadron collisions are usually performed
in two different factorization schemes of collinear singularities, the
$ \, \msbar \,$ scheme, used so far in this paper, and the  DIS scheme
[\ref{AEM}]. After having regularized the collinear singularities
in the parton matrix elements, the DIS-scheme parton densities
$ \, {\tilde f}_a^{(DIS)} \,$ are defined by enforcing the constraint
that the DIS structure function $ \, F_2 (x , Q^2) \,$ has the same
expression as in the na\"{\i}ve parton model. In particular, in the
one-photon
approximation to deep inelastic lepton-hadron scattering, the relation
 (\ref{parmod2}) is true to all orders in perturbation theory:
\begin{equation}
\label{f2disdef}
F_2(x,Q^2) = \sum_{i=1}^{N_f} e_i^2 \;\left[
{\tilde f}_{q_i}^{(DIS)}(x,Q^2) +
{\tilde f}_{{\bar q}_i}^{(DIS)}(x,Q^2) \right] \;\;.
\end{equation}
Equivalently, one can say that in the DIS scheme the DIS coefficient
functions are\footnote{Note that Eqs.~(\ref{f2disdefbis}) hold true only for
a factorization scale $ \, \mu^2_F = Q^2 $.}
\begin{eqnarray}
\label{f2disdefbis}
C_2^{NS \; (DIS)}(z; \as ( Q^2) , {Q^2/\mu^2_F} = 1 ) &=&
C_2^{S \; (DIS)}(z; \as ( Q^2) , {Q^2/\mu^2_F} = 1 ) =
\delta ( 1 - z) \;,
\nonumber\\
C_2^{g \; (DIS)}(z; \as ( Q^2) , {Q^2/\mu^2_F} = 1 ) &=&
0    \;\;.
\end{eqnarray}

In order to evaluate  higher-order contributions in the small-$x \, $
regime, the DIS scheme offers some
computational and phenomenological advantages [\ref{QAD}]. The former
 amounts to the fact that in the DIS scheme
one can explicitly resum the corrections  $ \, \as \, ( \as / N )^k \,$
for the quark anomalous dimensions to all orders in $ \, \as $, as we shall
illustrate below.
As to the latter, we notice that the next-to-leading contributions
  $ \, \as \, ( \as / N )^k \,$ to the gluon
anomalous dimensions are still unknown. Therefore, the knowledge of the
quark anomalous dimensions in the DIS scheme may facilitate
phenomenological investigations of the small-$ x \, $ behaviour of the
structure function $ \, F_2 ( x , Q^2 ) $.

The comment above applies once the DIS scheme has been defined
to all orders in perturbation theory.  The point is that Eq.~(\ref{f2disdef})
(or, equivalently, (\ref{f2disdefbis})) fixes
only the quark
densities
unambiguously.
The relation between the singlet quark density in the DIS
scheme and the $\msbar$-scheme parton densities is
\begin{equation}
\label{qdis}
{\tilde f}_{S , \, N}^{(DIS)}(\mu^2) =
C_{2, \, N}^{S} (\as(\mu^2),1) \;{\tilde f}_{S, \, N}(\mu^2) +
C_{2, \, N}^{g} (\as(\mu^2),1) \;{\tilde f}_{g, \, N}(\mu^2)
\end{equation}
or, in terms of the matrix $ \, U_{a b } \,$ in Eq.~(\ref{invtbl}),
\begin{equation}
\label{qdisbis}
U_{SS , \, N } (\as) =
C_{2, \, N}^{S} (\as,1) \;\;\;, \;\;\;
2 \, N_f \, U_{qg  , \, N } (\as) =
C_{2, \, N}^{g} (\as,1) \;\;.
\end{equation}
The DIS-scheme gluon density, instead, still remains ambiguous and is
given by an arbitrary combination of gluon and singlet quark densities in the
$ \, \msbar \,$ scheme
\begin{equation}
\label{gdis}
{\tilde f}_{g, \,N}^{(DIS)}(\mu^2) =
U_{g q, \, N}(\as(\mu^2)) \;{\tilde f}_{S, \, N}(\mu^2) +
U_{gg, \,N}(\as(\mu^2)) \;{\tilde f}_{g, \, N}(\mu^2)
\;\;,
\end{equation}
with the only constraint (\ref{longcons}), which reads
\begin{equation}
\label{longconsbis}
U_{gg , \, N=1 } (\as) =
1 - C_{2, \, N=1}^{g} (\as,1) \;\;\;, \;\;\;
 U_{gq  , \, N=1 } (\as) =
1 - C_{2, \, N}^{S} (\as,1) \;\;.
\end{equation}

The convention introduced in Ref.~[\ref{AEM}] for defining $ \,
{\tilde f}_g^{(DIS)} \,$ up to  ${\cal O}(\as) \,$ amounts to extending
Eq.~(\ref{longconsbis}) to any value of $ \, N \, $ in order $ \, \as $.
A natural generalization of this convention is to require
Eq.~(\ref{longconsbis}) to be valid for any  $ \, N \,$ and to
{\em all orders} in $ \, \as $. Doing that, the DIS-scheme gluon
density is completely defined.

Note, however, that
for the purposes of our all-order calculation, it is not necessary to specify
the actual form of the two matrix elements $U_{gq, \, N}$ and
$ U_{gg, \, N}$ in
Eq.~(\ref{gdis}). We just assume that they are chosen not to be
extremely
singular at high energies, i.e. they should not contain leading-order
contributions of the type $(\as/N)^k$ for $N \rightarrow 0$.
This is sufficient to ensure that most of the  $\msbar$-scheme results
obtained in the previous sections remain valid in the DIS scheme. In
particular the gluon anomalous dimensions $ \, \ga_{ g a }^{(DIS)} \, $
and the longitudinal coefficient functions $ \, C_L^{(DIS)} \,$
are
\begin{eqnarray}
\label{holddis}
\ga_{g a, \, N}^{ (DIS)} (\as) &=&
\ga_{g a, \, N} (\as)
+{\cal O}(\as^2 (\as/N)^k )
\;\;, \;\; (a = g , q) \;, \\
C_{L, \, N}^{A \; (DIS)}
\left(  \as ( \mu_F^2) , {Q^2 \over \mu^2_F}  \right) &=&
C_{L, \, N}^{A }
\left(  \as ( \mu_F^2) , {Q^2 \over \mu^2_F}  \right)
+ {\cal O}\left( \as^2 ( {\as \over N} )^k \right)
\;\;, \;\; (A = g , S) \;\;, \nonumber
\end{eqnarray}
where the resummed expressions for the $ \, \msbar  $-scheme anomalous
dimensions
$ \, \ga_{g a} \,$ and coefficient functions $ \, C_L^A \, $ are given in
Eqs.~(\ref{implbfkl}),(\ref{bfkliniq}),(\ref{CLNg}),(\ref{cpsL}).

The quark anomalous dimensions, instead, do not coincide any longer
(to this logarithmic accuracy) with the corresponding anomalous
dimensions in the $ \, \msbar \, $ scheme.
Using Eqs.~(\ref{newandim}) and (\ref{qdis}) we obtain
\begin{equation}
\label{nonholddis}
\ga_{qg , \, N}^{ (DIS)} (\as) =
\ga_{qg , \, N} (\as)   +
{1 \over { 2 \, N_f}} \,
C_{2, \, N}^{g }
(  \as  , 1  )
\, \ga_{gg , \, N} (\as)   +
{\cal O}(\as^2 ({\as/N} )^k )
 \;\;.
\end{equation}
On the other hand, the expression on the r.h.s. has been
computed with logarithmic accuracy $ \, \as \, ( \as / N )^k \, $
in Sect.~5.2.
Inserting Eqs.~(\ref{lincomb}),(\ref{2res})
into (\ref{nonholddis}),
we find the following resummed expression for the quark anomalous
dimension in the DIS scheme:
\begin{equation}
\label{resdis}
\ga_{qg,N}^{(DIS)}(\as)= \frac{\as}{2\pi} \;T_R \;\frac{2+3\ga_N-3\ga_N^2}
{3-2\ga_N} \;\frac{\Gamma^3(1-\ga_N) \,\Gamma^3(1+\ga_N)}{\Gamma(2+2\ga_N)
\,\Gamma(2-2\ga_N)} R_N(\as) +
{\cal O}\!\left(\as^2 ({\as}/{N})^k \right)
\;\;.
\end{equation}
The colour charge relation (\ref{qqq}) is still true in the DIS scheme:
\begin{equation}
\label{qqqdis}
\ga_{qq,\, N}^{S \, (DIS)} (\as) = \frac{C_F}{C_A} \left[
\ga_{qg,\, N}^{(DIS)} (\as) - \frac{\as}{2\pi}
\,T_R \,\frac{2}{3} \right] +
{\cal O}(\as^2 \left( \as/N)^k \right) \;\;.
%\nonumber
\end{equation}

Some comments are in order. The DIS-scheme result (\ref{resdis}) for the
quark anomalous dimensions  has to be contrasted with the results
discussed in Sect.~4 for the $ \, \msbar \, $ scheme. The algebraic
complications of the $ \, \msbar \, $ scheme prevented us from
obtaining resummed expressions in closed form for $ \, \ga_{ q g } \, $ and
$ \, C_2^g \, $ separately.  We were able to explicitly resum only the
combination in Eq.~(\ref{lincomb}), which turns out to be equivalent to the
anomalous dimensions in the DIS scheme. This computational simplification
has a more physical origin. The singlet sector of the deep inelastic
lepton-hadron scattering is characterized by four physical observables, which
can be studied in QCD perturbation theory: the structure functions
$ \, F_2^S $,
$ \, F_L^S \,$ and their first derivatives with respect to $ \, Q^2 $.
The $ \, \msbar \, $ scheme describes these observables in terms of eight
different quantities: four coefficient functions $ \, C_i^A \,$ ($i = 2, L $,
$ \, A = g , S$) and the four matrix elements of the singlet anomalous
dimensions. Obviously, only some linear combinations of them have to be
regarded as physical observables. The DIS scheme, reducing to two the
non-trivial coefficient functions ($C_2^S = 1 , \, C_2^g = 0 $), limits
the number of arbitrary unphysical quantities necessary to describe the
scattering process.  The ensuing anomalous dimensions are more easily
computable to all orders in $ \, \as \, $ because they are more directly
related to observable scaling violations.

Using the expansion (\ref{pertbfkl}) for the BFKL anomalous
dimension $ \, \ga_N \, $ and the expression (\ref{rn}) for $ \, R_N (\as) $,
the first perturbative terms of the quark anomalous dimension (\ref{resdis})
can be readily computed:
\begin{eqnarray}
\label{qadper}
\ga_{qg,N}^{(DIS)} &=& \frac{\as}{2\pi} \,T_R \,\frac{2}{3}
\left\{1 + \frac{13}{6} \abn +
\left(\frac{71}{18}-\zeta(2)\right) \left(\abn
\right)^2 +
\left[\frac{233}{27}-\frac{13}{6}\zeta(2)
+\frac{8}{3}\zeta(3)\right]
\left(\abn \right)^3
\right. \nonumber \\
&+&
%\left.
\left[\frac{1276}{81}-\frac{71}{18}\zeta(2)
+\frac{91}{9}\zeta(3) - 6 \zeta (4) \right]
\left(\abn \right)^4
% \nonumber \\
+
\left[  \frac{8384}{243}- {233 \over 27} \, \zeta (2) \right.
\nonumber\\
&+& \left. \left.
{710 \over 27} \, \zeta (3) - {20 \over 3} \, \zeta (2) \, \zeta(3) - 13
 \, \zeta (4)  + {22 \over 5} \, \zeta (5)   \right] \,
 \left(\abn \right)^5
+ {\cal O}\left(\left(\abn \right)^6\right) \right\}
\nonumber\\
&\simeq& \frac{\as}{2\pi} \,T_R \,\frac{2}{3}
\left\{1+2.17 \abn +2.30 \left(\abn \right)^2 +8.27
\left(\abn \right)^3 +
\right.
\nonumber\\
&+& \left. 14.92
\left(\abn \right)^4 +
29.23 \left(\abn \right)^5 +
{\cal O}\left(\left(\abn \right)^6\right) \right\} \;\;.
\end{eqnarray}
The coefficients of the first two terms in the curly bracket agree with the
 one- and two-loop calculations in the DIS scheme
[\ref{FP},\ref{Floratos},\ref{NDIS}].
Moreover, all the coefficients in Eq.~(\ref{qadper}) are systematically
larger than the corresponding $ \, \msbar$-scheme coefficients in
Eq.~(\ref{qms}). This behaviour is due to the additional contribution   of the
coefficient function $ \, C_2^g \, $ in Eq.~(\ref{nonholddis}), and it is
likely to persist in higher orders.

Note also that the all-order expression (\ref{resdis}) is analytic for
$ 0 \leq \ga_N < 1 / 2 $. Thus, independenly of the value of $ \, \as $,
the leading trajectory in $ \, N$-moment space is still given by
the BFKL pomeron. As the BFKL anomalous dimension $ \, \ga_N \, $
increases towards its saturation value at $ \, \ga_N = 1 / 2 $, the quark
anomalous dimension quickly increases, approaching a singularity due to the
pomeron normalization factor $ \, R_N ( \as ) \, $ (see Eq.~(\ref{rsing})).
This increase of $\ga_{qg}^{(DIS)}$ leads to strong scaling violations,
although the singularity at $\ga_N=1/2$ is cancelled  in
physical observables  by analogous contributions to the resummed coefficient
functions in Eq.~(\ref{holddis}).

\vskip 0.8 true cm

\setcounter{equation}{0}
\setcounter{sect}{6}
\setcounter{footnote}{0}

\noindent {\bf 6. Summary  }
\vskip 0.3 true cm

In the present paper we have shown how the high-energy factorization theorem
[\ref{CCH}] can be extended beyond the leading logarithmic  accuracy
in a manner which is consistent with the all-order factorization of collinear
singularities. Much effort has been devoted to investigating the issue of the
dependence on the factorization scheme of parton densities and coefficient
functions.
This analysis has led to the (off-shell)
$k_{\perp}$-factorization in dimensional regularization represented
schematically by Eq.~(\ref{AgG}) (see also Eqs.~(\ref{quagrektfaceps1}) and
(\ref{ktfacdis})). Eq.~(\ref{AgG}) has then been used to study the
high-energy (or small-$x$) behaviour of deep inelastic scattering
processes.

A first general consequence of Eq.~(\ref{AgG}) is that flavour non-singlet
observables are regular at small $x$ order by order in perturbation theory.

As regards  the singlet sector, $ k_{\perp}$-factorization allows one to sum
classes of logarithmic corrections to all orders in $ \, \as$. To this
end, one has to evaluate 2GI kernels (see Eq.~(\ref{AgG})) in fixed-order
perturbation theory and use the master equations
(\ref{maeq}),(\ref{maeqiniq}) for the gluon Green functions.

Eqs.~(\ref{maeq}) and (\ref{maeqiniq}) are the generalization of the BFKL
equation [\ref{BFKL}] to the case of $ \, n = 4 + 2 \, \varepsilon \,$
space-time dimensions. Their solution is discussed in Sect.~3. In particular,
the calculation of the gluon anomalous dimensions
$ \, \ga_{g g , \, N} (\as) $,
$  \ga_{g q , \, N} (\as) $
(see Eqs.~(\ref{glulead}),(\ref{implbfkl}),(\ref{bfkliniq}))
has been carried out to the leading logarithmic accuracy $\, (\as / N)^k \,$
in the context of dimensional regularization and the BFKL results on the
pomeron trajectory have been re-derived. Besides this, we have been able
to compute (to the same accuracy) the normalization factor $ \, R_N (\as) \,$
(see Eq.~(\ref{rn})) of the perturbative QCD pomeron in the $ \, \msbar $
factorization scheme.

The quark sector enters the QCD evolution equations to next-to-leading
logarithmic order $ \, \as \, (\as \ln x)^k $. Using high-energy
factorization, in Sect.~4 we have evaluated the corresponding quark Green
functions. We have shown that the result in Eq.~(\ref{quagresersol})
originates from an integral equation whose kernel is related to a
generalized (off-shell) Altarelli-Parisi splitting function (see
Eq.~(\ref{K0res1})). We have also discussed how
Eq.~(\ref{quagresersol}) can be used for evaluating the small-$N \,$ limit
of the quark anomalous dimensions
$ \, \ga_{q g , \, N} (\as) $ and
$  \ga_{q q , \, N}^{S} (\as) $.
The result of our  explicit calculation up to six-loop order is given in
Eq.~(\ref{qms}).

As an example of application of high-energy factorization to a specific hard
process, in Sect.~5 we have considered deep inelastic lepton-hadron
scattering (for the case of heavy-flavour production see
Refs.[\ref{CCH}-\ref{LRSS}] and Sect.~2.4). Resummed expressions to
next-to-leading accuracy
 $\,\as \, (\as / N)^k \,$
for the DIS coefficient functions $ \, C_2 \, $ and $ \, C_L \, $ are
presented in
Eqs.~(\ref{CLNg}),(\ref{lincomb}),(\ref{C2N}),(\ref{cpsL}),(\ref{cps2}).
These results are given in the $ \, \msbar $ factorization scheme.
In Sect.~5.3, we have also introduced an all-order generalization of the
DIS factorization scheme first proposed in Ref.~[\ref{AEM}]. Within
this scheme, where most of the DIS coefficient functions are trivial
(see Eqs.~(\ref{f2disdefbis}) and (\ref{holddis})), we have obtained the
next-to-leading resummed expressions (\ref{resdis}),(\ref{qqqdis}) for the
quark anomalous dimensions.

Quantifying precisely  the phenomenological consequences of the results
presented here is a matter of detailed numerical investigations. However, the
size of the next-to-leading-order coefficients in the perturbative
expansions (\ref{qms}),(\ref{CMSL}),(\ref{CMS2}) and (\ref{qadper}) suggests
that these contributions may have phenomenological relevance in accurate
analyses of scaling violations, already at the values of $ \, x \,$
($x \sim 10^{-3} \div 10^{-4} $) accessible at present hadron colliders.
In particular, since the first leading-order  coefficients of the gluon
anomalous dimensions are vanishing (see Eqs.~(\ref{pertbfkl})), the
next-to-leading corrections in  the quark sector computed in this paper
may be quite important for the study of the proton structure functions
being
measured at HERA. A fully consistent analysis to next-to-leading logarithmic
order obviously requires also the computation of the still unknown (to this
accuracy) gluon anomalous dimensions. We hope to report progress on this
subject in the near future.

\vskip 0.8 true cm

\noindent {\bf Acknowledgments.  } We would like to thank Marcello Ciafaloni
for a very long and intense collaboration on high-energy factorization and
small-$x$ physics. Useful discussions with Pino Marchesini and Bryan Webber
are also acknowledged.

\newpage

\renewcommand{\theequation}{A.\arabic{equation}}

\setcounter{equation}{0}

%\vskip 3 true cm
\noindent {\bf Appendix A }
\vskip 0.3 true cm

The rescaled parton densities $\, {\tilde f}_{a} (x, \mu^2) = x \,
f_a(x , \mu^2) \,$   introduced in Eq.~(\ref{tildef}) fulfil the
evolution equations
\begin{equation}
\label{evapp}
\frac{d \;{\tilde f}_a(x,\mu^2)}{d\ln \mu^2} = \sum_b \int_0^1 {dz} \;
P_{ab}(\as(\mu^2),z) \;{\tilde f}_b({x/z},\mu^2) \;\;,
\end{equation}
where $P_{ab}(\as,z)$ are the generalized Altarelli-Parisi splitting functions
in Eq.~(\ref{Pab}). According to our normalization, the splitting functions
have the following one-loop expressions
\begin{eqnarray}
\label{apspliapp}
P_{g g}^{(0)} (z ) &=& 2 \, C_A \, \left[ \left( {1 \over {1 - z}} \right)_{+}
- 1 + {{1-z} \over z} + z (1 - z) \right] + \left( {11 \over 6} \, C_A -
{2 \over 3} \, T_R \, N_f \right) \, \delta ( 1 - z) \;\;,
\nonumber\\
P_{g q_i}^{(0)} (z ) &=&
P_{g {\bar q}_i}^{(0)} (z ) =
 C_F \,
 {{1+ (1-z)^2} \over z} \;\;,
\nonumber\\
P_{ q_i g}^{(0)} (z ) &=&
P_{ {\bar q}_i g}^{(0)} (z ) = T_R \, \left[ z^2 + (1-z)^2 \right] \;\;, \\
%\nonumber\\
P_{ q_i q_j}^{(0)} (z ) &=&
P_{ {\bar q}_i {\bar q}_j}^{(0)} (z ) =
 C_F \, \left( {{1+z^2} \over {1 - z}} \right)_{+}  \, \delta_{i j} \;\;,
\;\;\;\;
P_{ { q}_i {\bar q}_j}^{(0)} (z ) =
P_{ {\bar q}_i { q}_j}^{(0)} (z ) = 0 \;\;, \nonumber
\end{eqnarray}
in terms of the $SU(N_c)$ colour factors ($N_c =3 $ is the  number of colours)
\begin{equation}
\label{colourfactor}
C_A = N_c \;\;\;, \;\;\;\;\; C_F = {{N_c^2 - 1} \over { 2 \, N_c}} \;\;\;,
\;\;\;\;\;
{\mbox {\rm Tr}} \,(t^a t^b) =
\delta_{ a b} \, T_R  = {1 \over 2} \,
\delta_{ a b}  \;\;\;.
\end{equation}

The leading-order splitting functions $P_{ab}^{(0)}(z)$ (which have been known
for a long time [\ref{GLAP}]) are factorization theorem invariants, i.e. they
do not depend on the explicit  procedure to factorize  collinear
singularities.
The physical reason for this is that they are directly related to observable
scaling violations in deep inelastic scattering  processes. On the
contrary, splitting functions and anomalous dimensions beyond one-loop order
do depend on the regularization and factorization schemes of collinear
singularities. Nonetheless, due to charge conjugation invariance and
$SU(N_f)$ flavour symmetry of QCD, they satisfy the following
scheme-independent properties
\begin{eqnarray}
\label{adpr}
\ga_{q_ig}=\ga_{{\bar q}_ig} \equiv \ga_{qg} \;\;\;\;&,& \;\;\;\;\;
\ga_{gq_i}=\ga_{g{\bar q}_i} \equiv \ga_{gq} \nonumber\\
\ga_{q_iq_j}=\ga_{{\bar q}_i{\bar q}_j} \equiv \ga_{qq}^{NS} \delta_{ij} +
\ga_{qq}^{S} \;\;&,& \;\;\;
\ga_{q_i{\bar q}_j}=\ga_{{\bar q}_iq_j} \equiv \ga_{q{\bar q}}^{NS}
\delta_{ij} + \ga_{q{\bar q}}^{S} \;\;.
\end{eqnarray}

The symmetry properties (\ref{adpr}) imply that the anomalous dimensions
matrix $\ga_{ab}$ has only seven independent components. Correspondingly,
three flavour non-singlet
(${\tilde f}^{(V)}$, ${\tilde f}_{q_i}^{(-)}$,  ${\tilde f}_{q_i}^{(+)}$)
and two flavour singlet
(${\tilde f}_S$, $ {\tilde f}_g$)
parton densities can be introduced
so that the evolution
equations
 (\ref{evapp}))
(or (\ref{fracd}))
are completely diagonalized (in the partonic space) for
the non-singlet sector. One explicitly finds (we drop the overall dependence
on $ \, N $, $\, \mu$ and $ \, \as$)
%\begin{eqnarray}
%\label{nseq}
%\frac{d \ln f^{(V)}}{d\ln \mu^2} &=& \ga^{(V)} \nonumber \\
%\frac{d \ln f_{q_i}^{(-)}}{d\ln \mu^2} &=& \ga^{(-)} \\
%\frac{d \ln f_{q_i}^{(+)}}{d\ln \mu^2} &=& \ga^{(+)} \;\;, \nonumber
%\end{eqnarray}
\begin{equation}
\label{nseq}
\frac{d \ln {\tilde f}^{(V)}}{d\ln \mu^2} = \ga^{(V)} \;\;, \;\;\;\;
\frac{d \ln {\tilde f}_{q_i}^{(-)}}{d\ln \mu^2} = \ga^{(-)} \;\;, \;\;\;\;
\frac{d \ln {\tilde f}_{q_i}^{(+)}}{d\ln \mu^2} = \ga^{(+)} \;\;,
\end{equation}
where
\begin{equation}
\label{fns}
{\tilde f}^{(V)} \equiv \sum_{j=1}^{N_f}
 ({\tilde f}_{q_j}-{\tilde f}_{{\bar q}_j}) \;\;, \;\;\;\;\;
{\tilde f}_{q_i}^{(\pm)} \equiv {\tilde f}_{q_i}
 \pm {\tilde f}_{{\bar q}_i} - \frac{1}{N_f}
\sum_{j=1}^{N_f} ({\tilde f}_{q_j} \pm {\tilde f}_{{\bar q}_j}) \;\;,
\end{equation}
and the non-singlet anomalous dimensions are given by
\begin{equation}
\label{adns}
\ga^{(V)} = \ga_{qq}^{NS} - \ga_{q{\bar q}}^{NS} + N_f (\ga_{qq}^{S} -
\ga_{q{\bar q}}^{S}) \;\;, \;\;\;\;\;
\ga^{(\pm)} = \ga_{qq}^{NS} \pm \ga_{q{\bar q}}^{NS} \;\;.
\end{equation}
The evolution equations are instead still coupled in the singlet sector:
\begin{eqnarray}
\label{sieq}
\frac{d \;{\tilde f}_S}{d\ln \mu^2} &=&
[ \ga_{qq}^{NS} + \ga_{q{\bar q}}^{NS} + N_f
(\ga_{qq}^{S} + \ga_{q{\bar q}}^{S}) ] {\tilde f}_S +
2 N_f \,\ga_{qg} {\tilde f}_g \;\;, \nonumber \\
\frac{d \;{\tilde f}_{g}}{d\ln \mu^2} &=&
\ga_{gq} {\tilde f}_S + \ga_{gg}
{\tilde f}_g \;\;,
\end{eqnarray}
where the quark singlet density is defined by
${\tilde f}_S = \sum_{i=1}^{N_f} ({\tilde f}_{q_i}+
{\tilde f}_{{\bar q}_i})$.

{}From Eq.~(\ref{apspliapp}) we see that all the three non-singlet
anomalous dimensions are degenerate in one-loop order. The anomalous
dimensions in two-loop order were computed in
Refs.~[\ref{CFP},\ref{FP},\ref{Floratos}]. In this order the degeneracy
mentioned above is partially removed because $ \ga_{q \bar q}^{NS} \not= 0$.
However we still have $ \ga^{(V)} (\as) =
\ga^{(-)} (\as) + O (\as^3)$  since
$ \ga_{q  q}^{S} $ and
$ \ga_{q \bar q}^{S} $ coincide in $  O(\as^2)$. The equality between
$ \ga_{q  q}^{S} $ and
$ \ga_{q \bar q}^{S} $ is expected to be violated starting from
 $  O(\as^3)$.

The high-energy power counting in Sect.~2.3 implies that the non-singlet
anomalous dimensions in Eq.~(\ref{adns}) (and hence
$ \ga_{q  q , \, N }^{NS} $,
$ \ga_{q \bar q , \, N}^{NS} $,
$ \ga_{q  q , \, N }^{S} -
 \ga_{q \bar q , \, N }^{S} $)
are  regular  for $N \rightarrow 0$ order by order in $\as$.
In other words, the corresponding $n$-loop splitting functions $P^{(n-1)}$
in Eq.~(\ref{Pab}) are less singular than $1/x$ for $x \rightarrow 0$.
All the high-energy contributions $\as^n/N^k
\;(n \geq k \geq 1)$ are thus associated
 with the gluon anomalous dimensions $\ga_{gg,N},\ga_{gq,N}$ and with the
quark anomalous dimensions $ \ga_{qg,N} , \ga_{qq,N}^S \simeq
\ga_{q{\bar q},N}^S$.

\vskip 2 true cm

\newpage

\renewcommand{\theequation}{B.\arabic{equation}}

\setcounter{equation}{0}

%\vskip 3 true cm
\noindent {\bf Appendix B }
\vskip 0.3 true cm

In order to obtain the solution of the master equation (\ref{maeq}) for the
$ \, k_{\perp}$-distribution $ \, {\cal F}^{(0)} $, it is convenient to
introduce the dimensionless distribution $ \, {\bar {\cal F}} $, as follows
\begin{equation}
\label{fbarapp}
{\cf}^{(0)}_N(\bk; \as, \mu, \varepsilon) \equiv
\delta^{(2+2\varepsilon)}(\bk) +
{ {\Gamma (1+ \varepsilon)} \over
{ ( \pi \, \bk^2 )^{1 + \varepsilon}} } \;
{\bar {\cf}}_N(\as ({ \bk^2/\mu^2})^\varepsilon,
\varepsilon ) \;\;.
\end{equation}
Note that
 $ \, {\bar {\cal F}} $ does not depend on $ \, \varepsilon $, $ \as \, $ and
$ \, \bk \,$ independently. In fact, rewriting Eq.~(\ref{maeq}) in terms of
 $ \, {\bar {\cal F}} $, we see that it only depends on $ \, \varepsilon \,$
and the combination $ \, \as ( \bk^2 / \mu^2 )^\varepsilon \;$:
\begin{eqnarray}
\label{maeqbarapp}
&~&
{\bar {\cf}}_N( \as ({ \bk^2/\mu^2})^\varepsilon,
\varepsilon ) =
\abn  \, S_\varepsilon \,
{{e^{ \varepsilon \, \psi (1 )}} \over {\Gamma ({1 +  \varepsilon})}}  \,
\left( {\bk^2 \over \mu^2} \right)^\varepsilon
\nonumber\\
&+& \frac{\abar}{N}
\left( {\bk^2 \over \mu^2} \right)^\varepsilon
\int
\frac{d^{2+2\varepsilon} \bq}{(2\pi )^{2\varepsilon}}
\;\frac{1}{\pi {\bq}^2 \,(\bk-\bq)^2} \,
\left[
 \frac{\bk^2}{ \left( (\bk-\bq)^2 \right)^\varepsilon} \,
{\bar {\cf}}_N( \as ({ {(\bk-\bq)^2}/\mu^2})^\varepsilon,
\varepsilon ) \right.
\nonumber\\
&-& \left. \frac{\bk \cdot (\bk - \bq)}{(\bk^2)^\varepsilon} \,
{\bar {\cf}}_N( \as ({ \bk^2/\mu^2})^\varepsilon,
\varepsilon ) \right] \;\;.
\end{eqnarray}
Then, we perform the shift $ \, \bq \to {\bq}^{\prime}= \bq - \bk \, $ of the
integration variable in Eq.~(\ref{maeqbarapp}),  and introduce the ratio
$ \, \tau = | \bq^{\prime} | / | \bk | \,$
and the angle $ \, \theta \,$ between $ \, {\bq}^{\prime} \, $ and $ \, \bk$.
The integral equation (\ref{maeqbarapp}) now reads:
\begin{eqnarray}
\label{maeqbarapp1}
{\bar {\cf}}_N( \as ({ \bk^2/\mu^2})^\varepsilon,
\varepsilon ) &=&
\abn  \, S_\varepsilon \,
{{e^{ \varepsilon \, \psi (1 )}} \over {\Gamma ({1 +  \varepsilon})}}  \,
\left( {\bk^2 \over \mu^2} \right)^\varepsilon \, \\
%\nonumber\\
&\cdot& \left\{ 1 +
\frac{2 \, \Gamma(1 + \varepsilon)}{\sqrt{\pi} \,
\Gamma({1\over 2} + \varepsilon)} \, \int_0^\infty \frac{d \tau}{\tau}
\int_0^\pi d \theta \, \frac{(\sin \theta)^{2 \varepsilon}}{1+ 2 \,
\tau \, \cos \theta + \tau^2} \,
\right.
\nonumber\\
&\cdot& \left.
 \left[
{\bar {\cf}}_N( \as ({ { \tau^2 \bk^2}/\mu^2})^\varepsilon,
\varepsilon )
+ \tau^{1 + 2 \varepsilon} \, \cos \theta \,
{\bar {\cf}}_N( \as ({ \bk^2/\mu^2})^\varepsilon,
\varepsilon )
\right] \right\} \;\;. \nonumber
\end{eqnarray}
The angular integration in Eq.~(\ref{maeqbarapp1}) can be performed
in terms of associated Legendre functions.
However we did not find the ensuing representation of any convenience.

The perturbative solution of Eq.~(\ref{maeqbarapp1}) has the following
expression:
\begin{equation}
\label{sersolapp}
{\bar {\cf}}_N( \as ({ \bk^2/\mu^2})^\varepsilon,
\varepsilon ) =
\sum_{k=1}^\infty
\, \left[
\abn  \, S_\varepsilon \,
{{e^{ \varepsilon \, \psi (1 )}} \over {\Gamma ({1 +  \varepsilon})}}  \,
\left( {\bk^2 \over \mu^2} \right)^\varepsilon
\, \right]^k \, c_k (\varepsilon) \;\;\;.
\end{equation}
Inserting Eq.~(\ref{sersolapp}) into Eq.~(\ref{maeqbarapp1}), we obtain the
recurrence relation (\ref{recurs})
for the perturbative coefficients $ \, c_k (\varepsilon) $. The recurrence
factor $ \, I_k (\varepsilon) \,$ is given by
\begin{equation}
\label{Irecur}
 I_k (\varepsilon) = I (\ga = k \varepsilon; \varepsilon) \;\;,
\end{equation}
where $ \, I ( \ga ; \varepsilon) \,$ is the following integral
\begin{eqnarray}
\label{Iinteg}
 I (\ga ; \varepsilon) &=&
\frac{2 \, \Gamma(1 + \varepsilon)}{\sqrt{\pi} \,
\Gamma({1\over 2} + \varepsilon)} \, \int_0^\infty \frac{d \tau}{\tau}
\int_0^\pi d\theta \, \frac{(\sin \theta)^{2 \varepsilon}}{1+ 2 \,
\tau \, \cos \theta + \tau^2} \,
(\tau^{2 \ga} + \tau^{1 + 2 \varepsilon} \, \cos \theta )
\nonumber\\
&=& \frac{2 \, \pi \, \Gamma(1 + \varepsilon)}{\sqrt{\pi} \,
\Gamma({1\over 2} + \varepsilon)} \,
\int_0^\pi d \theta  \, (\sin \theta)^{2 \varepsilon -1} \,
\left[ \frac{\sin \left( (1- 2 \ga) \theta \right)}{\sin (2 \, \pi \, \ga)}
+ \frac{ \cos \theta \, \sin  ( 2 \varepsilon \theta) }{\sin (2 \, \pi \,
\varepsilon)} \right]
\nonumber\\
&=&{ 1 \over \varepsilon} \,
\frac{\Gamma^2 (1 + \varepsilon)}{ \Gamma ({1} + 2 \, \varepsilon)}
\left[
\frac{\Gamma (1 + 2 \, \varepsilon) \, \Gamma ( \ga) \,
\Gamma (1 - \ga) }
{ \Gamma \left(
\varepsilon + \ga \right) \,\Gamma \left( 1 + \varepsilon - \ga \right)}
-  \Gamma ( 1 + \varepsilon)  \, \Gamma (1 - \varepsilon) \right]  \;\;.
\end{eqnarray}
Using Eqs.~(\ref{Irecur})
and (\ref{Iinteg}) we obtain
Eq.~(\ref{calcI}).

The integral  $ \, I ( \ga ; \varepsilon) \, $ in Eq.~(\ref{Iinteg})
represents the action of the kernel of the master equation (\ref{maeqbarapp})
 (or, equivalently, (\ref{maeq})) on a test function $ \, {\cal F} (\bk) \,$
behaving like $ \, (\bk^2 )^\ga$. Therefore the fact that
  $ \, I ( \ga ; \varepsilon) \, $ is finite for
  $ \, \varepsilon \to 0  \, $ and $ \, \ga $ fixed is a consequence of the
collinear regularity of the kernel of the BFKL equation in four dimensions.
Note also that this regularity is achieved through the cancellation of
collinear singularities which are present separately in the real and virtual
contributions, i.e. in the first and second terms in the square bracket of
Eq.~(\ref{maeqbarapp}) or (\ref{Iinteg}).

As discussed in Sect.~3, the lack of scale invariance in the master equation
(\ref{maeqbarapp}) does not allow us to find a general solution for $ \,
{\bar {\cal F}} \, $
for any value of $ \, \varepsilon$. This means that we are not
able to resum the formal power series expansion in Eq.~(\ref{sersolapp}).
Nevertheless, we can obtain an explicit all-order solution in the relevant
limit of small $ \, \varepsilon \,$ values. This limit is sufficient to
compute the transition function $ \, \Gamma_{g g } \, $ (i.e. the gluon
anomalous dimensions) and the associated normalization factor $ \, R_N \, $
in Eqs.~(\ref{renggcap2}),(\ref{Rcoeff}).

To do this, let us perform the derivative of Eq.~(\ref{maeqbarapp1})
with respect to $ \, \ln \as \,$ and then divide both sides by $ \, {\bar
{\cal F}} $. We thus obtain
\begin{eqnarray}
\label{maeqderiv}
&~& \varepsilon \,{\partial \over { \partial \ln \as}} \ln {\bar {\cf}}_N
( \as ,  \varepsilon ) =
\varepsilon +
\abn  \, S_\varepsilon \,
{{e^{ \varepsilon \, \psi (1 )}} \over {\Gamma ({1 +  \varepsilon})}}  \,
\frac{2 \, \Gamma(1 + \varepsilon)}{\sqrt{\pi} \,
\Gamma({1\over 2} + \varepsilon)}
\nonumber\\
&~& \cdot
\int_0^\infty \frac{d \tau}{\tau}
\int_0^\pi d \theta  \,\frac{(\sin \theta)^{2 \varepsilon}}{1+ 2 \,
\tau \, \cos \theta + \tau^2} \\
%\nonumber\\
&~& \cdot  \left[
{
{ {\bar {\cf}}_N ( \as \, \tau^{ 2 \varepsilon},  \varepsilon )}
\over
{{\bar {\cf}}_N ( \as ,  \varepsilon )}
} \,
\varepsilon \,{\partial \over { \partial \ln \as}} \ln {\bar {\cf}}_N ( \as
\, \tau^{ 2 \varepsilon},  \varepsilon )
%}
%,  \varepsilon )
+ \tau^{1 + 2 \varepsilon} \, \cos\theta \,
\varepsilon \,{\partial \over { \partial \ln \as}} \ln {\bar {\cf}}_N ( \as ,
\varepsilon )
\right]  \;\;. \nonumber
\end{eqnarray}
Unlike Eq.~(\ref{maeqbarapp1}), Eq.~(\ref{maeqderiv}) is homogenous with
respect to $ \, {\bar {\cal F}} $. Therefore we can easily factorize the
singular transition function $ \, \Gamma_{g g,N} (\as, \varepsilon)$.

More precisely, let us notice that the distribution
$ \, {\bar {\cal F}} \,$
introduced in Eq.~(\ref{fbarapp})
is related to the integrated gluon Green function (\ref{integrG}) as follows
\begin{eqnarray}
\label{integrGapp}
{\bar {\cf}}_N ( \as \, ({Q^2/\mu^2})^\varepsilon ,  \varepsilon ) &=&
Q^2 {\partial \over { \partial Q^2}}
G^{(0)}_{gg , \, N}( \as ({Q^2/\mu^2})^\varepsilon ,  \varepsilon )
\nonumber\\
&=&
\varepsilon \,{\partial \over { \partial \ln \as}}
G^{(0)}_{gg , \, N}( \as ({Q^2/\mu^2})^\varepsilon ,  \varepsilon )  \;\;.
\end{eqnarray}
Using the factorization formula (\ref{renggcap2}),
\begin{eqnarray}
\label{renggapp}
G^{(0)}_{gg , \, N}   ( \as  ,  \varepsilon) &=&
G_{gg , \, N}   ( \as  ,  \varepsilon) \,
\Gamma_{gg , \, N}   ( \as  ,  \varepsilon)
\nonumber\\
&=&
G_{gg , \, N}   ( \as  ,  \varepsilon) \,
\exp \left( {1 \over \varepsilon} \int_0^{\as \, S_\varepsilon} {{d \alpha}
\over \alpha} \, \ga_N ( \alpha) \right) \;\;,
\end{eqnarray}
we thus have
\begin{equation}
\label{renfbarapp}
{\bar {\cf}}_N    ( \as  ,  \varepsilon) = \left[
\ga_N ( \as \, S_\varepsilon ) +
\varepsilon {\partial \over { \partial \ln \as}}
\ln G_{gg , \, N}
( \as  ,  \varepsilon) \right]
\, G_{gg , \, N}   ( \as  ,  \varepsilon) \,
\Gamma_{gg , \, N}   ( \as  ,  \varepsilon) \;\;.
\end{equation}
Since the anomalous dimensions
$\, \ga_N ( \as \, S_\varepsilon ) \,$ and the renormalized Green function
\linebreak
$\, G_{gg , \, N}   ( \as  ,  \varepsilon) \,$ are regular  for
$ \, \varepsilon \to 0 $, it follows from Eq.~(\ref{renfbarapp}) that
the relevant functions involved in
Eq.~(\ref{maeqderiv}) and
Eq.~(\ref{maeqderiv}) itself  can be expanded in
$ \, \varepsilon \ $ around
$ \, \varepsilon = 0 $.
A straightforward calculation gives:
\begin{eqnarray}
\label{expandapp}
&\varepsilon& {\partial \over { \partial \ln \as}} \ln {\bar {\cf}}_N ( \as \,
\tau^{2 \, \varepsilon } ,
\varepsilon )
= \left[
  \ga_N ( \as \, S_\varepsilon ) + \varepsilon
\left( {{\partial \ln \ga_N (\as)} \over { \partial \ln \as}} +
{{\partial \ln R_N (\as)} \over { \partial \ln \as}} \right)
\right]
\nonumber\\
&\cdot& \left[
1 +  2 \, \varepsilon \, \ln \tau \,
 {{\partial  \ln \ga_N (\as)} \over { \partial \ln \as}} +
{\cal O}( \varepsilon^2) \,
\right] \;\;,
\nonumber\\
&~&{
{{\bar {\cf}}_N ( \as \, \tau^{ 2 \varepsilon},  \varepsilon )}
\over
{{\bar {\cf}}_N ( \as ,  \varepsilon )}
}
= \left[  1 + 2 \, \varepsilon \, \ln \tau \,
\left( {{\partial \ln \ga_N (\as)} \over { \partial \ln \as}} +
{{\partial \ln R_N (\as)} \over { \partial \ln \as}} \right)  \right. \\
%\nonumber\\
&+& \left.
 2 \, \varepsilon \, {\ln}^2 \tau \,
 {{\partial  \ga_N (\as)} \over { \partial \ln \as}} +
{\cal O}( \varepsilon^2) \,
\right]  \, \exp \left\{
2 \, \ga_N ( \as \, S_\varepsilon ) \, \ln \tau \right\}
\;\;, \nonumber
\end{eqnarray}
where $ \, R_N (\as ) =
 G_{gg , \, N}   ( \as  ,  \varepsilon= 0 ) $.
Correspondingly, Eq.~(\ref{maeqderiv}) reads as follows (we drop the overall
explicit dependence on $ \as $)
\begin{eqnarray}
\label{maeqderiv1}
 \ga_N +  \varepsilon \,
\left( {{\partial \ln \ga_N } \over { \partial \ln \as}} +
{{\partial \ln R_N } \over { \partial \ln \as}} \right)
+{\cal O}( \varepsilon^2) &=&
 \varepsilon + \left[
 \ga_N +  \varepsilon \,
\left( {{\partial \ln \ga_N } \over { \partial \ln \as}} +
{{\partial \ln R_N } \over { \partial \ln \as}} \right)
\right]
\nonumber\\
&\cdot&
\abn \, \Omega (\ga_N ; \varepsilon)
\, I (\ga_N ; \varepsilon) \;\;,
\end{eqnarray}
where we have introduced the (second-order) differential operator $ \, \Omega $
\begin{equation}
\label{Omega}
 \Omega (\ga_N ; \varepsilon) =  1 +
   \varepsilon \, \left[
\left( 2 \, {{\partial \ln \ga_N } \over { \partial \ln \as}} +
{{\partial \ln R_N } \over { \partial \ln \as}} \right)
\,  {{\partial } \over { \partial \ga_N}}
+ {1 \over 2 }\,
 {{\partial  \ga_N } \over { \partial \ln \as}}
\, \left( {{\partial } \over { \partial \ga_N}} \right)^2 \right]
+ {\cal O}( \varepsilon^2)
\end{equation}
acting on the integral
$\, I (\ga_N ; \varepsilon) \, $ in Eq.~(\ref{Iinteg}).
Equating the
$ \, {\cal O}(\varepsilon^0) \,$ and
$ \, {\cal O}( \varepsilon) \,$
terms in Eq.~(\ref{maeqderiv1}) we thus find an implicit equation for
$ \, \ga_N (\as ) \,$ and a differential equation for $ \, R_N ( \as ) $.

To order $\, \varepsilon^0 \,$, we obtain
\begin{equation}
\label{implbfklapp}
1 = \abn \, \chi \left( \ga_N (\as) \right) \;\;,
\end{equation}
i.e. the result (\ref{implbfkl}) for the BFKL anomalous dimension, $ \, \chi
(\ga) \,$ being the characteristic function in Eq.~(\ref{cfmo57}). To order
$ \, \varepsilon \,$, we have
\begin{eqnarray}
\label{ordereps}
&~& \left( 2 \, {{\partial \ln \ga_N } \over { \partial \ln \as}}  +
{{\partial \ln R_N } \over { \partial \ln \as}} \right)
\, \chi^{\prime} (\ga_N)
+ {1 \over 2 }\,
 {{\partial  \ga_N } \over { \partial \ln \as}}
\, \chi^{\prime \prime} (\ga_N) +
\nonumber\\
&+&
{1 \over 2 }\, \left[
2 \, \psi^{\prime} (1)
- \psi^{\prime} (\ga_N)
- \psi^{\prime} (1- \ga_N)
+ \chi^{2} (\ga_N)
\right] = - { N \over { \abar \ga_N}} \;\;,
\end{eqnarray}
where
$\, \chi^{\prime} (\ga)$ , $
\, \chi^{\prime \prime} (\ga) \,$ are the first and second derivatives of
$\, \chi (\ga) \,$ with respect to $ \, \ga $. In order to solve
Eq.~(\ref{ordereps}), it is convenient to consider $ \, \ga_N \, $ as the
independent variable
\begin{equation}
\label{chainruleapp}
{{\partial \ln R_N } \over { \partial \ln \as}} =
 {{\partial  \ln \ga_N } \over { \partial \ln \as}}
\; {{\partial \ln R_N } \over { \partial \ln \ga_N}} \;\;,
\end{equation}
and use Eq.~(\ref{implbfklapp}):
\begin{equation}
\label{orderepszero}
 { N \over { \abar }} =
 \chi (\ga_N)
\;\;, \;\;\;\; {{\partial  \ln \ga_N } \over { \partial \ln \as}}
= - { {\chi (\ga_N) } \over { \ga_N \,  \chi^{\prime} (\ga_N) }}
\;\;.
\end{equation}
Inserting Eqs.~(\ref{chainruleapp}),(\ref{orderepszero}) into
Eq.~(\ref{ordereps}) we get the differential equation
\begin{equation}
\label{eqR}
{{\partial \ln R_N } \over { \partial \ln \ga_N}} =
 {1 \over 2 } \, \ga_N
\, \left[ - { 2 \over \ga_N}
 - {{\chi^{\prime \prime} (\ga_N)} \over
{ \chi^{\prime} (\ga_N)}}
+ { \chi (\ga_N)} +
{ {
2 \, \psi^{\prime} (1)
- \psi^{\prime} (\ga_N)
- \psi^{\prime} (1- \ga_N) } \over
 { \chi (\ga_N)} }
\right]
\;\;,
\end{equation}
whose solution is given by Eq.~(\ref{rn}).

\vskip 2 true cm

\renewcommand{\theequation}{C.\arabic{equation}}

\setcounter{equation}{0}

%\vskip 3 true cm
\noindent {\bf Appendix C }
\vskip 0.3 true cm

The evaluation of the quark anomalous dimensions starting from the
$ \, k_{\perp}$-factorization formula (\ref{quagrektfaceps2}) requires the
explicit computation of the off-shell kernel $ \, {\hat K }_{q g}$.
Inserting
Eq.~(\ref{quagrektfaceps1})
into Eq.~(\ref{Gdritto}) and comparing the latter with
Eq.~(\ref{quagrektfaceps2}) we find
\begin{equation}
\label{Kidentiapp}
 {\hat K}_{ q g , \, N }( {\bk^2/Q^2}, \as ( {Q^2/\mu^2} )^\varepsilon;
\varepsilon ) = \int_0^1 \, d z \, z^{N-1}
\, {\hat K}_{ q g }( z , {\bk^2/Q^2}, \as ( {Q^2/\mu^2} )^\varepsilon;
\varepsilon ) \;\;,
\end{equation}
where
$\, {\hat K}_{ q g } ( z )\,$
is obtained from
$\, {\hat K}^{(0) } (q, k) \,$ (see Fig.~7b), after  integration over $ q $, as
follows
\begin{equation}
\label{Kdefapp}
 {\hat K}_{ q g }\!\left( z , {\bk^2 \over  Q^2}, \as\!\left( {Q^2 \over
 \mu^2}\right)^\varepsilon; \varepsilon \right) =
\int \frac{dq^{2} \, d^{2 + 2 \varepsilon} \bq}{2(2\pi)^{4+2\varepsilon}}
\, \Theta\!\left( Q^2 - |q^2| \right) \,
\left(\frac{\bar p\!\!\!/}{2{\bar p}\cdot q} \right) _{\alpha\beta}
{\hat K}^{(0) \; \al \beta}_{\mu \nu } (q,k)
{ \frac{ k_{\perp}^{\mu} \, k_{\perp}^{\nu} }{\bk^2} } \;\;.
\end{equation}

In Eq.~(\ref{Kdefapp}) the following Sudakov parametrization for the
momenta $ k$ and $ q $ is understood:
\begin{equation}
\label{sudapp}
k^{\mu} = y \, p^{\mu} + k_{\perp}^{\mu}
\;\;, \;\; q^{\mu} =  x \, p^{\mu} + q_{\perp}^{\mu} + { { q^2 + \bq^2} \over
{2 \,  x \, p \cdot {\bar p} }} \, {\bar p}^{\mu} \;\;,\;\;
z \equiv {x \over y}
\;\;.
\end{equation}
It is also convenient to introduce the boost-invariant (along the
$ \, k$-direction) transverse momentum
 $  {\tilde \bq} $
\begin{equation}
\label{qtilde}
  \, {\tilde \bq}
= \bq - z \, \bk \,  \;\;.
\end{equation}
Performing the Dirac and colour algebra we get
\begin{eqnarray}
\label{K01}
&~& \left(\frac{\bar p\!\!\!/}{2{\bar p}\cdot q} \right) _{\alpha\beta}
{\hat K}^{(0) \; \al \beta}_{\mu \nu } (q,k)
{ \frac{ k_{\perp}^{\mu} \, k_{\perp}^{\nu} }{\bk^2} } =
g_s^2
\,\left( \mu^2 \right)^{- \varepsilon}
\, { \delta^{a b} \over {N_c^2 - 1} } {\mbox {\rm Tr}} \left( t^a t^b \right)
\nonumber\\
&~& \cdot \; 2 \, \pi \, \delta_{+}\!\left( (k-q)^2 \right) \,
%\Theta \left( (k - q)_{0} \right) \,
%\Theta \left( Q^2 - | q^2 |  \right) \, \cdot
\, {z \over { 2 \, {\bar p} \cdot q}} \, {1 \over {({q^2})^2}} \,
{1 \over \bk^2} \, {\mbox {\rm Tr}} \left[  \,
\rlap/q \,
{\bar {\rlap/p}} \,
\rlap/q \,
{\rlap/k}_{\perp} \,
(\rlap/q - \rlap/k ) \,
{\rlap/k}_{\perp}
\right]
\nonumber\\
&~& =
16 \, \pi^2 \, \as
\,\left( \mu^2 \right)^{- \varepsilon}
\,  T_R \; { z^2 \over {1-z}} \; \Theta (1 - z)
\, {1 \over {({q^2})^2}} \, \delta\!\left( q^2 +
{ { {\tilde {\bq}}^2} \over {1-z}} + z \, \bk^2 \right)
\nonumber\\
&~& \cdot \left[
{ { {\tilde {\bq}}^2} \over {z (1-z)}} - { 4 \over \bk^2}
\left( \bk \cdot  {\tilde {\bq}} \right)^2
+ 4 \, (1 - 2 \, z)
\, \bk \cdot  {\tilde {\bq}} + 4 \, z \, (1 -z) \, \bk^2
\right] \;\;,
\end{eqnarray}
and,  inserting Eq.~(\ref{K01}) into Eq.~({\ref{Kdefapp}),
the azimuthal average over $  {\tilde {\bq}} $ and the integration over
$q^2$ can easily be performed, thus leading to
\begin{eqnarray}
\label{K0resapp}
&~& {\hat K}_{ q g}\!\left( z, {{\bk^2} \over  Q^2}, \as\!\left(
{Q^2 \over \mu^2}\right)^\varepsilon; \varepsilon \right) =
{\as \over {2 \, \pi}}
\, S_\varepsilon \,
{{e^{\varepsilon \, \psi(1)}} \over {\Gamma(1 + \varepsilon)}} \,
T_R \; z \; \Theta (1-z) \,
\int_{0}^\infty
{d {\tilde \bq}^2} \,
 \left( {{{\tilde \bq}^2} \over \mu^2} \right)^{ \varepsilon} \\
%\nonumber\\
&\cdot&
\Theta\!\left( Q^2 -
{ { {\tilde {\bq}}^2} \over {1-z}}
- z {\bk}^2 \right)
{{ {\tilde \bq}^2} \over
{\left[ {\tilde \bq}^2 + z \, (1-z) \, \bk^2 \right]^2 }}
 \left[ 1 -
{ {2 \, z \, (1 - z) } \over {1+ \varepsilon}}  +
4 \, z^2 \, (1-z)^2 \,
 { \bk^2 \over {{ \tilde \bq}^2}}
\right] \;\;. \nonumber
\end{eqnarray}
Eq.~(\ref{K0resapp}) is precisely the result in Eq.~(\ref{K0res}), expressed
in terms of the off-shell splitting function (\ref{K0res1}).

The $ \, {\tilde {\bq}}^2$-integration in Eq.~(\ref{K0resapp}) is not
elementary and provides a representation of the off-shell kernel $ \,
{\hat K}_{q g} \, $ in terms of hypergeometric functions. However, in order
to obtain the power series expansion
(\ref{quagresersol})
for the Green function
$\, { G}^{(0)}_{ q g } $, it is
more convenient to carry out first the $ \, \bk$-integration in
Eq.~(\ref{quagrektfaceps2}). Inserting the expansion (\ref{sersol}) into
Eq.~(\ref{quagrektfaceps2}), we get
\begin{eqnarray}
\label{quagresersolapp}
&~& {G}^{(0)}_{ q g , \, N}( \as ({Q^2/\mu^2})^\varepsilon,
\varepsilon ) =
 {\as \over {2 \, \pi}} \, T_R \, S_\varepsilon \,
{{e^{\varepsilon \, \psi(1)}} \over {\Gamma(1 + \varepsilon)}} \,
\left( {Q^2 \over \mu^2} \right)^{ \varepsilon} \,
 h_{q g , \, N } (\ga = 0; \varepsilon)
\nonumber\\
&\cdot& \left\{
1 + \sum_{k=1}^\infty
\, \left[ \abn
\, S_\varepsilon \,
{{e^{\varepsilon \, \psi(1)}} \over {\Gamma(1 + \varepsilon)}} \,
\left( {Q^2 \over \mu^2} \right)^\varepsilon
\, \right]^k \, {1 \over {k \, \varepsilon}} \, c_k (\varepsilon)
\; {{h_{q g , \, N } (\ga = k \, \varepsilon; \varepsilon) } \over
{h_{q g , \, N } (\ga = 0; \varepsilon) }}
\,  \right\} \;\;,
\end{eqnarray}
where the function
$ \, h_{q g , \, N } (\ga ; \varepsilon) \, $ is defined by the following
 $\, k_{\perp}$-transform of the off-shell kernel $\, {\hat K}_{q g} \,$
\begin{equation}
\label{hqgapp}
 {\as \over {2 \, \pi}}
\, T_R \, S_\varepsilon \,
{{e^{\varepsilon \, \psi(1)}} \over {\Gamma(1 + \varepsilon)}} \,
\left( {Q^2 \over \mu^2} \right)^{ \varepsilon} \,
h_{q g , \, N } (\ga; \varepsilon) \equiv \ga \, \int_0^\infty
\frac{d \bk^2}{\bk^2} \, \left( {\bk^2 \over Q^2} \right)^\ga \,
 {\hat K}_{ q g , \,N }\!\left( {{\bk^2} \over  Q^2}, \as\!\left(
{Q^2 \over \mu^2}\right)^\varepsilon; \varepsilon \right) \;\;.
\end{equation}
The evaluation of
$ \, h_{q g  } \, $ is straightforward. We insert Eq.~(\ref{K0resapp}) into
Eq.~(\ref{hqgapp}) and perform first the trivial integration over $ \, \bk^2
\,$ with $ \, {\tilde \bq}^2 / ( z (1-z) \bk^2 ) = \tau \,$ fixed. Then,
the integrations over $  \tau  $ and $ z$ decouple and can be carried out in
terms  of Euler gamma functions. Note also that the $ \, N \to 0 \,$ limit of
$ \, h_{q g, N} \, $ is regular (i.e.,
$\,  {\hat K}_{ q g } (z) \sim z $, modulo $ \, \ln z$-terms, for
$\, z \to 0 \, $ and any value of
$\,  {\bk^2} /  Q^2 $) . Therefore, to the logarithmic accuracy we are
interested in, we can limit ourselves to computing
$ \, h_{q g , \, N=0 } $. The final result is:
\begin{equation}
\label{hqgres}
 \, h_{q g , \, N=0 } (\ga; \varepsilon) =
 { { 4 + \varepsilon - 3 \, \ga} \over
 {\ga  + \varepsilon}} \,
{ { \Gamma (1 + \ga) \, \Gamma(1 - \ga) \,
\Gamma ( 1 + \varepsilon) \,
\Gamma( 2 + \, \varepsilon) } \over { \Gamma ( 1 + \varepsilon +
\ga) \, \Gamma
(4 + \varepsilon - \ga) } } \,
 \;\;\;.
\end{equation}
Using Eq.~(\ref{hqgres}), we recover Eqs.~(\ref{quagresersol}) and
(\ref{dneps}) by the identification
\begin{equation}
\label{ratiodc}
 \,{{ d_k (\varepsilon)} \over
 { c_k (\varepsilon) }} =
 {{ h_{q g , \, N=0 } (\ga = k \, \varepsilon; \varepsilon) } \over
 { h_{q g , \, N=0 } (\ga = 0 ; \varepsilon)} }
\;\;.
\end{equation}

As discussed in Sect.~4, we have not been able to use the power series
expansion (\ref{quagresersol}) for explicitly resumming all the
next-to-leading logarithmic corrections in the quark anomalous dimensions
$ \, \ga_{q g , \, N} $. In general, to this accuracy we can write
\begin{equation}
\label{gaqgapp}
  \ga_{q g , \, N } (\as) = {\as \over { 2 \pi}}\, T_R \, {2 \over 3} \,
 \left\{ 1 +
\sum_{k = 1}^\infty \, a_k \left( \abn \right)^k
\right\} \;\;,
\end{equation}
and the values of the coefficients
$ \, a_k \, $ for $ \, k \leq 5 \, $ are
given in Eq.~(\ref{qms}). The calculation of the higher-order coefficients
 is much more  cumbersome.
Here, we present only the result for the rational part of $ a_k $. In other
words, let us split $ a_k$ as follows
\begin{equation}
\label{ratirrat}
a_k = r_k + b_k \;\;,
\end{equation}
where $ b_k$ is an irrational number given in terms of powers of Riemann
zeta functions $ \zeta (n ) \, $  ($n \geq 3$) and $ r_k $ is the residual
rational contribution to $ a_k $. We find the following expression for
$ r_k $ in the $ \msbar $ scheme:
\begin{equation}
\label{gaqgrat}
r_k = { {2^{k - 2}} \over {k!}} \, \left[ 3 + \left( { 1\over 3} \right)^k
\right] \;\;,
\end{equation}
or, equivalently,
\begin{equation}
\label{gaqgrat1}
1 + \sum_{k = 1}^{\infty} \, r_k  \left( \abn \right)^k = {3 \over 4} \,
\left[ \exp \left( 2 \, \abn\right) + {1 \over 3} \, \exp \left( {2 \over 3}
\, \abn\right) \right] \;\;.
\end{equation}
One can easily check that Eq.~(\ref{gaqgrat}) reproduces the values of $ r_k$
in Eq.~(\ref{qms}), i.e. $
r_1 = 5/3 , \,
r_2 = 14/9 , \,
r_3 = 82/81 , \,
r_4 = 122/243 , \,
r_5 = 146/729 $.
The derivation of the result (\ref{gaqgrat}) is left as an exercise for the
reader.

\newpage

{\large \bf References}
\begin{enumerate}
%\normalsize

\item \label{EPS}
      S.\ Catani, preprint DFF 194/11/93, plenary talk at the EPS Conference on
High Energy Physics, Marseille, July 1993 (to appear in the Proceedings), and
references therein.

\item \label{FAD}
      E.A.\ Kuraev, L.N.\ Lipatov and V.A.\ Fadin, Phys.
      Lett. B60 (1975) 50.

\item \label{Ciaf}
      M.\ Cia\-fa\-lo\-ni, \np{296}{249}{87}.

\item \label{GDTK}
      L.V.\ Gribov, Yu.L.\ Dokshitzer, S.I.\ Troyan and V.A.\ Khoze, Sov. Phys.
      JETP 67 (1988) 1303.

\item \label{CFM}
      S.\ Catani, F.\ Fiorani and G.\ Marchesini, \pl{234}{339}{90},
      \np{336}{18}{90}.

\item \label{Muel}
      A. H. Mueller, preprint CU-TP-609 (1993).

\item \label{MW}
      G.\ Marchesini and B.R. Webber, \np{349}{617}{91}, \np{386}{215}{92}.

\item \label{CCH}
      S. Catani, M. Ciafaloni and F. Hautmann, Phys. Lett. B242
      (1990) 97, Nucl. Phys. B366 (1991) 135.

\item \label{CE}
      J.C. Collins and R.K. Ellis, Nucl. Phys. B360 (1991) 3.

\item \label{LRSS}
      E.M.\ Levin, M.G.\ Ryskin, Yu.M.\ Shabel'skii and A.G.\ Shuvaev, Sov. J.
      Nucl. Phys. 53 (1991) 657.

\item \label{CSS}
      J.C.\ Collins, D.E.\ Soper and G.\ Sterman, in {\it Perturbative Quantum
      Chromodynamics}, ed. A.H. Mueller (World Scientific, Singapore, 1989)
      and references therein.

\item \label{BFKL}
      L.N.\ Lipatov, Sov. J. Nucl. Phys. 23 (1976) 338; E.A.\ Kuraev,
      L.N.\ Lipatov and V.S.\ Fadin, Sov. Phys. JETP  45 (1977) 199; Ya.\
      Balitskii and L.N.\ Lipatov, Sov. J. Nucl. Phys. 28 (1978) 822.

\item \label{uni1}
      L.V.\ Gribov, E.M.\ Levin and M.G.\ Ryskin, \prep{100}{1}{83};
      E.M. Levin and M.G. Ryskin, Phys. Rep. 189 (1990) 268.

\item \label{uni2}
      A.H. Mueller and J. Qiu, Nucl. Phys. B268 (1986) 427;
      A.H. Mueller, Nucl. Phys. B335 (1990) 115.

\item \label{uni3}
      J.\ Bartels, preprint DESY-91-074, \pl{298}{204}{93}, preprint
      DESY-93-028 .

\item \label{uni4}
      E.M.\ Levin, M.G.\ Ryskin and A.G.\ Shuvaev,
      Nucl. Phys. B387 (1992) 589;
      E.\ Laenen, E.M.\ Levin and A.G.\ Shuvaev, preprint
      Fermilab-PUB-93/243-T.

\item \label{FL}
      V.S.\ Fadin and L.N.\ Lipatov, \np{406}{259}{93}.

\item \label{CK}
      J.C.\ Collins and J.\ Kwiecinski, Nucl. Phys. B316 (1989) 307;
      J. Kwiecinski, A. D. Martin and P.J. Sutton,
      Phys. Rev. D44  (1991) 2640.

\item \label{Dur}
       A. J. Askew, J. Kwiecinski, A. D. Martin and P. J. Sutton,
       Phys. Rev. D47 (1993) 3775, preprint DTP-93-28.

\item \label{EKL}
      R.K. Ellis, Z. Kunszt and E. M. Levin, preprint Fermilab-PUB-93/350-T.

\item \label{HEF}
      S.\ Catani, M.\ Ciafaloni and F.\ Hautmann, \pl{307}{147}{93}.

\item \label{QAD}
      S.\ Catani and F.\ Hautmann, \pl{315}{157}{93}.

\item \label{Regge}
      See, for instance, P.D.B.\ Collins, {\it An introduction to Regge
      theory and high energy scattering} (Cambridge University Press,
Cambridge,
      1977); {\it Regge theory of low-$p_T$ hadronic interactions}, ed. L.
      Caneschi (North-Holland, Amsterdam, 1989).

\item \label{old}
      D. Amati, R. Petronzio and G. Veneziano, Nucl. Phys. B140 (1978)
      54, Nucl. Phys. B146 (1978) 29;
      R.K. Ellis, H. Georgi, M. Machacek, H. D. Politzer and
      G. G. Ross, Phys. Lett. 78B (1978) 281;
      C.T.\ Sachrajda, \pl{73}{281}{78}, \pl{76}{100}{78};
      S. Libby and G. Sterman, Phys. Rev. D18 (1978) 3252, 4737;
      A. H. Mueller, Phys. Rev. D18 (1978) 3705;
      A. V. Efremov and A. V. Radyushkin,
      Theor. Math. Phys. 44 (1981) 573, 664, 774.

\item \label{CFP}
      G.\ Curci, W.\ Furmanski and R.\ Petronzio,  \np{175}{27}{80}.

\item \label{GLAP}
      V.N.\ Gribov and L.N.\ Lipatov, Sov. J. Nucl. Phys. 15 (1972) 438,
      675; G.\ Altarelli and G.\ Parisi,
      \np{126}{298}{77}; Yu.L.\ Dokshitzer, Sov. Phys. JETP  46 (1977) 641.

\item \label{FP}
      W.\ Furmanski and R.\ Petronzio, \pl{97}{437}{80}.

\item \label{Floratos}
      E.G.\ Floratos, D.A.\ Ross and C.T.\ Sachrajda, \np{129}{66}{77}
      (E \np{139}{545}{78}), \np{152}{493}{79}; A.\ Gonzalez-Arroyo, C.\ Lopez
      and F.J.\ Yndurain, \np{153}{161}{79}; A.\ Gonzalez-Arroyo and C.\ Lopez,
      \np{166}{429}{80};
      E.G.\ Floratos, P.\ Lacaze and C.\ Kounnas,
      \pl{98}{89}{81}, 225.

\item \label{Stirl}
      See, for instance, W.J.\ Stirling, in Proceedings of the Aachen
      Conference {\it QCD -- 20 years later}, eds. P.M. Zerwas and H.A.
      Kastrup (World Scientific, Singapore, 1993), pag.~387.

\item \label{NDIS}
      E.B.\ Zijlstra and W.L.\ van Neerven, \np{383}{525}{92}.

\item \label{DY}
      T. Matsuura, R. Hamberg and W.L.\ van Neerven, \np{359}{343}{91};
      W.L.\ van Neerven and E.B.\ Zijlstra, \np{382}{11}{92}.

\item \label{HERA}
      S.\ Catani, M.\ Ciafaloni and F.\ Hautmann, in Proceedings of the
      HERA Workshop, eds. W.\ Buchm\"{u}ller and G.\ Ingelman (DESY Hamburg,
      1991), pag.~690,
%\item \label{teupitz}
%      S.\ Catani, M.\ Ciafaloni and F.\ Hautmann,
      Nucl. Phys. B (Proc. Supp.)
      29A (1992) 182.

\item \label{2pi}
      R.K. Ellis, H. Georgi, M. Machacek, H. D. Politzer and
      G. G. Ross, Nucl. Phys. B152 (1979) 285.

\item \label{CFMO}
      S.\ Catani, F.\ Fiorani, G.\ Marchesini
      and G.\ Oriani, \np{361}{645}{91}.

\item \label{BCM}
      A. Bassetto, M. Cia\-fa\-lo\-ni and G. Mar\-che\-si\-ni,
      Phys. Rep. 100 (1983) 201.

\item \label{jaro}
       T. Jaroszewicz, Phys. Lett. B116 (1982) 291.

\item \label{DATA}
      ZEUS Coll., M.\ Derrick et al., \pl{316}{412}{93};
      H1 Coll., I.\ Abt et al., \np{407}{515}{93}, \pl{321}{161}{94}.

\item \label{DR}
      W.\ Siegel, \pl{84}{193}{79}; I.\ Antoniadis and E.G.\ Floratos,
      \np{191}{217}{81}.

\item \label{CDIS}
      E.B.\ Zijlstra and W.L.\ van Neerven, \pl{383}{377}{92}.

\item \label{Sanchez}
      D.I.\ Kazakov, A.V.\ Kotikov, G.\ Parente, O.A.\ Sampayo and J.\
      S\'{a}nchez-Guill\'{e}n, \prl{65}{1535}{90};
      J.\ S\'{a}nchez-Guill\'{e}n, J.L.\ Miramontes, M.\ Miramontes,
      G.\ Parente and O.A.\ Sampayo, \np{353}{337}{91}.

\item \label{AEM}
      G.\ Altarelli, R.K.\  Ellis and G.\ Martinelli,
      \np{157}{461}{79}.

\end{enumerate}

\newpage

\begin{figcap}

\item \label{1} Factorization
of the physical cross section $ F$ in terms of
 partonic cross sections
($F^{(0)}$, $F^{(0)}_4$, $\dots$)
and parton distributions
(${\tilde f}^{(0)}$, ${\tilde f}^{(0)}_4$, $\dots$). In the case  of
on-shell partons $(p_i^2=0)$, the first term on the r.h.s. and those
in parenthesis represent respectively the
leading- and higher-twist contributions.

\item \label{2} Expansion of the  partonic cross section $ F^{(0)}$
in two-particle irreducible (2PI) kernels.

\item \label{3}
Action of the collinear projector $ {\cal P}_C $ on
the spin indices  of two kernels $A$ and $B$
in the cases of an intermediate (a) quark  and (b) gluon state.

\item \label{4}  Expansion in two-gluon irreducible (2GI) kernels
at high energy for
 (a) the   partonic cross section $ F^{(0)} $
and (b) the (singlet) quark  Green functions $ {\cal G}_{q a} $,
$ {\cal G}_{{\bar q} a} $.

\item \label{5}  Quark-antiquark contribution to the lowest-order absorptive
part $  A_{\mu \nu}  $ of the scattering amplitude $  \ga \, g \to \ga
\, g$.

\item \label{6}  The BFKL characteristic function $ \chi ( \gamma) $ for
$ 0 < \gamma < 1$. $  \ga_N  $ is the BFKL anomalous dimension.

\item \label{7} (a) High-energy factorization of the gluon $\to$ quark
Green function  $ {\cal G}_{q g} $
and (b) the corresponding off-shell kernel $ {\hat K}_{q g} $.
\end{figcap}

\end{document}